\documentclass[review,3p]{elsarticle}

\usepackage[T1]{fontenc}
\usepackage[utf8]{inputenc}
\usepackage[english]{babel}
\usepackage{multirow}
\addto\captionsenglish{}

\usepackage{graphicx}
\usepackage[caption=false]{subfig}
\usepackage[dvipsnames]{xcolor}
\usepackage[colorlinks]{hyperref}

\usepackage{amsthm}
\usepackage{amssymb}
\usepackage{amsmath}
\usepackage{mathtools}
\usepackage{textcomp}
\usepackage{siunitx}

\usepackage{lineno}

\usepackage{adjustbox}
\usepackage{url}
\usepackage{longtable}

\makeatletter
\providecommand{\doi}[1]{%
  \begingroup
    \let\bibinfo\@secondoftwo
    \urlstyle{rm}%
    \href{http://dx.doi.org/#1}{%
      doi:\discretionary{}{}{}%
      \nolinkurl{#1}%
    }%
  \endgroup
}
\makeatother

\usepackage{tikz,tikzscale,pgfplots,pgfplotstable}
\usetikzlibrary{shapes,arrows,calc}
\pgfplotsset{compat=1.14}

\graphicspath{{figures/}}

\renewcommand{\u}{\mathbf{u}}

\newcommand{\de}{\partial}

\newcommand{\pd}[1]{\partial_{#1}}
\newcommand{\dex}{\partial_x}
\newcommand{\dey}{\partial_y}

\renewcommand{\arraystretch}{1.5}
\newcommand{\BM}{\textsc{Basement}}
\newcommand{\mytitle}{\BM\ v3: a modular freeware for river process modelling over multiple computational backends}
\newcommand{\rom}[1]{\romannumeral #1)}
\journal{Environmental Modelling \& Software}

\begin{document}

\begin{frontmatter}



\title{\mytitle}


\author[eth,eawag]{Davide Vanzo\corref{cor1}}
\cortext[cor1]{corresponding author: vanzo@vaw.baug.ethz.ch}
\author[axpo]{Samuel Peter}
\author[tk]{Lukas Vonwiller}
\author[eth]{Matthias Buergler}
\author[sis]{Manuel Weberndorfer}
\author[unitn]{Annunziato Siviglia}
\author[eth]{Daniel Conde}
\author[eth]{David F. Vetsch}

\address[eth]{Laboratory of Hydraulics, Hydrology and Glaciology. ETH, Swiss Federal Institute of Technology, Z\"urich, Switzerland.}
\address[eawag]{Dept. of Surface Waters - Research and Management. Eawag, Swiss Federal Institute of Aquatic Science and Technology, Kastanienbaum, Switzerland.}
\address[unitn]{Dept. Civil, Environmental and Mechanical Engineering. University of Trento, Trento, Italy.}
\address[sis]{ID Scientific IT Services. ETH, Swiss Federal Institute of Technology, Z\"urich, Switzerland.}
\address[axpo]{Axpo Group, Baden, Switzerland.}
\address[tk]{TK CONSULT AG , Z\"urich, Switzerland.}


\begin{abstract}
Modelling river physical processes is of critical importance for flood protection, river management and restoration of riverine environments. Developments in algorithms and computational power have led to a wider spread of river simulation tools. However, the use of two-dimensional models can still be hindered by complexity in the setup and the high computational costs. Here we present the freeware \BM\ version 3, a flexible tool for two-dimensional river simulations that bundles solvers for hydrodynamic, morphodynamic and scalar advection-diffusion processes. \BM\ leverages different computational platforms (multi-core CPUs and graphics processing units GPUs) to enable the simulation of large domains and long-term river processes. The adoption of a fully costless worflow and a light GUI facilitate its broad utilization. We test its robustness and efficiency in a selection of benchmarks. Results confirm that \BM\ could be an efficient and versatile tool for research, engineering practice and education in river modelling.
\end{abstract}

\begin{keyword}
GPU-CUDA \sep river modelling \sep unstructured grid \sep shallow water \sep sediment transport \sep pollutant transport



\end{keyword}

\end{frontmatter}

\section*{Software availability}
\begin{description}
    \item[Name of software:] \BM, version 3 (v3)
    \item[Website:] www.basement.ethz.ch
    \item[E-mail:] basement@ethz.ch
    \item[Developer:] Numerical Modelling division at the Laboratory of Hydraulics, Hydrology and Glaziology (VAW), ETH Z\"urich
    \item[Language:] C++, C, CUDA
    \item[Interface:] graphical user interface (GUI), command-line interface (CLI)
    \item[Hardware:] CPUs, CUDA-enabled GPUs (optional)
    \item[OS:] Windows, Linux (Ubuntu)
    \item[Availability:] Freeware
\end{description}


\section{Introduction}
In the last decades the usage of numerical tools widely spread in several subjects of the environmental sciences. River science \cite[\textit{sensu}][]{Gilvear2016} is no exception in this trend, with a number of tools been developed to address variegate research questions \cite[e.g.][]{Brewer_2018,Shimizu2019}. Modelled river physical processes span from flood simulation, hydraulic and sediment dynamics, pollutant and temperature transport, to vegetation and flow interactions, just to mention a few \cite[e.g.][]{Crosato2011, Sharma2012, Williams2016, Dugdale2017, Teng2017}. Such river processes occur at different spatial and temporal scales, hence influencing the development and choice of suitable modelling tools.  

Advances in computational power, numerical algorithms and optimization routines that occurred in the last decades allowed for the spread of more and more sophisticated numerical tools. In the context of river science, two-dimensional depth-averaged (hereinafter 2D) models are nowadays of common use in research and engineering practice. This is particularly true for some applications such as flood modelling and river morphodynamics \cite[e.g.][]{Shimizu2019,zischg2018}. The increasing usage of 2D river models is also closely bonded with the growing availability of high-resolution river datasets. In particular, advances in LiDAR, UAV-Photography and others remote-sensed survey technologies enable river topographic scans at an unprecedented level of detail \cite[e.g.][]{Marcus_2010, thomas2016}.

The increased computational capabilities and refined datasets open the gates for near-census \cite[\textit{sensu}][]{Pasternack2011} numerical modelling of several river processes. Indeed, 2D river models have the capability to simulate fine spatial (centimeters to meters) and temporal scales (seconds to days). At such scales, relevant hydro-morphodynamic processes (e.g. bar formation) and also ecohydraulic processes (e.g. habitat dynamics) can hence be modelled \cite[e.g.][]{Maddock_1999, Siviglia2013, Wyrick2014, guan2016}. Nevertheless, 2D river models can still be computationally demanding, with simulations lasting several days. This is particularly true when complex physical processes, such as morphodynamics, are accounted for \cite{Siviglia2016}. Moreover, large-scale or near-census applications (i.e. with millions of computational cells) and/or long-term simulations (i.e. years) all concur to increase overall computational costs. Such drawbacks particularly apply for the investigation of highly unsteady river processes such as artificial or natural flood waves, where explicit numerical schemes are preferable. In such cases the overall computational time scales exponentially with the number of computational cells due to stability constraints \cite[e.g.][]{Toro2001}.

Increasing the efficiency and the computational performance of river models represents yet a challenge. Pitfalls arise with the number of computational cells, but also with the inherent complexity of 2D models. For example, challenges are to be found in the setup of the computational domain \cite{Nahorniak2018} but also in the definition of particular boundary conditions \cite{Costabile2015b, dazzi2020}. Increasing the computational performance is also sought by developing alternative numerical solution strategies for the underlying physical governing equations. Examples are variegate, spanning from the adoption of a local timestep for the numerical integration \cite[e.g.][]{Sanders2008, Dazzi2018}, the automatic adaptation of the computational mesh \cite[e.g.][]{Powell1993}, the use of acceleration factors for the hydro-morphodynamic problem \cite[e.g][]{Carraro2018, Morgan2020}, to the reformulation of the governing mathematical equations for water quality simulations \cite[e.g.][]{Vanzo2016}, to mention a few.

Parallel computing solutions are the most popular strategies to reduce computational time. They historically benefit from the continuous improvements of computational performances of both single CPUs and clusters. The general aim of parallelization techniques is to split the total computational load into tasks that can be executed simultaneously by different computational units \cite[e.g.][]{Afzal2016}. The use of the Graphics Processing Unit (GPU) as a general-purpose computational resource developed rapidly in the last decade \cite{Owens2008}. For many parallelizable workloads, offloading work to GPUs is a relatively cheap and efficient high-performance computing strategy that is also easily upgradable in standard desktop workstations.

By means of GPU parallelization, numerical models can potentially be accelerated by an order of magnitude and more. The efficiency of such a parallelization depends on the data exchange between the main memory and the processors, with a complex memory hierarchy and bandwidth bottlenecks \cite{Mudalige2012a}. These low-level constraints can limit computational speedup and depend on the model data/memory handling, the underlying model complexity and the nature of the governing equations to be solved. In applications such as 2D river models the type of computational mesh, i.e. structured or unstructured, has a significant influence on the final computational speedup.

In the last decade, river simulation models have benefited from GPU parallelization. Specific and \textit{ad-hoc} implementations of GPU-based models for 2D hydrodynamic \cite[e.g.][]{Brodtkorb2012, Smith2013, Vacondio_2014, Vacondio2017}, and occasionally morphodynamic simulations \cite[e.g.][]{Hou2020} have become available. The vast majority of these models are based on structured grids which allow for an easier implementation and for relatively higher computational speedups. This is due to the fact that, for structured meshes, the data structure is inherently simpler, which reduces the need for mappings and indirections. To the best of the Authors' knowledge, few hydrodynamic models implement GPU-acceleration on unstructured meshes \cite{Castro2011,Lacasta2014b,Lacasta2015b,Petaccia2016}, with very limited \textit{ad-hoc} implementations for transient flows morphodynamics \cite[e.g.][]{Juez2016}.

Bundled river modelling software that supports GPU acceleration is available for commercial use (e.g. RiverFlow2D (hydronia.com/riverflow2d), TUFLOW (tuflow.com), but costless ones are still few \cite[see][]{GarciaFeal2018}. An increase in availability of freeware GPU-based river models would be beneficial for environmental modelers in academic research and education, but also in consultancy and engineering offices.  

In this paper we introduce the \BM\ software (version 3), a freeware application developed at the Laboratory of Hydraulics, Hydrology and Glaciology of ETH Z\"urich. The software can simulate two-dimensional hydrodynamic, morphodynamic, and scalar advection-diffusion processes of scientific and practical interest. It can seamlessly run on GPU-enabled workstations, as well as on more standard multi-core CPUs. This flexibility in the choice of the computational backend is achieved by integrating the OP2 framework \cite{Mudalige2012a,Regulyb,Giles2012}. This framework provides an additional abstract layer for the acceleration of numerical models on unstructured computational meshes, and has been successfully implemented in similar modelling context \citep{reguly2018}.
The obtained parallelization performance alleviates the computational limitations when simulating high resolution (or large) computational domains and/or long term processes \citep[e.g.][]{giles2020}. This is particularly relevant when aiming at the calibration \citep{beckers2020} or at the uncertainty evaluation \citep{jung2015, thomas2016} of deterministic models. As proof of concept, a flood wave uncertainty propagation analysis with \BM\ has been proposed in \citep{Peter2017}.

In the current version \BM\ is available for both Windows and Linux-based (Ubuntu) environments. It is provided with a Command Line Interface (CLI) to easily perform batch simulations, but also with a light Graphical User Interface (GUI). The \BM\ software aims to enable a broad range of potential users to skilfully simulate river processes in the domain of river engineering and research on state-of-the-art computational hardware. Moreover, with accompanying scholar programs and extended documentation material and tutorials, the software is designed to be a valuable didactic tool for engineering and river science students. 

The paper is structured as follows: \S\ref{sec:application_context} provides the software application context that justifies the adopted mathematical and numerical strategies. \S\ref{sec:math_models} to \S\ref{sec:IC_BC} report the mathematical basis, the numerical strategies and main features of the basic modules of \BM. The software design, the modelling workflow and the parallelization solutions are presented in \S\ref{sec:software_design}. A selection of benchmarks are reported in \S\ref{sec:testcases}, whilst conclusions and outlooks are drawn in \S\ref{sec:conclusions}.

\section{Application context}\label{sec:application_context}
One of the main goals of the novel software design of version 3 is the capability to tackle river processes at different spatial and temporal scales. For example, \BM\ can be used to simulate large scale (i.e. basin scale) flood propagation, but also reach scale morphodynamic processes such as formation and evolution of fluvial bars. Moreover, it can be applied together with high-resolution topographies (in the order of centimeters) to simulate ecohydraulic processes at different ecological scales (e.g. habitat modelling). This range of application possibilities is enabled by specific characteristics of the software. In particular:
\renewcommand{\descriptionlabel}[1]{\hspace{\labelsep}\textit{#1}}
\begin{description}
    \item[unsteady and transitional flows:] \BM\ can deal with strongly unsteady flows and different flow regimes (sub- and super-critical). For this reason, \BM\ is particularly suitable for simulations of river flows in Alpine contexts, the propagation of natural flood waves as well as hydropeaking events. This is ensured by the adoption of a robust and accurate shock-capturing explicit solver for the hydrodynamic problem (\S\ref{sec:numerics_hydro});
    \item[accurate front propagation:] it is possible to simulate extreme events such as dam-break induced floods, but also ecologically-relevant processes such as the wetting-drying of riparian areas and in-channel morphologies due to artificial flow alterations. This is achieved by an implemented shock-wave capturing numerical scheme complemented with a robust treatment of wet-dry interfaces (\S\ref{sec:numerics_hydro});
    \item[complex river topographies:] the use of unstructured grid for the computational domain discretization enables for an accurate description of complex river morphologies and riverrine structures (\S\ref{sec:numerics_discretization}). The adoption of an unstructured mesh also reduces the strong anisotropy of structured meshes, which can be crucial for particular applications. 
    \item[large problems:] the software adopts a parallelization strategy tailored to the acceleration of problems on unstructured meshes (\S\ref{sec:parallelization_strategy}). Moreover, \BM\ simulations can efficiently be executed on different computational backends. Those backends inlude GPU cards, therefore allowing for the simulation of large domains (millions of computational cells) on standard workstations, having a limited cost.
    \item[multiple river processes:] the software is designed in a modular way, so different river processes such as hydrodynamics, sediment or advection-diffusion of a scalar (e.g. a non-reactive pollutant) can be simulated by activating specific modules at setup time (\S\ref{sec:software_design}). Different types of boundary conditions (\S\ref{sec:IC_BC}) and closure relationships are available to simulated, for example, simple hydraulic structures (e.g. weirs) or flow inputs/outputs (e.g. water intakes).
    The modular design (\S\ref{sec:software_design}) allows to retain good parallelization performances in the simulation of different river processes, as shown in  (\S\ref{sec:performance}).  
\end{description}

The basic modules available in \BM\ are \rom{1} hydrodynamics, \rom{2} morphodynamics and \rom{3} advection-diffusion of scalar quantities. Each module is composed by different sets of hyperbolic equations describing the conservation and evolution of the water flow (hydrodynamics), the fluvial sediment (morphodynamics) and the concentration of passive solutes (scalar advection-diffusion). The governing equations represent a so-called Initial-Boundary Value Problem \citep{Toro2001}, where process-specific initial and boundary conditions are required to be set. The following Sections presents the main governing equations and closure relationships (\S\ref{sec:math_models}), the numerical strategies (\S\ref{sec:num_sol}) and finally the initial and boundary conditions (\S\ref{sec:IC_BC}) for the three basic modules. The main module features are also listed in Table~1 of the Supplementary Material.

\section{Mathematical formulation}\label{sec:math_models}

\subsection{Hydrodynamics}
The hydrodynamic module solves the so-called \textit{shallow water} equations (hereinafter SWE) \cite[e.g.][]{Toro2001}. The two-dimensional SWE are of practical interest with regard to water flows with a free surface under the influence of gravity.

\begin{figure}[tbp]
    \centering
    \def\svgwidth{0.5\columnwidth}
    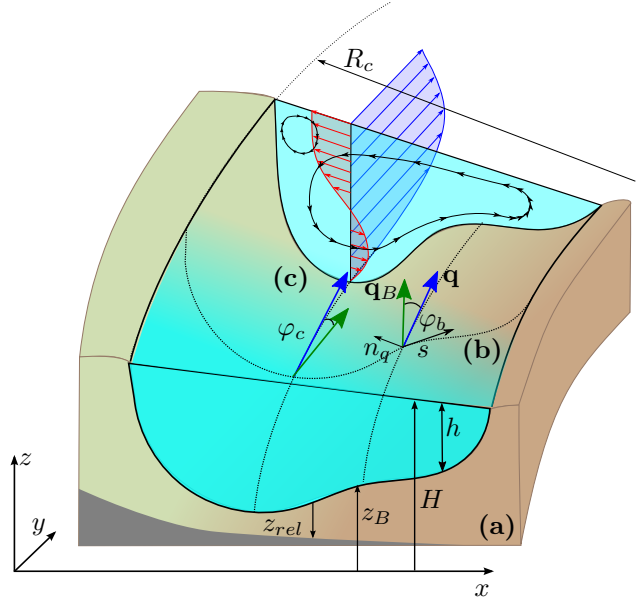
    \caption{\textbf{Notation of scalar and vectorial quantities.} \textbf{(a)} Reference system $(x,y,z)$ with water surface elevation $H$, water depth $h$, bed elevation $z_B$ and non-erodible fixed bed depth $z_{rel}$. \textbf{(b)} Bed load transport ($\mathbf{q_B}$) deviation angle $\varphi_b$ from the flow direction $\mathbf{q}$ due to gravitational effects caused by the local lateral slope $s$. \textbf{(c)} Bed load transport ($\mathbf{q_B}$) deviation angle $\varphi_c$ from the flow direction $\mathbf{q}$ due to the spiral flow motion caused by the curvature of radius $Rc$.}
    \label{fig:variables_effects}
\end{figure}

Considering a Cartesian reference system $(x,y,z)$ where the $z$ axis is vertical and the $(x,y)$ plane is horizontal (Fig.~\ref{fig:variables_effects}a), the system of governing equations can be expressed as:
\begin{equation}\label{eq:Hyd_gov_equations}
\begin{cases}
\pd{t} H +\dex q_x + \dey q_y =S_h \\
\pd{t} q_x + \dex \left( \dfrac{q_x^2}{h}+\dfrac{1}{2}gH^2 -gH z_B \right)+ \dey \left( \dfrac{q_xq_y}{h} \right) =  -gH \dex z_B -ghS_{fx} \\
\pd{t} q_y + \dex \left( \dfrac{q_xq_y}{h} \right)+ \dey \left( \dfrac{q_y^2}{h}+\dfrac{1}{2}gH^2 -gHz_B\right) = -gH \dey z_B -ghS_{fy}\\
\end{cases}
\end{equation}
where the system unknowns are the water surface elevation $H$ [m], and the two directional components of ${\mathbf q}=(q_x, q_y)$ [\si{m^2/s}], representing the flow discharge per unit width. With $z_B$ [m] we indicate the bottom elevation, whilst  $h=(H-z_B)$ [m] is the water depth,  and $g$ [\si{m/s^{2}}] the acceleration due to gravity. Note that the depth-averaged velocity vector can be consequently expressed as ${\mathbf u}=(u, v)=(q_x/h,q_y/h)$ [\si{m/s}]. Finally $S_{fx}$ and $S_{fy}$ [-] represent the dimensionless friction terms in $x$ and $y$ direction, whilst $S_h$ [\si{m/s}] represents potential external contribution/subtraction of flow discharge to the mass conservation equation. 

\subsubsection{Hydrodynamic closure relationships}\label{sec:hyd_closure_relations}
To solve the system \eqref{eq:Hyd_gov_equations}, closure relationships for the friction terms $S_{fx}, S_{fy}$ and the contribution of external inflow/outflow discharge $S_h$ must be provided. 

\paragraph{Friction terms}
Under the hypothesis of turbulent flow, hence under the assumption that the energy line slope is proportional to the square of the flow velocity, the friction terms $S_{fx}, S_{fy}$ can be written as:
\begin{equation}\label{eq:friction_closure}
S_{fx}=\frac{u \lVert{\mathbf u}\rVert}{ghc_f^2} \; ; \; S_{fy}=\frac{v \lVert{\mathbf u}\rVert}{ghc_f^2},
\end{equation}
where $c_f$ is the dimensionless friction coefficient. Several formulae are available for $c_f$. \BM\ implements four well know formulations of power or logarithmic type, given in Table~\ref{tab:friction_closure}.

\begin{table}[tbp]
    \centering
    \caption{\textbf{Friction closure relationships for the hydrodynamic problem}. Formulations for the dimensionless friction coefficient $c_f$; for both Ch\'{e}zy and Bezzola entries, $d_{90}$ is the 90\textsuperscript{th} percentile of the sediment grain size distribution.}\label{tab:friction_closure}
     \noindent\adjustbox{max width=\textwidth}{
    \begin{tabular}{lllll}\hline 
        \textbf{Closure} & \textbf{Expression} & \textbf{Parameters} & \textbf{Range} & \textbf{Ref}\\
        \hline
        Strickler & $c_f=k_{str}h^{1/6}/\sqrt{g}$ & $k_{str}$ [\si{m^{1/3}s^{-1}}] & 7-40 & \citep{armanini2018}\\\hline
        Manning & $c_f=h^{1/6}/(n\sqrt{g})$ & $n$ [\si{m^{-1/3}s}]  & 0.025-0.143 & \citep{armanini2018}\\\hline
        \multirow{2}{*}{Ch\'{e}zy} &
        \multirow{2}{*}{$\left.
        \begin{array}{lccl} 
        c_f=6.25 + \log{\left(h/K_{s}\right)} & \quad \text{for} & \quad & h > K_{s}\\ c_f=6.25                            & \quad \text{for} & \quad & h \leq K_{s}\end{array}\right.$
        }& \multirow{2}{*}{$K_s$ [m]} & $K_s=n_k d_{90}$, & \citep{graf1966}\\
        &&&with $n_k=2\div3$&\\\hline
        Bezzola & 
        $\left.
        \begin{array}{lll}
        c_f=2.5\sqrt{1-y_{R}/h}\; \ln{\left(10.9h/y_{R}\right)}, & \quad \text{for} & \quad h/y_R > 2 \\
        c_f=1.25\sqrt{h/y_R}\; \ln{\left(10.9h/y_{R}\right)}, & \quad \text{for} & \quad 0.5 \leq h/y_R \leq 2 \\
        c_f=1.5, & \quad \text{for} & \quad h/y_R < 0.5
        \end{array}
        \right.$
        & $y_R$ [m] & $y_R = n d_{90}$, $n \approx 1$ & \cite{Bezzola2002}\\\hline
    \end{tabular}}
\end{table}

\paragraph{External inflow/outflow discharge}
The term $S_h$ [\si{m/s}] is used to represent additional sources of water like rainfall and springs, or water abstraction (sink), and can be defined over subsets of the computational domain. The external water source can be provided by the user as total discharge [\si{m^3/s}] or as intensity [\si{mm\per\hour}], per squared meter. Different behaviour can be imposed for each external source/sink area:

\begin{itemize}
    \item Exact (source/sink): the exact given water volume is added (source) or extracted (sink) from the surface. This is the only option for water addition. In case of water abstraction, the simulation might end abruptly if the available water volume is smaller than the volume prescribed for subtraction. This option allows to have the full control on the water entering/leaving the computational domain, and is useful to simulate e.g. managed hydraulic structures, such as regulated water intakes.     
    \item Available (sink): the given water volume to extract is limited by the available water volume in the single element (i.e. computational cell). With this abstraction option, the simulation proceeds with no interruptions because the water conservation is ensured. This option is useful to simulate particular unmanaged hydraulic structures, such as diversion spillways.
    \item Infinity (sink): all available water will be abstracted from the computational domain.
\end{itemize}

\subsection{Morphodynamics}
The basic morphodynamic module solves the so-called Exner equation \cite{Exner1925}. It describes the bed evolution due to erosion or deposition, which results in the elevation change of the actual bed level $z_B$. Assuming the same coordinate reference of Fig.~\ref{fig:variables_effects}a, it reads
\begin{equation}\label{eq:exner}
(1-p)\pd{t} z_B + \dex q_{B_x}+ \dey q_{B_y} =S_{b},
\end{equation}
where $p$ [-] is the bed sediment porosity, assumed constant in space and time, $S_{b}$ [\si{m/s}] is an external source term specifying local inputs or outputs of sediment material (e.g. slope collapse or excavation) and ${\mathbf q_B} = (q_{B_x},q_{B_y})$ [\si{m^2/s}] is the specific sediment transport flux.

The morphodynamic module in its basic form accounts only for sediment transport occurring in the form of bed-load or total-load \citep{armanini2018,parkerEbook}. The simulation of sediment transport as suspended load is delegated to a specific module, planned in future versions of the software. 

\subsubsection{Morphodynamic closure relationships}\label{sec:morpho_closures}
Two closure relationships are needed to numerically solve the problem of \eqref{eq:exner}: a sediment transport formula and the external source/sink of sediments.

\paragraph{Sediment transport formula}
\BM\ implements three different types of sediment transport formulae as given in Table~\ref{tab:sedtransport_closure}. The first two, Meyer-Peter and M\"{u}ller like (MPM-like), and Grass like (Grass-like) expressions are adequate to simulate bed-load dominated sediment transport condition. The Engelund and Hansen formulation allows for the estimation of total sediment transport (i.e. suspended and bed-load), whilst Smart and J\"{a}ggi account for bedload transport in steep channels.

The expressions and the typical parameter values to calculate the specific sediment transport magnitude $\lVert{\mathbf q_{B}}\rVert$ are given in Table~\ref{tab:sedtransport_closure}. In the MPM-like formulation, $\theta$ is the dimensionless bed shear stress (i.e. Shields parameter \citep{armanini2018}), $\theta_{cr}$ is the critical dimensionless bed shear stress, $d_m$ is the representative grain diameter, $s=\rho_s/\rho$ is the relative density of the sediment with respect to water. The coefficients $\alpha$, $m$ and the critical threshold $\theta_{cr}$ can be assigned by the user or adopted from literature (see Table~\ref{tab:sedtransport_closure}).
The Grass-like model proposes a simple bedload transport formula, where $\lVert{\mathbf q_{B}}\rVert$ is a function of the flow velocity magnitude, with $u_{cr}$ as critical threshold velocity. The coefficients $\alpha$, $m$ and the critical threshold $\u_{cr}$ can be assigned by the user or adopted from literature (Table~\ref{tab:sedtransport_closure}).
Engelund and Hansen \cite{Engelund1972} proposed a total transport formula for uniform bed material without a threshold condition for incipient motion.

\begin{table}[tbp]
    \centering
    \caption{\textbf{Sediment transport closure relationships for the morphodynamic problem.} Expressions provide an estimation of the specific sediment transport magnitude $ \lVert{\mathbf q_{B}}\rVert$ [\si{m^2/s}].
        }\label{tab:sedtransport_closure}
    \noindent\adjustbox{max width=\textwidth}{
    \renewcommand{\arraystretch}{2}
        \begin{tabular}{llll}\hline 
            \textbf{Type} & \textbf{Expression} & \textbf{Parameters}  & \textbf{Ref}\\
            \hline
            \multirow{2}{*}{MPM-like}& \multirow{2}{*}{$\alpha(\theta -\theta_{cr})^{m}\sqrt{(s-1)gd_m^3} $} & $\alpha=8,m=1.5,\theta_{cr}=0.047$ & \cite{MPM1948}\\
             & & $\alpha=4.93,m=1.6,\theta_{cr}=0.047$ &\cite{wong2006}\\\hline
            Grass-like & $\alpha (  \lVert{\mathbf u}\rVert - u_{cr})^{m} $ & $\alpha\approx\mathcal{O}(-2;-3),m=3,u_{cr}=0.0$ & \cite{Grass1981}\\\hline
            Engelund and Hansen & $0.05 c_f^2 \theta^{5/2} \sqrt{(s-1) gd_m^3}$ & $-$ & \cite{Engelund1972}\\\hline
            Smart and Jäggi & $\alpha\left(\frac{d_{90}}{d_{30}}\right)^{0.2}J^{0.6}\lVert{\mathbf u}\rVert(\theta-\theta_{cr})d_m$ & $\alpha=8,\theta_{cr}=0.05$ & \cite{Smart1983}\\\hline
        \end{tabular}}
\end{table}

\paragraph{Local corrections of the sediment transport}
The morphodynamic module implements three corrections to the basic Exner equation~\eqref{eq:exner} to account for the influence of local characteristics of the flow and the bottom on the sediment transport. Namely, i) the influence of local slope on incipient motion, ii) the effect of lateral bed slope and iii) of the flow curvature on the sediment transport direction.

The threshold condition for incipient motion of grains, by Shields \cite{Shields1936}, is valid for an almost horizontal bed. In case of a sloped bed in flow direction or transverse to it, the stability of grains is either increased or reduced due to the gravity. The critical shear stress value can be adapted consequently to account for the influence of local longitudinal and transversal slopes. A common approach is to scale the critical shear stress for almost horizontal bed $\theta_{cr}$ with a correction factor $k$:
\begin{equation} \label{eq:shields_localslope}
\theta^*_{cr} = k\theta_{cr}.
\end{equation}
\BM\ implements the correction factor $k$ as proposed in \cite{VanRijn1989} and \cite{Chen2010}. Implementation details are given in the \href{https://basement.ethz.ch/download/documentation/docu3.html}{official documentation}.

The bedload direction can be corrected to account for two relevant morphodynamic processes linked to the slope of the bed and the curvature of the flow. The deviation of the bedload direction from the flow direction can thus be modelled as a deviation angle $\varphi=\varphi_b+\varphi_c$, sum of the correction angle for bed slope ($\varphi_b$) and curvature ($\varphi_c$), as depicted in Fig.~\ref{fig:variables_effects}b and c. The bedload vector is then rotated with the rotation matrix ${\bf T}(\varphi)$, being
\begin{equation}\label{eq:rotation_matrix}
{\bf T}=\begin{bmatrix}
cos\varphi & -sin\varphi\\
sin\varphi & cos\varphi\\
\end{bmatrix},
\end{equation}
where the angle is positive counterclockwise.

The angle $\varphi_b$ is estimated with the approach proposed in \cite{Ikeda82r} and \cite{Talmon1995} for the effect of the local transversal bed slope. In particular, the bedload direction deviates from the flow direction in presence of a local transversal bed slope, due to the gravity acting on the bedload sediment particles (Fig.~\ref{fig:variables_effects}b). The bed load deviation $\varphi_b$ with respect to the flow is therefore evaluated as
\begin{equation}\label{eq:lateral_slope_correction}
\tan \varphi_{b} = -N_l\sqrt{\dfrac{\theta_{cr}}{\theta}}\cdot \mathbf{s}\cdot \mathbf{n}_{q}, \; \text{for} \; \mathbf{s}\cdot\mathbf{n}_{q} < 0,
\end{equation}
where $N_l$ is an experimental lateral transport factor ($0.75 \leq N_l \leq 2.63$), $\mathbf{s}=\left(\partial_x z_B,\partial_y z_B\right)$ is the local bed slope and $\mathbf{n}_q$ is the unit vector perpendicular to  $\mathbf{q}$ in downhill direction (Fig.~\ref{fig:variables_effects}b).

The angle $\varphi_c$ accounts for the effect of a marked flow curvature. Due to three dimensional spiral flow motion that establishes in curved flows, the bed load direction tends to point towards the inner side of the curve, while the flow direction points towards the outer side (Fig.~\ref{fig:variables_effects}c). This curvature effect is taken into account according to an approach proposed in \cite{Engelund1974}, where the deviation angle $\varphi_{c}$ is determined as
\begin{equation}\label{eq:spiral_flow_correction}
\tan{\varphi_{c}} = -N_{*} \dfrac{h}{R},
\end{equation}
where $h$ is the water depth, $N_{*}$ is a curvature factor, and $R$ denotes the radius of the river bend, positive for curvature in counterclockwise direction. The curvature factor $N_{*}$ mainly depends on bed roughness and assumes values $N_{*} \approx 7$ for natural streams \cite{Engelund1974}, and values up $N_{*} \approx 11$ for laboratory channels \cite{Rozovskii1961}.

\paragraph{External sediment input/output}
The source term $S_b$ represents additional sediment mass input or output (sink) that can be defined on subsets of the computational domain. The source can be specified as total volume flux including porosity [\si{m^3/s}]. Similarly to the hydrodynamic case (\S\ref{sec:hyd_closure_relations}), different approaches are adopted for the sediment sink, namely \textit{exact}, \textit{available} and \textit{infinity}.

\subsubsection{Fixed bed concept}
Morphodynamic simulations generate deposition and erosion patterns of the riverbed. Erosion processes, if not limited, can proceed indefinitely in the vertical direction. To account for the presence of non-erodible river bottom, as in case of bedrock or concrete cover, a non-erodible fixed bed depth $z_{rel}$ (Fig.~\ref{fig:variables_effects}a) can be set. This threshold also determines the volume of sediment available for transport. The fixed bed elevation is defined relative to the initial bottom elevation $z_B$ with $z_{rel} \leq 0$.

                

\subsection{Scalar advection-diffusion}\label{sec:scalar_transport}
A number of environmental processes, such as pollutant, temperature or nutrient transport, can be modelled assuming the passive advection and diffusion of a scalar quantity, in the form of dissolved or particulated particles \cite[e.g.][]{Vanzo2016}. The scalar advection-diffusion module allows for the simultaneous simulation of multiple passive species, up to a maximum of 5. The transport of a generic species $c$ can be described by the following advection-diffusion equation:
\begin{equation}\label{eq:scalar_goveq}
\pd{t} q_c + \dex \left[\dfrac{q_xq_c}{h} 
-h \left(K_{xx} \dex \phi_c + K_{xy} \dey \phi_c \right)\right]
+\dey\left[ \dfrac{q_yq_c}{h} -h \left(K_{yx} \dex \phi_c + K_{yy} \dey \phi_c \right)\right]=S_{\phi_c}, \text{with}\, c=[1,5],
\end{equation}
where the unknown is $q_c$, the specific mass of the species $c$. It can be expressed as $q_c=h\phi_c$, with $\phi_c$ the volumetric concentration and $h$ the water depth, as usual. The term $S_{\phi_c}$ is a net source of $c$ and $K_{ij}$ [\si{m^2/s}] are the components of the 2D diffusion tensor.


\subsubsection{Scalar advection-diffusion closure relationships}\label{sec:scalar_closures}
For the scalar advection-diffusion module, the closure relationships are used to model the contribution of external scalar input and output. In particular, the term $S_{\phi_c}$ represents an additional scalar mass flux that is added within a set of elements defined by regions. The source can be specified either as an imposed concentration value or a total volumetric flux [\si{m^3/s}]. The behavior is analogous to the case of hydro- and morphodynamics, \S\ref{sec:hyd_closure_relations} and \S\ref{sec:morpho_closures}.

The terms $K_{ij}$ of the diffusion tensor vary considerably with respect to the physical nature of the transported species. Diffusive transport is modelled in terms of both molecular diffusion $K^m$ and turbulent dispersion $K_{ij}^t$, such that $K_{ij} = K^m I_{ij} + K_{ij}^t$, with $I_{ij}$ the identity matrix. The molecular diffusion is assumed as an isotropic Fickian process with constant coefficient $K^m$. Turbulent dispersion is non-isotropic ($K^t_{ij}$) and scale with the friction velocity $u_*= \lVert{\mathbf u}\rVert/c_f$ and water depth via a longitudinal $\alpha_L$ and transversal $\alpha_T$ non-dimensional coefficient. Suitable values for open channel flows in natural environments are $\alpha_L$=13 and $\alpha_T$=1.2 \citep{Vanzo2016}.

\section{Numerical solution}\label{sec:num_sol}
The numerical solution of the governing equations \eqref{eq:Hyd_gov_equations}, \eqref{eq:exner} and \eqref{eq:scalar_goveq} is sought in a finite volume framework, with a spatial discretization based on unstructured meshes (\S\ref{sec:numerics_discretization}). For the temporal integration, an explicit first order Euler scheme is used. In its basic configuration, the temporal integration proceeds in a synchronous-decoupled way for all the modules, meaning that the modules are independently integrated in time with the same timestep (\S\ref{sec:CFL_condition}).
The following sections detail the domain discretization strategy and the adopted numerical solver for the fluxes calculation of the three basic modules. The interested reader should refer to the provided references for specific implementation details.

\subsection{Domain discretization}\label{sec:numerics_discretization}
\begin{figure}[tbp]
    \centering
    \includegraphics[angle=0,width=0.5\columnwidth]{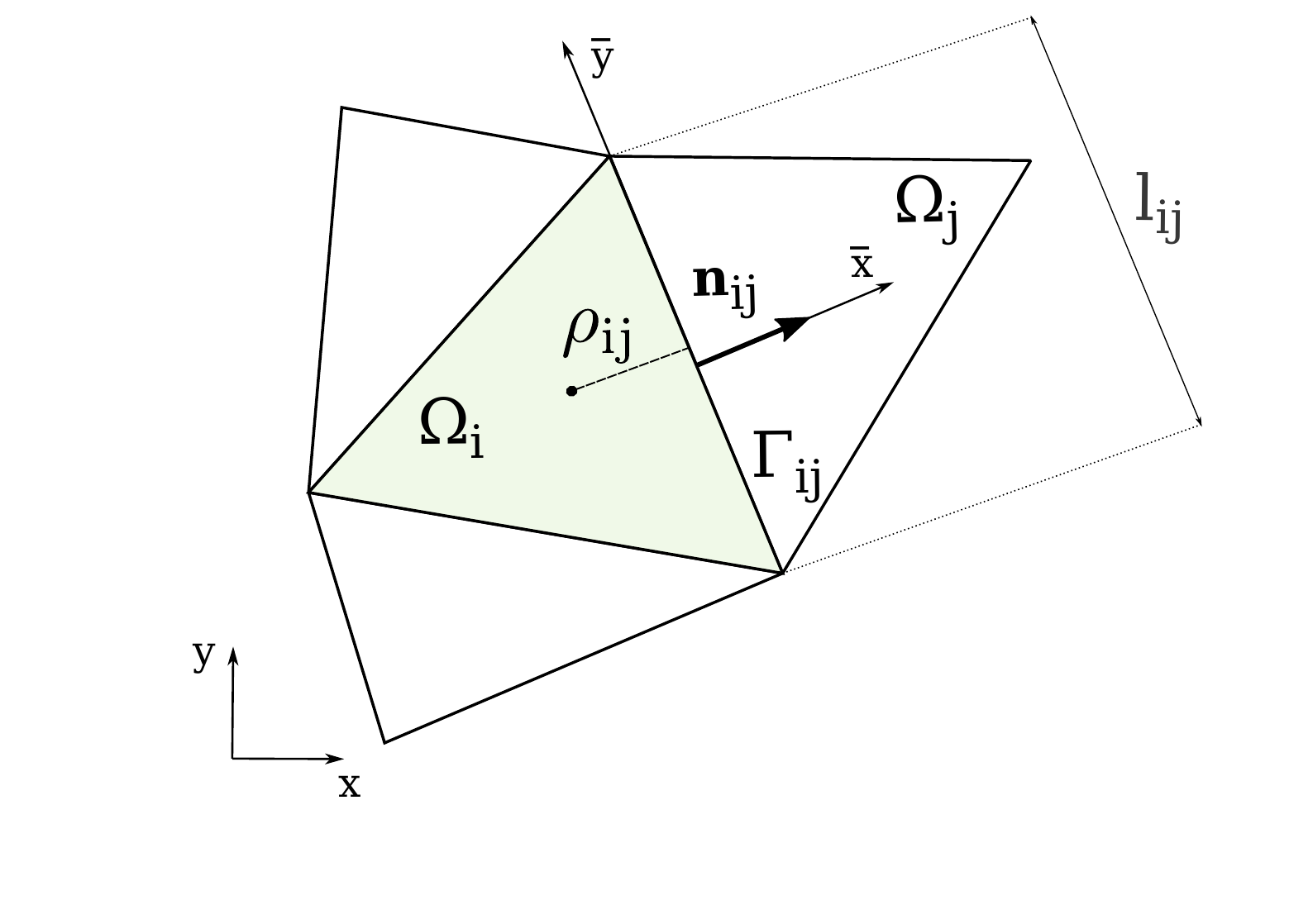}
    \caption{\textbf{Sketch of the triangular discretization}. Main notations adopted for the generic computational cell $i$ and its $j$-th neighbour (with $j$=1,2,3).}
    \label{fig:unstructured_discretization}
\end{figure}

The problem is discretised adopting a finite volume approach over unstructured triangular meshes. A conforming triangulation $T_{\Omega}$ of the computational domain $\Omega \subset \mathbb{R}^2$ by elements $\Omega_i$ such that $T_{\Omega}=\bigcup \Omega_i$, is assumed. 
Given a finite volume element $\Omega_i$ (Fig.~\ref{fig:unstructured_discretization}), $j=1, 2, 3$ is the set of indexes such that $\Omega_j$ is a neighbour of $\Omega_i$. $\Gamma_{ij}$ is the common edge of two neighbour cells $\Omega_i$ and $\Omega_j$, and $l_{ij}$ its length. ${\mathbf{n}}_{ij}=(n_{ij,x},n_{ij,y})$ is the unit vector which is normal to the edge $\Gamma_{ij}$ and points toward the cell $\Omega_j$.

\subsection{Hydrodynamics}\label{sec:numerics_hydro}
The system of governing equations \eqref{eq:Hyd_gov_equations} can be cast in vectorial form as
\begin{equation}\label{eq:vectorialformshallowwater}
\pd{t} \boldsymbol{U} + \dex \boldsymbol{F}_x + \dey \boldsymbol{F}_y = \boldsymbol{S},
\end{equation}
where left-handside terms of \eqref{eq:vectorialformshallowwater} are
\begin{equation}\label{eq:swe_vectorial_form}
\begin{array}{*{3}ccc}
\boldsymbol{U}=\begin{bmatrix}
H \\ q_x \\ q_y 
\end{bmatrix}
&,&\boldsymbol{F}_x=\begin{bmatrix}
q_x \\ \dfrac{q_x^2}{h}+\dfrac{1}{2}gH^2 -gH z_B \\ \dfrac{q_xq_y}{h}
\end{bmatrix}
&,& \boldsymbol{F}_y=\begin{bmatrix}
q_y \\ \dfrac{q_xq_y}{h} \\ \dfrac{q_y^2}{h}+\dfrac{1}{2}gH^2 -gH z_B
\end{bmatrix}
\end{array}.
\end{equation}
The vector of source terms can be written as $\mathbf{S}(\boldsymbol{U})=\mathbf{S}_{h}+\mathbf{S}_{fr}(\boldsymbol{U})+\mathbf{S}_{bed}(\boldsymbol{U})$, where
\begin{equation}\label{eq:swe_vectorial_form_sourceterms}
\begin{array}{*{3}ccc}
\boldsymbol{S}_h=\begin{bmatrix}
S_h \\ 0 \\ 0
\end{bmatrix}
&,& \boldsymbol{S}_{fr}=\begin{bmatrix}
0 \\ -gh S_{fx} \\ -gh S_{fy}
\end{bmatrix}
&,& \boldsymbol{S}_{bed}=\begin{bmatrix}
0 \\ -gH \dex z_B \\ -gH \dey z_B
\end{bmatrix}
\end{array}.
\end{equation}
By integrating the governing system of equations \eqref{eq:vectorialformshallowwater} in the control volume $V=[\Omega_i]\times[t^n,t^{n+1}]$, we obtain the general update formula for the triangular element $i$:
\begin{equation}\label{eq:hydrodynamicupdateformula}
\boldsymbol{U}_i^{n+1}=\boldsymbol{U}_i^n-\frac{\Delta t}{\left|\Omega_i\right|}\sum_{j=1}^3 l_{ij} \left[ \boldsymbol{F} _{ij} 
\right] +\Delta t  \boldsymbol{S}_{i}.
\end{equation}
Problem unknowns at cell $i$ and discrete time $n$ are represented by cell averages $\boldsymbol{U}_i^n$; the numerical solution sought at time $t^{n+1}=t^n+\Delta t$ is denoted by $\boldsymbol{U}_i^{n+1}$. In \eqref{eq:hydrodynamicupdateformula}, $\boldsymbol{F}_{ij}$ are the hydrodynamic fluxes estimated at the cell interface $ij$, as in Fig.~\ref{fig:unstructured_discretization}.

To compute the fluxes $\boldsymbol{F}_{ij}$ for the hydrodynamic system \eqref{eq:vectorialformshallowwater}, several well-established solvers are available. Here we adopt the well-known HLLC approximate Riemann solver \cite{Toro1994} which is a modification of the basic HLL scheme to account for the influence of intermediate contact waves. Further details on the HLLC approach are available in Chapter~10 of \cite{Toro2001}. The solver is proved to be robust and efficient in simulating unsteady flows and the advection of passive tracers \citep{Vanzo2016}.

The numerical discretization of the three terms of the source vector $\mathbf{S}$ \eqref{eq:swe_vectorial_form_sourceterms} is conducted separately, according to the nature of each term. The external inflow/outflow contribution $\mathbf{S}_{h}$ is added explicitly to the continuity equation, as it is not a function of the problem unknowns. The stiff friction source terms $\mathbf{S}_{fr}(\boldsymbol{U})$ are integrated with Runge-Kutta 2 \citep[e.g.][]{Toro2009a} in a semi-implicit fashion after adopting a splitting technique. The implementation is analogous to the ones proposed in \cite{Siviglia2013,Vanzo2015,Vanzo2016}. The topographical terms $\mathbf{S}_{bed}(\boldsymbol{U})$ are discretized using the modified-state approach proposed by \cite{Duran2013}. This results in an easy and robust treatment of complex topographies and wetting and drying problems \cite{Vanzo2016}.

\subsection{Morphodynamics}\label{sec:numerics_morpho}
The Exner equation is solved in a synchronous-decoupled way with respect to the shallow water problem (\S\ref{sec:numerics_hydro}), meaning that the numerical integration of the Exner equation~\eqref{eq:exner} adopts the same integration timestep $\Delta t$ of the hydrodynamic problem. The general update formula for the Exner problem reads:
\begin{equation}\label{eq:exnerupdateformula}
z_{Bi}^{n+1}=z_{Bi}^n-\frac{1}{1-p}\left[\frac{\Delta t}{\left|\Omega_i\right|}\sum_{j=1}^3 l_{ij} \left[ q _{Bij} \right] +\Delta t  S_{bi}\right], 
\end{equation}
with the same symbols introduced for \eqref{eq:exner}. The term $q _{Bij}$ represents the normal sediment flux at the cell interface $ij$ (Fig.~\ref{fig:unstructured_discretization}).

For the numerical estimation of the term $q_{Bij}$, a number of approaches are available in literature. In the current version, \BM\ implements an Approximate Riemann Solver of HLL-type \cite[\textit{sensu}][]{Toro2001}, as in \cite{Soares2011}. The sediment flux is thus calculated as
\begin{equation}\label{eq:exner_hll}
q _{Bij} = \frac{\lambda_s^+q _{Bi}-\lambda_s^-q _{Bj}+\lambda_s^+\lambda_s^-(z_{Bj}-z_{Bi})}{\lambda_s^+-\lambda_s^-},
\end{equation}
where pedix $i$ ($j$) refers to quantities evaluated at the corresponding cell (Fig.~\ref{fig:unstructured_discretization}), and $\lambda_s^+$, $\lambda_s^-$ are speed estimation of the morphological problem. We adopt the following closure \cite{Soares2011}:
\begin{equation}\label{eq:exner_wave_assign}
\lambda_s^-=min(\lambda_{1i},\lambda_{1j})\;,\lambda_s^+=max(\lambda_{2i},\lambda_{2j}).
\end{equation}
The expression for the terms $\lambda_{1}$ and $\lambda_{2}$, calculated for both cell $i$ or $j$ reads:
\begin{equation}\label{eq:exner_wave_estimation}
\lambda_{1,2} = \frac{1}{2} \left( u_n - c \pm \sqrt{(u_n-c)^2+4\frac{\de q_{B,n}}{\de q_n}c^2} \right),
\end{equation}
where $u_n$ is the normal velocity at the cell interface $ij$ (Fig.~\ref{fig:unstructured_discretization}) and $c=\sqrt(gh)$ is the so-called wave celerity.

\subsection{Scalar advection-diffusion} \label{sec:numerics_scl}
The scalar advection-diffusion problem is solved in a synchronous-decoupled way with respect to the shallow water problem (\S\ref{sec:numerics_hydro}). We reformulate the governing equation~\eqref{eq:scalar_goveq} via a Cattaneo-type relaxation technique, as proposed by \cite{Vanzo2016}. Two additional scalar conservation equations are then added to \eqref{eq:scalar_goveq}:
\begin{equation}\label{eq:relaxation_terms}
\de_t \psi_x^c - \de_x \frac{\phi_c}{\zeta} = - \frac{\psi_x^c}{\zeta} \,, \;\;\;\;
\de_t \psi_y^c - \de_y \frac{\phi_c}{\zeta} = - \frac{\psi_y^c}{\zeta}
\end{equation}
where the new symbols are $\zeta$, a positive and small relaxation time, and $\psi_x^c$ and $\psi_y^c$ are two auxiliary variables that recover $\de_x \phi_c$ and $\de_y \phi_c$, respectively, for a sufficiently small $\zeta$ \cite{Vanzo2016}. After a trivial substitution of $\psi_x^c \approx \de_x \phi_c$ and $\psi_y^c \approx \de_y \phi_c$ into \eqref{eq:scalar_goveq}, the system composed by \eqref{eq:scalar_goveq} and \eqref{eq:relaxation_terms} can be rewritten in vectorial form as
\begin{equation} \label{eq:vecform_scalar}
\pd{t} \boldsymbol{U} + \dex \boldsymbol{A}_{x} + \dey \boldsymbol{A}_{y} + \dex \boldsymbol{D}_{x} + \dey \boldsymbol{D}_{y} = \boldsymbol{S}_c + \boldsymbol{S}_{rel}
\end{equation}
where the vectors $\boldsymbol{U}$, $\boldsymbol{A}_x$, $\boldsymbol{D}_x$, $\boldsymbol{S}_c$ and $\boldsymbol{S}_{rel}$ read
\begin{equation}\label{eq:vecform_scalar2}
\begin{array}{*{4}cccc}
\boldsymbol{U} =
\begin{bmatrix} q_c \\ \psi_x^c \\ \psi_y^c \end{bmatrix}
,&
\boldsymbol{A}_{x} =
\begin{bmatrix}   \dfrac{q_c q_x}{h} \\ 0 \\ 0 \end{bmatrix}
,& 
\boldsymbol{D}_{x} =
\begin{bmatrix} -h(K_{xx}\psi_x^c + K_{xy}\psi_y^c)\\ -\frac{q_c}{\zeta h} \\ 0 \end{bmatrix} 
,&
\boldsymbol{S}_c =
\begin{bmatrix} S_{\phi_c} \\ 0 \\ 0 \end{bmatrix}
,&
\boldsymbol{S}_{rel} =
\begin{bmatrix} 0 \\ -\frac{\psi_x^c}{\zeta} \\ -\frac{\psi_y^c}{\zeta} \end{bmatrix},
\end{array}
\end{equation}
with $\boldsymbol{Q}$ representing the conserved scalar quantities, $\boldsymbol{A}_x$ is the advective fluxes vector, $\boldsymbol{D}_x$ is the diffusive-relaxed fluxes vector, both in x direction. The scalar source terms are $\boldsymbol{S}_c$, whilst the source terms arising form the relaxation are $\boldsymbol{S}_{rel}$. For brevity, we omit the formulation for the y direction ($\boldsymbol{A}_y$ and $\boldsymbol{D}_y$), which is analogous. The interested reader can refer to \cite{Vanzo2016} for a step-by-step derivation.

The scalar fluxes in \eqref{eq:vecform_scalar2} are solved through the SVT solver introduced by \cite{Vanzo2016}. The scheme presents a flux-splitting approach combining the advective and diffusive-relaxed fluxes and employs different solvers for each. The HLLC solver, applied also for the hydrodynamic fluxes  (\S\ref{sec:numerics_hydro}) provides the advective component of the scalar fluxes at the cell interface $\boldsymbol{A}_{ij}$. For the diffusive-relaxed component, the SVT technique derives the fluxes at the interface $\boldsymbol{D}_{ij}$ directly from the Riemann invariants of a two non-linear wave Riemann problem.

Similarly to the hydro- and morphodynamic problems, the control volume $V=[\Omega_i]\times[t^n,t^{n+1}]$ is used to integrate the governing system \eqref{eq:vecform_scalar}, in order to obtain the following scalar update formula at the element $i$
\begin{equation}
\boldsymbol{U}_i^{n+1}=\boldsymbol{U}_i^n-\frac{\Delta t}{\left|\Omega_i\right|}\sum_{j=1}^3 l_{ij} \left[ \boldsymbol{A} _{ij} + \boldsymbol{D}_{ij}
\right] + \Delta t  (\boldsymbol{S}_c + \boldsymbol{S}_{rel}),
\label{eq:scl:updateformula}
\end{equation}
where the fluxes $\boldsymbol{A}_{ij}$ and $\boldsymbol{D}_{ij}$ are  computed at each cell interface $ij$ (Fig.~\ref{fig:unstructured_discretization}).

The numerical integration of the two source terms vectors is conducted separately, according to the nature of the terms. The simple scalar sources $\boldsymbol{S}_c$ are computed with a first-order Euler scheme, while the stiff relaxation source terms $\boldsymbol{S}_{rel}$ are integrated by means of a locally implicit Euler method.

\subsection{Stability condition}\label{sec:CFL_condition}
Numerical integration proceeds with a dynamic timestep $\Delta t$, evaluated at each time loop (Fig.~\ref{fig:module_sequence}b) that fulfills the well-known Courant-Friedrichs-Lewy stability conditions \cite{Toro2001}. In the current implementation, the condition is expressed as:
\begin{equation}\label{eq:CFL_condition}
\Delta t = CFL \min_{1\leq i\leq N}\left(\min_{1\leq j\leq 3}\left(\frac{\rho_{ij}}{{\lambda_{ij}}}\right)\right)\;, 
\end{equation}
where $\rho_{ij}$ is twice the distance between the edge $j$ and the centroid of the cell $i$ (Fig.~\ref{fig:unstructured_discretization}), and $N$ is the total number of domain elements. The term $\lambda_{ij}$ is an estimation of the largest eigenvalue of the hydrodynamic problem \eqref{eq:Hyd_gov_equations}, namely  $\lambda_{ij}=|u_n|+\sqrt{gh}$ with the simbology already introduced. The CFL coefficient ranges between 0 and 1: by default it is set to 0.9, if not specified otherwise.

\section{Initial and boundary conditions}\label{sec:IC_BC}

\subsection{Initial conditions}\label{sec:ICs}
All modules require the user to define the initial conditions of the simulation. Two types of initial conditions are similarly available for all the modules:
\begin{itemize}
    \item region defined: user explicitly defines the initial values of the problem unknowns (e.g. water depth and specific discharge for hydrodynamics). Different values can be assigned to different region of the computational domain;
    \item continue: values are taken from the result file of previous simulations.
\end{itemize}
In addition, the hydrodynamic module allows also to set dry conditions (no water in the domain) as initial conditions. In this case, the domain will progressively fill with water, in relation to the assigned inflow boundary conditions or internal sources (\S\ref{sec:hyd_closure_relations}).

\subsection{Boundary conditions}\label{sec:BCs}
The boundary conditions (hereinafter BCs) have different specifications for each core module (see following Sections), but they all classify in three common types: external \textit{standard}, external \textit{linked} and \textit{internal} BCs.

\begin{figure}[tbp]
    \centering
    \includegraphics[angle=0,width=0.6\columnwidth]{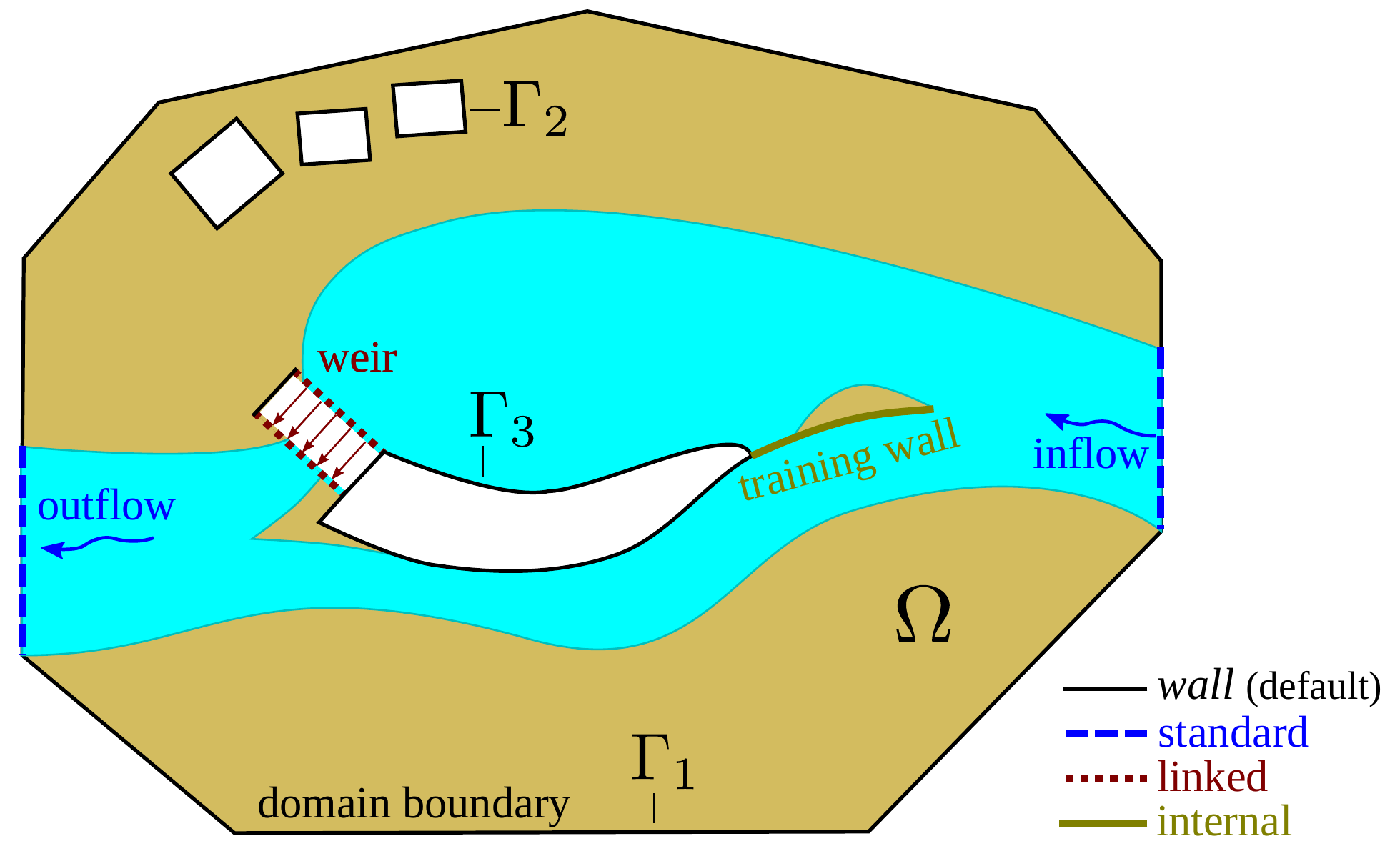}
    \caption{\textbf{Example of modelling domain with different types of boundary conditions.} The computational domain $\Omega$ can include the river channels but also the surrounding floodplains. The domain is delimited by the \textbf{external} BCs: impermeable walls (default type, $\Gamma_{1,2,3}$) are depicted in solid black, while standard inflow and outflow are in dashed blue. A weir (dotted brown) is modelled with an external linked BC. A training wall (thick solid green) is modelled with an internal BC.}
    \label{fig:BCs}
\end{figure}

Fig.~\ref{fig:BCs} exemplifies the main concepts adopted for the BCs. The computational domain $\Omega$ is defined by the domain boundaries, as  $\Gamma_{1,2,3}$. An external \textit{standard} BC is dependent only on the local flow conditions and on some user-defined rules. This represents the most common case, for example to define impermeable walls or river inflows and outflows. By default, all the external boundaries are all set as wall. The \textit{wall} BC consists of a fixed, frictionless, reflective impermeable wall. In external \textit{linked} BCs instead, the local BCs are defined also with information from a \textit{linked} boundary. Typical example is a weir, where the flow discharge at the downstream side of the weir depends on the water stage on the upstream side. 

The third type of BCs, \textit{internal}, are defined within the computational domain $\Omega$, and not at the edges (Fig.~\ref{fig:BCs}). This BC type comes in handy in case of very large domain application, because it allows to test different configurations of hydraulic structures (e.g. different locations of a weir or training wall), without the need of regenerate the entire computational mesh for every configuration.

A summary of the main feature of the BCs for the three core modules follows here. The interested reader can refer to the \href{https://basement.ethz.ch/download/documentation/docu3.html}{official documentation} for further details. 

\subsubsection{Hydrodynamic BCs}\label{sec:hyd_bc}
The hydrodynamic module implements different types of BCs, with a different level of customization. Depending on the BC type, user-assigned data is requested, as single constant value in time (e.g. lake level, constant discharge), as time series (e.g. hydrograph), or as set of parameters describing a dynamic behaviour (e.g. weir activation rule). In particular:

\begin{itemize}
    \item Standard BCs: in addition to \textit{wall} BC, inflows (upstream BCs) and outflows (downstream BCs) can be assigned. As standard inflows, three options are provided (namely \textit{uniform, explicit, zhydrograph}). For all cases, given a total volume discharge $Q$~[\si{m^3\per\second}] or water surface elevation [m] as input, the inflows condition for the mathematical unknowns of system~\eqref{eq:Hyd_gov_equations} is set. For the standard outflows a value for the water depth $h$ must be specified. Possible options are: uniform conditions, hydraulic weir, rating curve, hydrograph and zerogradient (i.e. Neumann BC). It is worth remarking that the specific type of upstream and downstream BCs should be selected depending on the local flow conditions (i.e. sub- or super-critical).
    \item Linked BCs: this type of boundaries establish a \textit{link} between two certain region of the domain where the governing equations are not solved. This type of BCs are particularly designed to simulate the behaviour of hydraulic structures within the river channel, such as weirs, gates, bridges, spillways.
    \item Internal BCs: they are fictitious boundaries defined as segments at the interfaces of some computational cells. On these segments, three different conditions can be enforced, instead of the solution of the SWE \eqref{eq:Hyd_gov_equations}. Options are: \textit{static} walls, \textit{dynamic} walls and \textit{rating curve}. With the static wall, the standard \textit{wall} condition is applied on both sides of the internal boundary. With the dynamic wall, the wall conditions are applied until reaching a given threshold value (time or water depth) after which the wall is removed, and the SWE are solved. With the rating curve option (or \textit{h-Q relation}) a give flow relation is applied on one side of the internal boundary, while on the other side, wall conditions apply (unidirectional flow). The internal BCs are particularly useful when simulating, for example, the collapse of hydraulic structures: after the collapse, the simulation can proceed by calculating the actual free unsteady flow (i.e. governing equations) over the (former) boundary.
\end{itemize}

\subsubsection{Morphodynamic BCs}\label{sec:morpho_bcs}
The sediment flow is defined as a specific bedload flux, which is averaged and evenly distributed at the domain boundary conditions over the boundary length. In analogy with the hydrodynamic module, the morphodynamic boundaries are of type external \textit{standard} and \textit{linked}. 

\begin{itemize}
    \item Standard BCs: for the upstream BCs, \BM\ implements three versions that allow to simulate: i) a given input of sediment as time series (i.e. sedimentograph), ii) a sediment input derived from the hydrodynamic conditions under transport capacity conditions or iii) bed equilibrium condition, where the upstream bed elevation is kept constant. Two downstream BCs are available, allowing the simulation of i) equilibrium condition and ii) check-dam. In this second option an equilibrium boundary condition is activated only if the bed level reaches a given threshold value, otherwise a wall type boundary is assumed.
    \item Linked BCs: one BC is available. It allows for the simulation of sediment transport through given hydrodynamic linked conditions, hence to ensure sediment continuity in the simulated channel.
\end{itemize}

\subsubsection{Scalar advection-diffusion BCs}\label{sec:scalar_bcs}
Scalar BCs are defined in terms of concentration of total volumetric rate [\si{m^3/s}], evenly distributed throughout the length of the relevant domain boundary. The implemented types are:
\begin{itemize}
    \item Standard BCs: three types are available. i) scalar inflow as a constant value; (ii) scalar inflow as a time-series and (iii) zerogradient (i.e. Neumann BC) outflow.
\end{itemize}

\section{Software design}\label{sec:software_design}

\subsection{Modelling workflow}\label{sec:user_workflow}
\begin{figure}[tbp]
    \centering
    \def\svgwidth{\columnwidth}
    \includegraphics[angle=0, width=0.6\columnwidth]{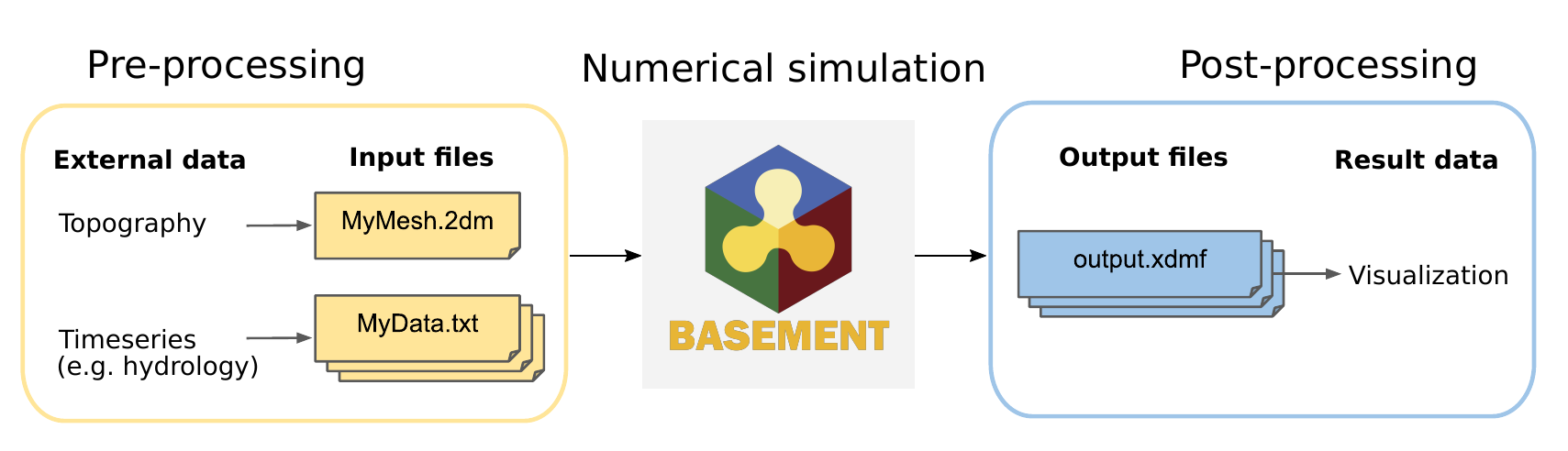}
    \caption{\textbf{Modelling workflow.} Pre-processing: generation of the computational mesh from topographical data and definition of input time series. Numerical simulation: \BM. Post-processing: elaboration and visualization of the results.}
    \label{fig:ModellingProcedure}
\end{figure}

The standard modelling procedure involves three phases: the pre-processing phase, the numerical simulation phase and the post-processing phase (Fig.~\ref{fig:ModellingProcedure}). \BM\ is designed to integrate into this workflow. Moreover, the entire workflow relies on open-source or freeware tools. In the following we list the phases and provide a short description of the different configuration and results file formats as used by \BM.
\begin{enumerate}
    \item Pre-processing: in this phase the user is required to define the model domain and the input data. The mesh file (customized \href{ https://www.xmswiki.com/wiki/SMS:Mesh_Generation}{2dm format}, "MyMesh.2dm" in Fig.~\ref{fig:ModellingProcedure}) contains the description of the triangular unstructured computational mesh. The file can be generated with BASEmesh, a Python script as well as a QGIS plugin (see \href{https://basement.ethz.ch/}{software website} for details), or via grid generator software that supports the 2dm format. In addition, further input data such as time series of water and sediment discharge (or other quantities) to be used as BCs can be provided (ASCII format, "MyData.txt" in Fig.~\ref{fig:ModellingProcedure}).
    \item Numerical simulation: the actual simulation can be run via either CLI or GUI. In \BM\ version 3, the numerical simulation is split into three steps (description follows in \S\ref{sec:simulation_steps}). The final simulation results are stored in a general purpose binary container (Hierarchical Data Format HDF5, www.hdfgroup.org). \BM\ generates also an XDMF file (eXtensible Data Model and Format, http://www.xdmf.org) which contains a machine-readable description of the data stored in the HDF5 file.
    \item Post-processing: the XDMF file ("output.xdmf" in Fig.~\ref{fig:ModellingProcedure}) can be opened with the \href{ https://plugins.qgis.org/plugins/crayfish/}{Crayfish} plugin for QGIS or with \href{https://www.paraview.org/}{Paraview} for final results visualization and further post-processing. In addition, \textit{ad-hoc} Python scripts can be used to manipulate results directly from the binary container (some scripts are provided at \href{https://basement.ethz.ch/}{software website}).
\end{enumerate}

\subsection{Simulation steps}\label{sec:simulation_steps}
\begin{figure}[tbp]
    \centering
    \def\svgwidth{\columnwidth}
    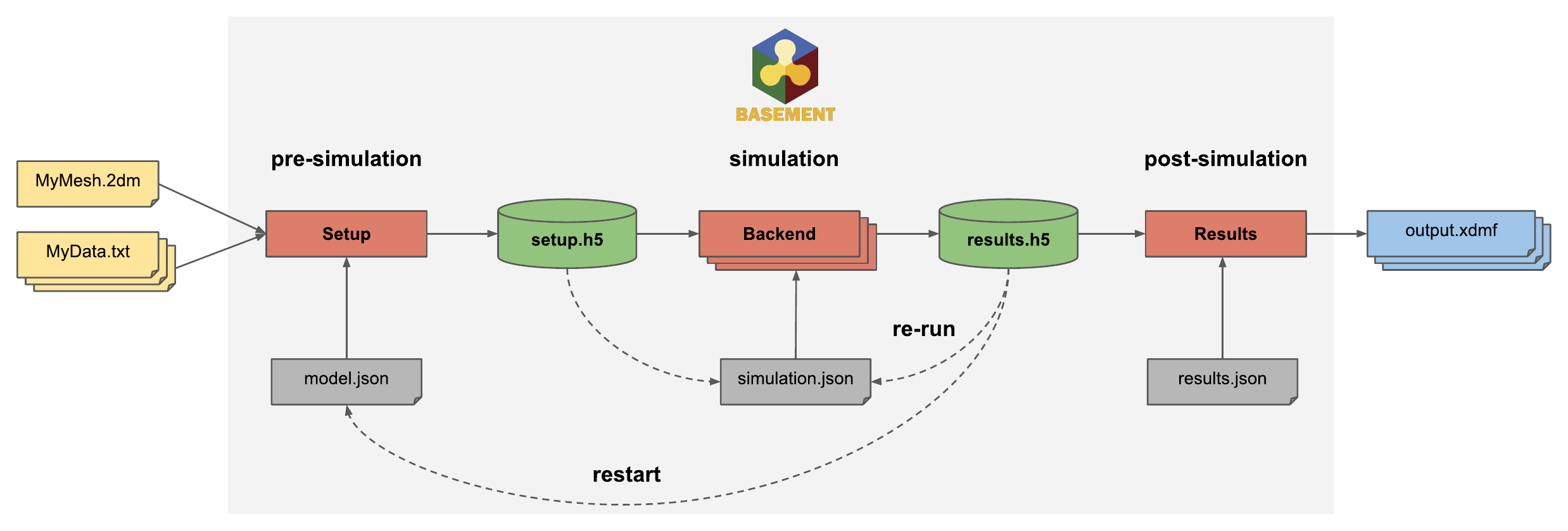
    \caption{\textbf{Software components and simulation steps.} The \BM\ software is composed of a set of executables (red rectangles) driven by JSON configuration files (grey labels). Data is stored in HDF5 containers (green cylinders). In dashed arrows: the special actions of simulation \textit{re-run} and \textit{restart}.}
    \label{fig:simulation_steps}
\end{figure}

The numerical simulation phase consists of three steps: the pre-simulation, the simulation, and the post-simulation (Fig.~\ref{fig:simulation_steps}). Each step can be completed by running a corresponding \BM\ executable via the Graphical User Interface or Command Line Interface. This modular design allows a customization of the simulation workflow by the user and an efficient batch processing of \BM\ steps. For instance, the programs can be run from a scripting language like Python. 

The different executables are configured using a dedicated command file, as detailed in the following list. These command files use the standardized JSON file format (JavaScript Object Notation) (Fig.~\ref{fig:simulation_steps}). The \BM\ GUI is designed to support the user with creating the command files, running and monitoring the three simulation steps. In particular, the GUI validates the configuration parameters and automatically adds required parameters where default values are available.

\begin{enumerate}
    \item The pre-simulation step focuses on the model definition. In particular, the "model.json" command file contains: i) physical properties, ii) initial conditions and iii) boundary conditions of the physical problem, and further iv) numerical parameters. The setup executable first reads the computational mesh "MyMesh.2dm", the external required data "MyData.txt" and the command file "model.json". Then it validates and stores the model inside the binary container "setup.h5".
    \item The simulation is carried out on a selected computational backend (\S\ref{sec:parallelization_strategy}). It is driven by the command file "simulation.json" that contains the simulation parameters (e.g. execution time, output time and desired output quantities). The program reads and executes the model "setup.h5" generated in the previous step. The results of the simulation are stored in a second binary container: "results.h5".
    \item The post-simulation step is configured using a command file "results.json" that contains the selected output format (currently only XDMF is supported). The output is then available for the post-processing phase. (\S\ref{sec:user_workflow}).
\end{enumerate}

When it is necessary to run a new simulation starting from the results of a previous one, two options are available: \textit{Restart} and \textit{Re-run}. When performing a \textit{Restart} the \textit{pre-simulation} step is executed again, i.e. a new model is generated from scratch with (potentially) a new set of parameters. The user indicated an existing "results.h5" file that is to be used to fetch the initial conditions for the new model. The \textit{Re-run} option does not generate a new model, but uses the existing model ("setup.h5") with initial conditions taken from the current results file. In this scenario the user can only modify the simulation and results parameters (i.e. duration, output timestep and output type), but not the the model parameters. This second option is particularly useful in the case of large models (i.e. millions of computational cells): in this situation the pre-simulation step can be computationally demanding (i.e. tens of minutes). If the user only needs to extend the simulation duration, for example, then the \textit{Re-run} option allows to skip the pre-simulation step.


\subsection{Modularity and Sequencing}
\BM\ aims to simulate different river processes with a high level of flexibility and efficiency: with this in mind, we designed the software adopting a modular approach. The two core concepts of this design are \textit{Modules} and \textit{Kernels}, described here below.

\begin{figure}[tbp]
    \centering
    \includegraphics[angle=0, width=0.8\columnwidth]{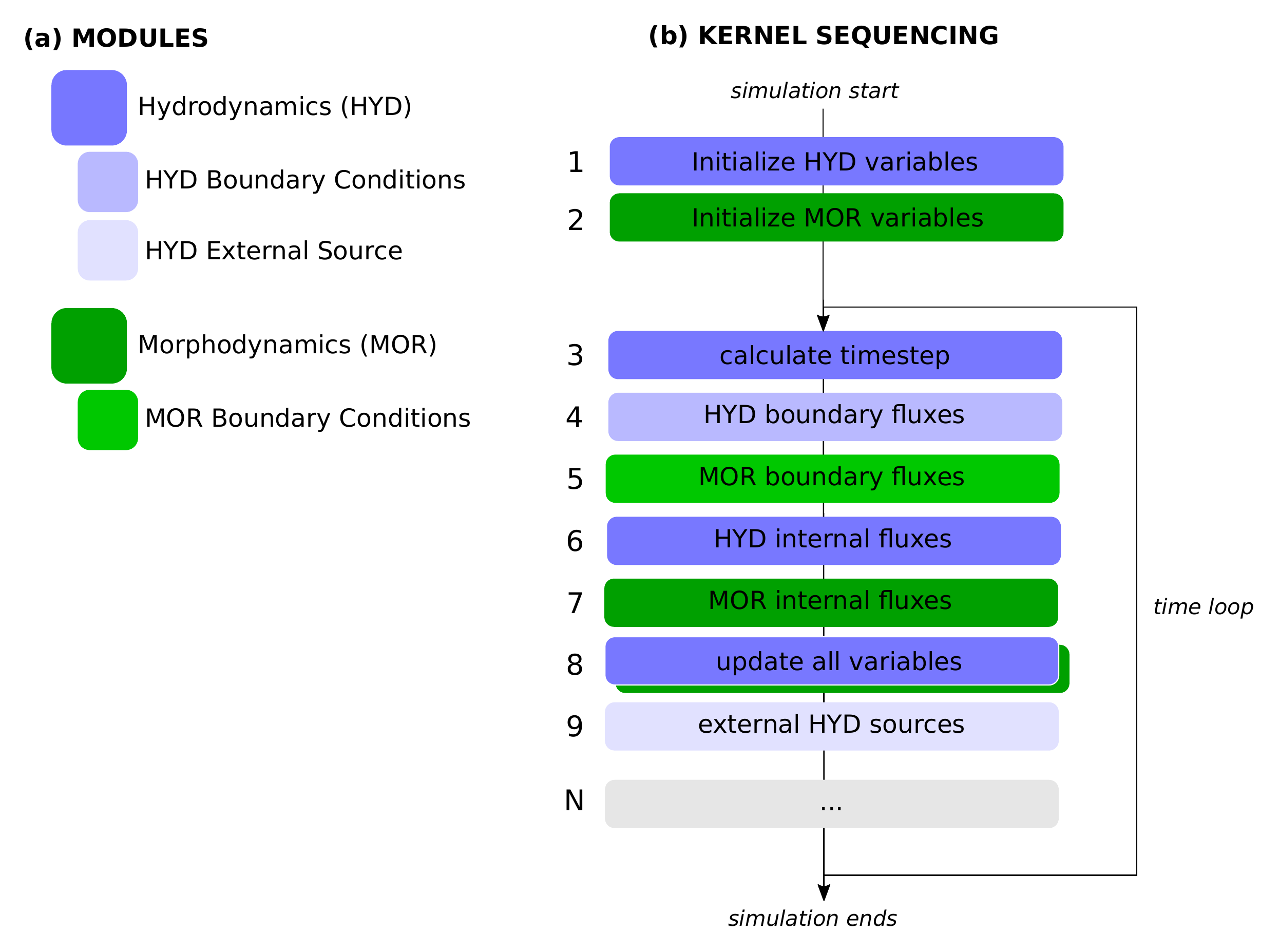}
    \caption{\textbf{Examples of modules and associated kernels.} \textbf{(a)} At model setup, depending on the simulated physical problem,  the user triggers the activation of a set of modules. \textbf{(b)} The active modules trigger a unique kernel sequence to be executed at simulation time to correctly simulate the requested processes. Each module corresponds to a different set of kernels.}
    \label{fig:module_sequence}
\end{figure}

\textit{Modules} take care of the simulation of specific river processes (e.g. module "Hydrodynamics" in Fig.~\ref{fig:module_sequence}a). They can be nested to simulate processes with an increasing level of detail/complexity (e.g. module "HYD External Source", Fig.~\ref{fig:module_sequence}a). Modules are activated by the user in the pre-simulation step. An activated module triggers the execution of a number of \textit{Kernels} throughout the simulation.

A \textit{Kernel} is a set of operations to be executed on each entity (e.g. a cell $i$ or an edge $j$, Fig.~\ref{fig:unstructured_discretization}) of the computational domain or a subset of it. Depending on the specific task, \textit{Kernels} can be scheduled for a single execution (i.e. initialization Kernels) or for repeated execution in each iteration of the integration time loop (Fig.~\ref{fig:module_sequence}b). While the time loop is executed with a timestep $\Delta t$ that satisfies certain stability conditions (\S\ref{sec:CFL_condition}), the Kernels can be scheduled for execution at different (larger) time intervals to reflect the nature of the simulated processes.

The architecture based on Modules and Kernels has two main advantages. First, it is \textit{flexible} in that it allows users (and software developers) to easily add or remove specific modules without interfering with other existing modules. In particular, this permits an integration of further modules as development continues (\S\ref{sec:conclusions}). Second, it is \textit{efficient}, because only the necessary Kernels are scheduled for execution at setup time (pre-simulation step).

\subsection{Parallelization Strategy and Computational Backends}\label{sec:parallelization_strategy}
The parallelization strategy of the \BM\ numerical core addresses two main aspects: i) the use of different technologies (i.e. computational backends) generated from the same, unique software source code. This allows for an easier source code maintenance and integration of future/different backends. ii) An efficient and heavy parallelization of the numerical core following the concept of data parallelism. To this end the numerical core of \BM\ integrates OP2 \cite{Mudalige2012a,Giles2012}, which is an open-source framework for the development of unstructured grid applications. Using source-to-source translation, OP2 generates the appropriate code for different target platforms by introducing an additional level of abstraction between the numerical algorithm and its execution. It supports multi-core CPUs, GPUs, and even clusters via MPI (Message Passing Interface, http://www.mpi-forum.org).

\BM\ currently supports multi-core CPUs and GPUs. When starting the simulation, the user can select to compute on the CPU, the GPU, or a combination of both. All the currently supported backends (Table~\ref{tab:backends}) are available for both Windows and Linux (Ubuntu) operating systems. It is important to note that the choice of graphics processing units is currently limited to Nvidia (CUDA) cards. The precise requirements are provided in the \href{https://basement.ethz.ch/download/documentation/docu3.html}{official documentation}. All the backends can execute the numerical simulations in double (default) or single precision, with different performance characteristics (\S\ref{sec:performance}).

\begin{table}[tbp]
    \centering
        \caption{\textbf{Description of available backend types for \BM\ v3.}}\label{tab:backends}
        \begin{tabular}{ll}\hline 
            \textbf{Type} & \textbf{Description}\\
            \hline
            \textbf{seq} & sequential execution on the CPU\\
            \textbf{omp} & multi-threading using OpenMP technology\\
            \textbf{cuda} & GPU \\ 
            \textbf{cudaC} & GPU with some kernels running sequentially on the CPU \\
            \textbf{cudaO} & GPU with some kernels running in parallel (OpenMP) on the CPU \\\hline
        \end{tabular}
\end{table}

\section{Results}\label{sec:testcases}
A set of selected test cases (T1-T6) are proposed here to test the robustness, accuracy and efficiency of the three basic modules. Table~\ref{tab:testcases} summarizes the key features of each test case. The interested reader can refer to the official software documentation for further examples. Finally, \S\ref{sec:performance} focuses on the software performance and scalability. All the test cases are freely available at (link provided after paper acceptance).

\begin{table}[tbp]
    \centering
    \caption{\textbf{Main features of the adopted benchmarks.}}\label{tab:testcases}
    \begin{tabular}{llll}\hline
        \textbf{ID} & \textbf{Module} & \textbf{Comparison} & \textbf{Key features}\\
        \hline
        \multirow{2}{*}{T1} & \multirow{2}{*}{hydrodynamic} &
        \multirow{2}{*}{field data} & \multicolumn{1}{p{8cm}}{\rom{1} highly unsteady shock wave generation and propagation, \rom{2} wet-and-dry processes, \rom{3} performance of the hydrodynamic module} \\
        \hline
        \multirow{2}{*}{T2} & \multirow{2}{*}{morphodynamic} &
        \multirow{2}{*}{lab data} & \multicolumn{1}{p{8cm}}{\rom{1} sediment transport with a transcritical 1D flow, \rom{2} upstream BCs: uniform inflow,  \rom{3} downstream BCs: imposed water level} \\
        \hline
        \multirow{2}{*}{T3} & \multirow{2}{*}{morphodynamic} &
        \multirow{2}{*}{lab data} & \multicolumn{1}{p{8cm}}{\rom{1} 2D dam-break in a complex domain, \rom{2} sediment transport with an advancing wet-and-dry front, \rom{3} downstream BCs: free outflow, \rom{4} performance of the morphodynamic module} \\
        \hline
        \multirow{2}{*}{T4} & \multirow{2}{*}{morphodynamic} &
        \multirow{2}{*}{lab data} & \multicolumn{1}{p{8cm}}{\rom{1} erosion and deposition in a channel bend, \rom{2} sediment transport direction correction, \rom{3} upstream BCs: unsteady hydrograph} \\
        \hline
        \multirow{2}{*}{T5} & \multirow{2}{*}{scalar advection-diffusion} &
        \multirow{2}{*}{numerical sol} & \multicolumn{1}{p{8cm}}{\rom{1} 1D strong rarefaction waves, \rom{2} conservation of a steady discontinuity of the scalar quantity} \\
        \hline
        \multirow{2}{*}{T6} & \multirow{2}{*}{scalar advection-diffusion} &
        \multirow{2}{*}{numerical sol} & \multicolumn{1}{p{8cm}}{\rom{1} 2D complex domain, \rom{2} advection and diffusion of two different scalar quantities, \rom{3} conservation and mixing of the scalars, \rom{4} performance of the scalar advection-diffusion module} \\
        \hline
    \end{tabular}
\end{table}

\subsection{T1: Malpasset dam collapse}\label{sec:malpasset_test}
\begin{figure}[tbp]
    \centering
    \includegraphics[angle=0, width=0.8\columnwidth]{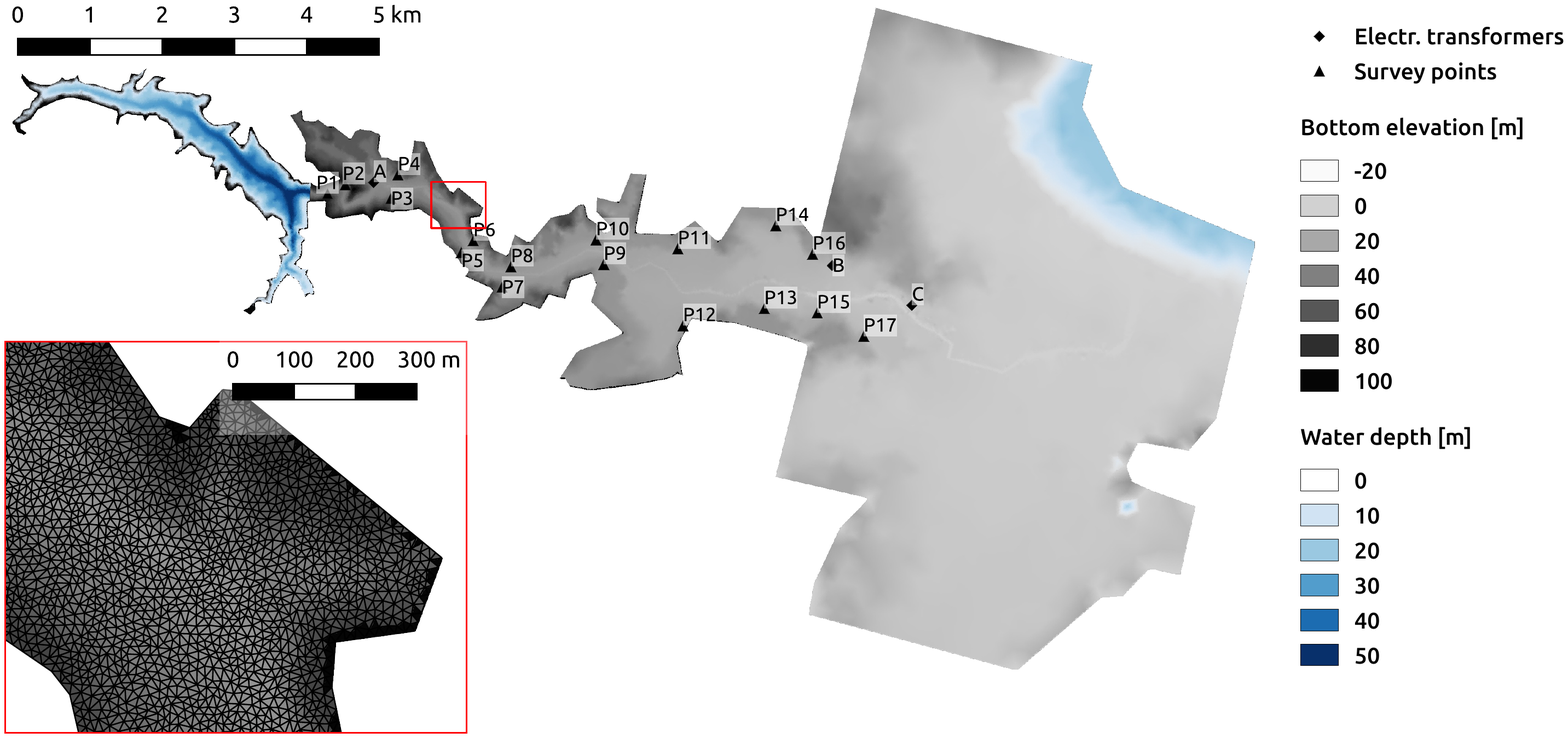}
    \caption{\textbf{T1: planar view of the Malpasset test case.} Visualization of the computational domain illustrating the bottom elevation (gray scale) and the initial water depth (blue scale). Letters indicate the location of the transformer stations (A-C), and the maximum water level survey points (P1-P17). The left box shows a magnification of the computational mesh, as reference.}
    \label{fig:malpassetDomain}
\end{figure}

The scope of this test is to assess the robustness and accuracy of the hydrodynamic solver when simulating a shock-type hydrodyamic wave travelling on a highly irregular and dry domain. The collapse of the Malpasset dam, in the Reyran River Valley (Fréjus, France), represents a well-established hydrodynamic benchmark for numerical models \citep[e.g.][]{hervouet1999,valiani2002,singh2011}. In 1959, the 66.5 m high dam collapsed almost instantaneously, generating an up to 40 m high flood wave that propagated down the Reyran valley, destroyed the two villages Malpasset and Bozon and reached the Mediterranean Gulf 21 minutes later \cite{valiani2002}. The propagation of the flood wave was reconstructed via the maximum water level and the flood arrival time, recorded at multiple locations. In particular, the maximum water level is available from a police survey for 17 survey points, marked as P1 to P17 in Fig.~\ref{fig:malpassetDomain} and the flood arrival time is known from three electric transformer stations which have been destroyed by the flood wave. The locations of the transformer stations are indicated as A, B and C in Fig.~\ref{fig:malpassetDomain}. Coordinates and recorded arrival times are listed in Table~\ref{tab:malpassetETtime}. We make use of such field data to test the performance of the hydrodynamic module. 

The computational domain is discretized with 499,059 triangular elements. The domain boundaries are set to \textit{walls} (\S\ref{sec:hyd_bc}), with exception of the downstream boundary located in the Mediterranean Gulf, where a fixed water level was set to 0 m. The initial conditions are a fixed water surface elevation of 100 m in the reservoir, and dry conditions in the rest of inland domain. The initial velocity was set to \SI{0.0}{m/s} in the entire domain. In accordance with \cite{hervouet1999}, the Manning's friction coefficient was set to \SI{0.033} {m^{-1/3}s} for the whole domain. The CFL number was set to 0.9.

\begin{figure}[tbp]
    \centering
    \includegraphics[angle=0, width=0.8\columnwidth]{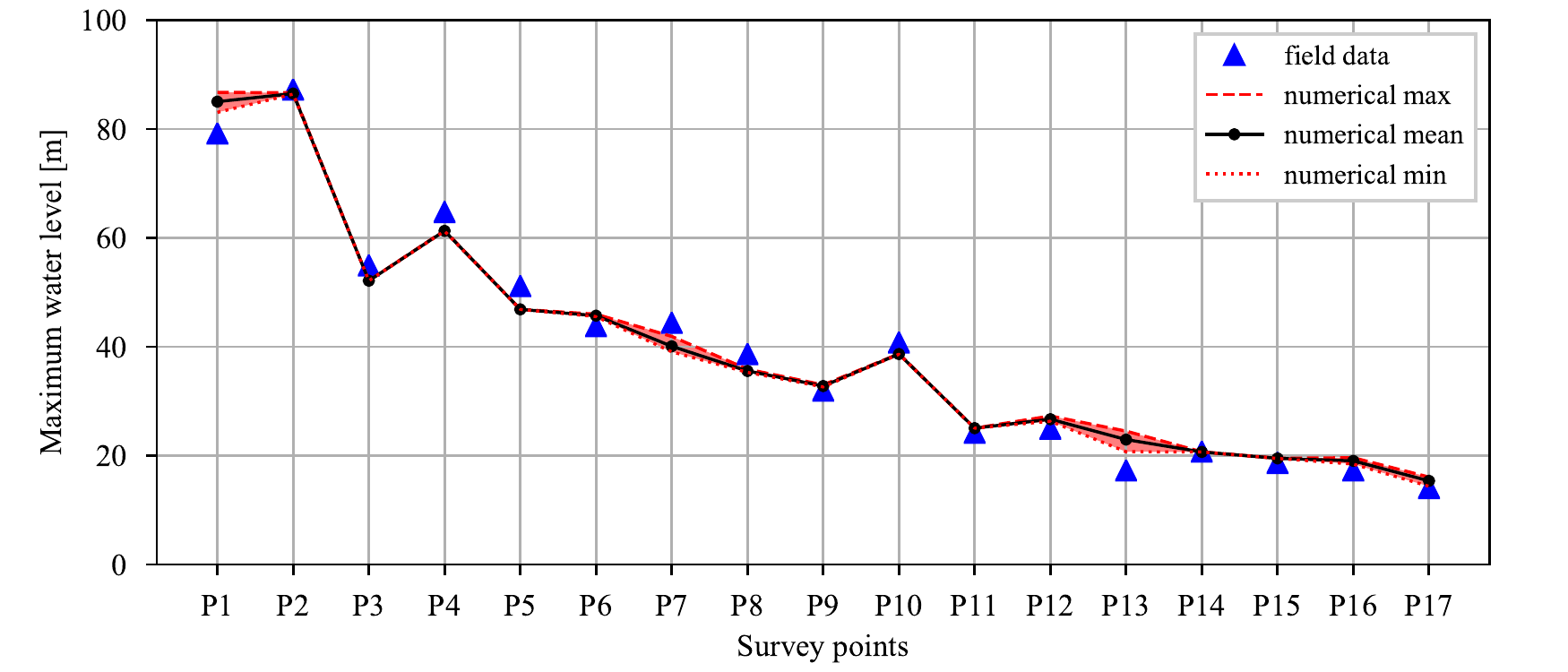}
    \caption{\textbf{T1: Malpasset dam-break wave maximum water level.} Numerical values are compared to the field data at the survey points P1 to P17: simulated values are given as maximum, minimum (red lines) and mean value (black line) of the computational cell containing the survey point and its three neighbours.}
    \label{fig:malpassetSurveyPoints}
\end{figure}

The simulated maximum water levels are compared to the 17 field observations in Fig.~\ref{fig:malpassetSurveyPoints}, with overall good agreement. The average relative error is 7.15\%, with the largest observed at P13, with an overestimation of 30.6\%. To highlight the effects of the topographical approximations of the Digital Elevation Model (and hence of the computational mesh), we compared recorder and simulated water level values as follows. For each punctual maximum water level recorded in field observations (blue triangles in Fig.~\ref{fig:malpassetSurveyPoints}), we compared the maximum simulated values of the spatial mean, maximum and minimum among the computational cell containing the observation point and its three neighbours (black and red series in Fig.~\ref{fig:malpassetSurveyPoints}). We expect lower discrepancies between recorded and simulated values where the numerical values, hence the topographical elevations, are spatially homogeneous. As a matter of fact, points P1, P7 and P13 (Fig.~\ref{fig:malpassetSurveyPoints}) have the large discrepancy between measured and simulated values, but also the largest spatial variability of the numerical values (red shaded area). This suggests that such discrepancies relates more to the local topographical approximations of the DTM rather than to the numerical model. 

Observed and simulated times of flood arrival are given in Table~\ref{tab:malpassetETtime}. Simulated values are in good agreement with measured ones for all the electrical transformer stations (ET). Simulated arrival times have a maximum relative error of 3.8\% for ET B, corresponding to an absolute delay of 47 s. It is worth mentioning that the friction value influences the simulated arrival times.

\begin{table}
\centering
\caption{\textbf{T1: Malpasset dam-break wave arrival times.} Observed and simulation time of flood arrival (TFA) and relative error (Err) for the three electrical transformer stations (ET) destroyed by the flood wave.}
\label{tab:malpassetETtime}
\begin{tabular}{ l r r r r r} 
 \hline
 ET & x & y & TFA\textsubscript{obs} & TFA\textsubscript{sim} & Err\\ 
0 [-] & [m] & [m] & [s] & [s] & [\%]\\ 
 \hline
 A & 5550 & 4400 & 100 & 103 & 3\\ 
 B & 11900 & 3250 & 1240 & 1287 & 3.8\\ 
 C & 13000 & 2700 & 1420 & 1435 & 1\\ 
 \hline
\end{tabular}
\end{table}

\subsection{T2: Propagation of a sediment bore}
Scope of the test is to assess the robustness of the de-coupled hydro-morphodynamic solver approach, particularly when simulating the sediment transport over a transcritical flow. This represents a critical test, especially when adopting de-coupled approaches \citep[e.g.][]{cordier2011}. Moreover, the simulation tests the morphological solver capability in well reproducing the dynamics of an advancing sediment bore.

In this test case, the flume experiment proposed in \citep[][run 2]{bellal2003} is reproduced numerically. The computational domain is a composed by a straight 6.9 by 0.5~m channel, representing the lower part of the original experimental flume, and is discretized with 24,612 triangular elements. The sediment has a characteristic diameter of 1.65~mm. The water and sediment discharge at the upstream boundary are set to \SI{0.012}{\cubic\meter\per\second} and \SI{0.196}{\cubic\meter\per\second} (with porosity) respectively. The flume is at initial uniform flow conditions, characterized by a supercritical flow. At $t$=0~s, a fixed water level of 0.2093~m  is imposed at the downstream boundary and sediment transport out of the domain is stopped. This results in the formation of an hydraulic jump moving upstream in the flume, and a subsequent downstream propagation of a sediment bore. The CFL number is set to 0.9, the bed porosity is assumed constant and equal to 0.42, and the simulation duration is 500~s.

\begin{figure}[tbp]
    \centering
    \includegraphics[angle=0, width=0.8\columnwidth]{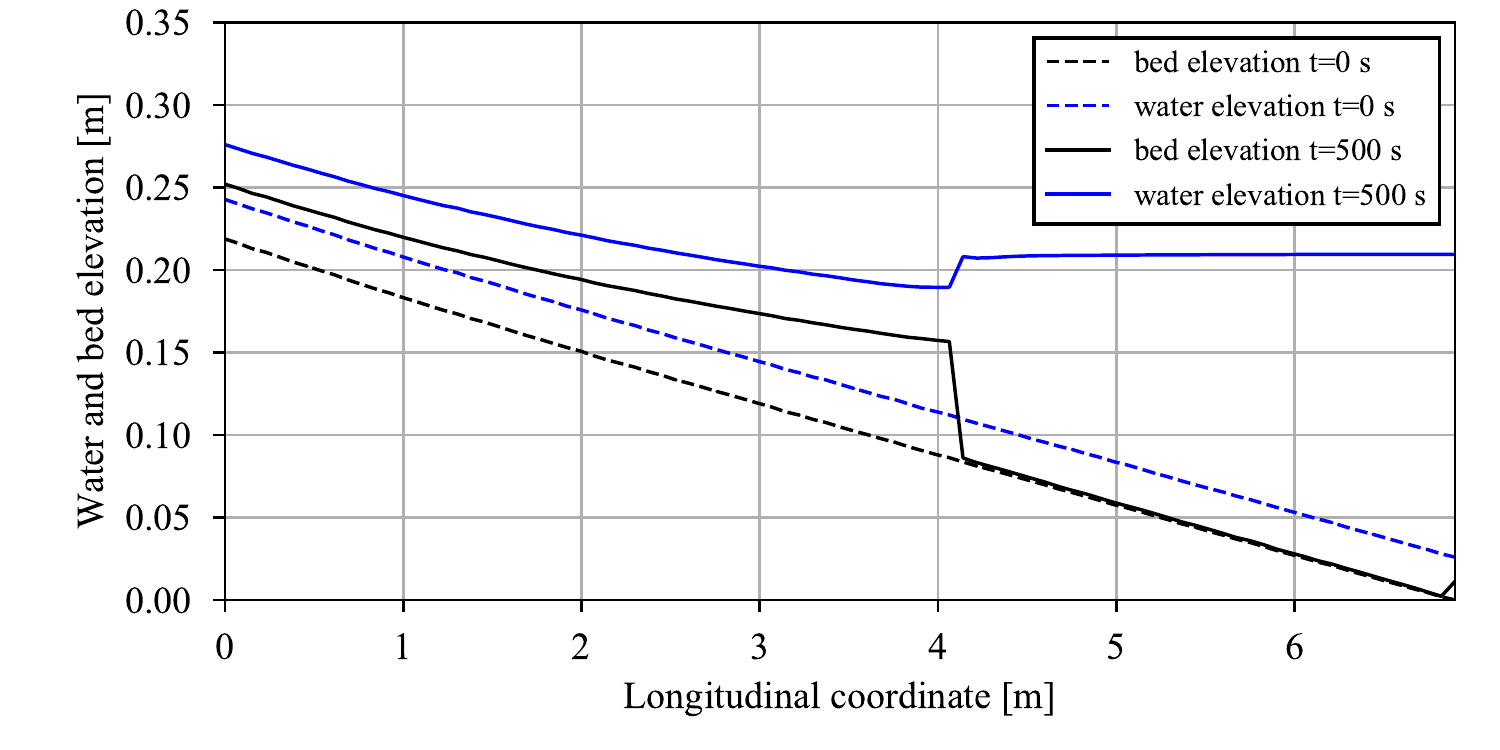}
    \caption{\textbf{T2: propagation of a sediment bore.} Initial (dashed lines) and final (solid lines) longitudinal profiles of bed elevation (black) and water elevation (blue) for the propagation of a sediment bore test.}
    \label{fig:Bellal_profile}
\end{figure}

\begin{figure}[tbp]
    \centering
    \includegraphics[angle=0, width=0.8\columnwidth]{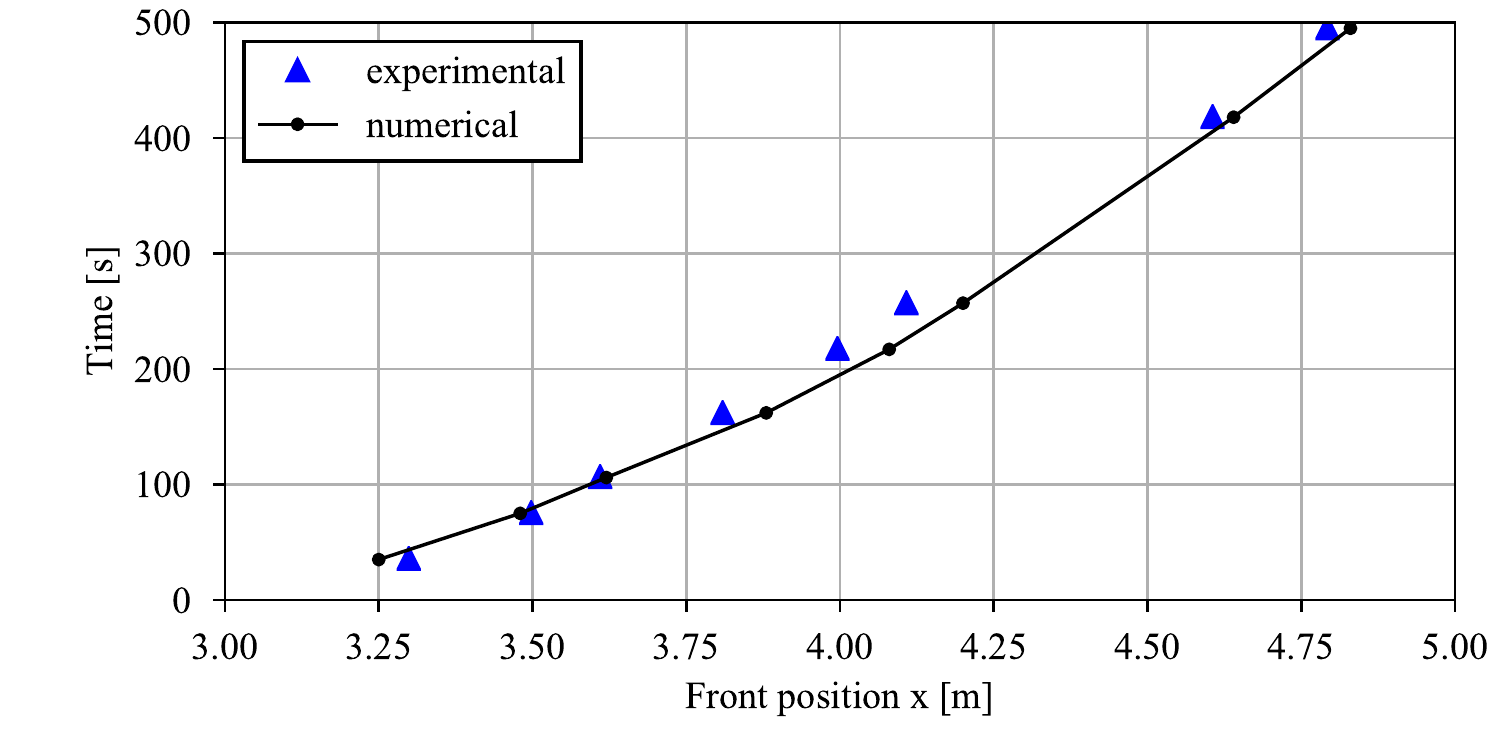}
    \caption{\textbf{T2: evolution in time of the sediment front position.} Blue triangles are the experimental values \citep{bellal2003}, while the black line is the numerical solution.}
    \label{fig:BellalFront}
\end{figure}

Fig.~\ref{fig:Bellal_profile} shows the initial and final profiles of the simulated bed and water elevations. The solver reproduces well the sharp transition between super- and sub-critical flow conditions. The position of the sediment front in time is shown in Fig.~\ref{fig:BellalFront}, with good agreement between simulated and experimental values.

\subsection{T3: Dam-break over a mobile bed with a sudden enlargement}\label{sec:xia_test}
\begin{figure}[tbp]
    \centering
    \includegraphics[angle=0, width=0.5\columnwidth]{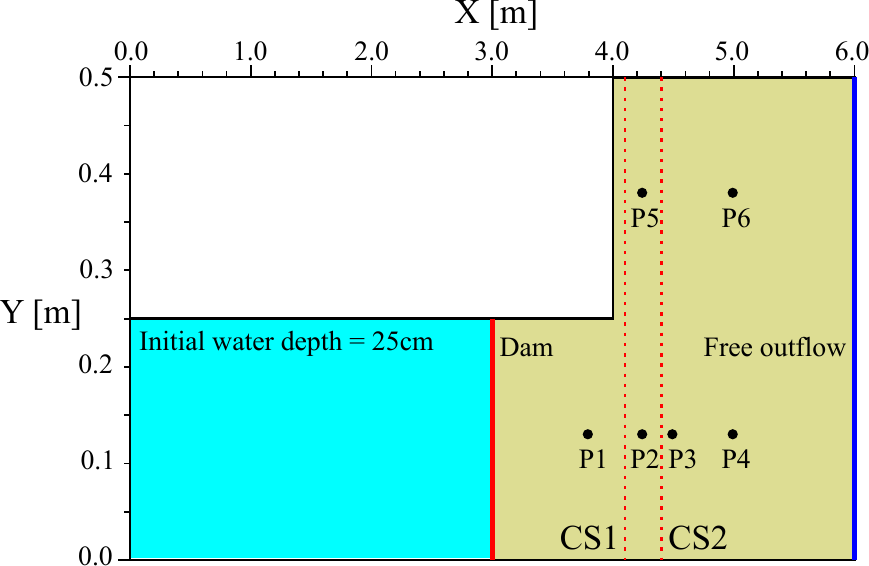}
    \caption{\textbf{T3: planar view of the dam-break over mobile bed setup} (deformed axis). Experimental and numerical results are compared at survey points P1 to P6 during the simulation, and at cross-sections CS1 and CS2 at the end of simulation.}
    \label{fig:XIAflume}
\end{figure}

Scope of the test is to assess code robustness in simulating sediment transport at wet-dry interface, and the accuracy in reproducing scour/deposition patterns. The experiment illustrated in \citep{Goutiere2011} represents a well-know morphodynamic test for numerical models \citep[e.g.][]{Siviglia2013,Soares2011,juez2014}. The domain consists of a flat flume with a non-symmetrical sudden enlargement (Fig.~\ref{fig:XIAflume}). The bed is composed of a coarse uniform sand with a median diameter of $d_m$=1.82~mm. The initial conditions are defined by an horizontal layer of fully saturated sand of thickness 0.1~m over the whole domain and an initial water storage of depth 0.25~m upstream of the dam, located at section $x$=3.0~m. At time $t$=0~s, the dam is suddenly removed, resulting in the propagation of a dam break wave with consequent sediment transport.

The computational domain is discretized by unstructured triangular cells at different resolutions (follows in Tab.~\ref{tab:Ncells_performance}). Inviscid wall boundary conditions are set at the upstream and lateral domain boundaries, while a free-outflow condition is used at the downstream outlet. The Manning coefficient is set to \SI{0.0167}{m^{-1/3}s}, the sediment density and porosity are set to \SI{2680}{kg/m^3} and 0.47, respectively. The sediment transport is evaluated with the MPM-like formula (Table~\ref{tab:sedtransport_closure}), setting $\theta_{cr}$= 0.0495 for the critical Shields stress, $\alpha$ = 3.97 and $m$ = 1.5 for the remaining parameters \citep{wong2006}. The CFL number is set to 0.9. The numerical simulations last 12 seconds.

\begin{figure}[tbp]
    \centering
    \includegraphics[angle=0, width=0.9\columnwidth]{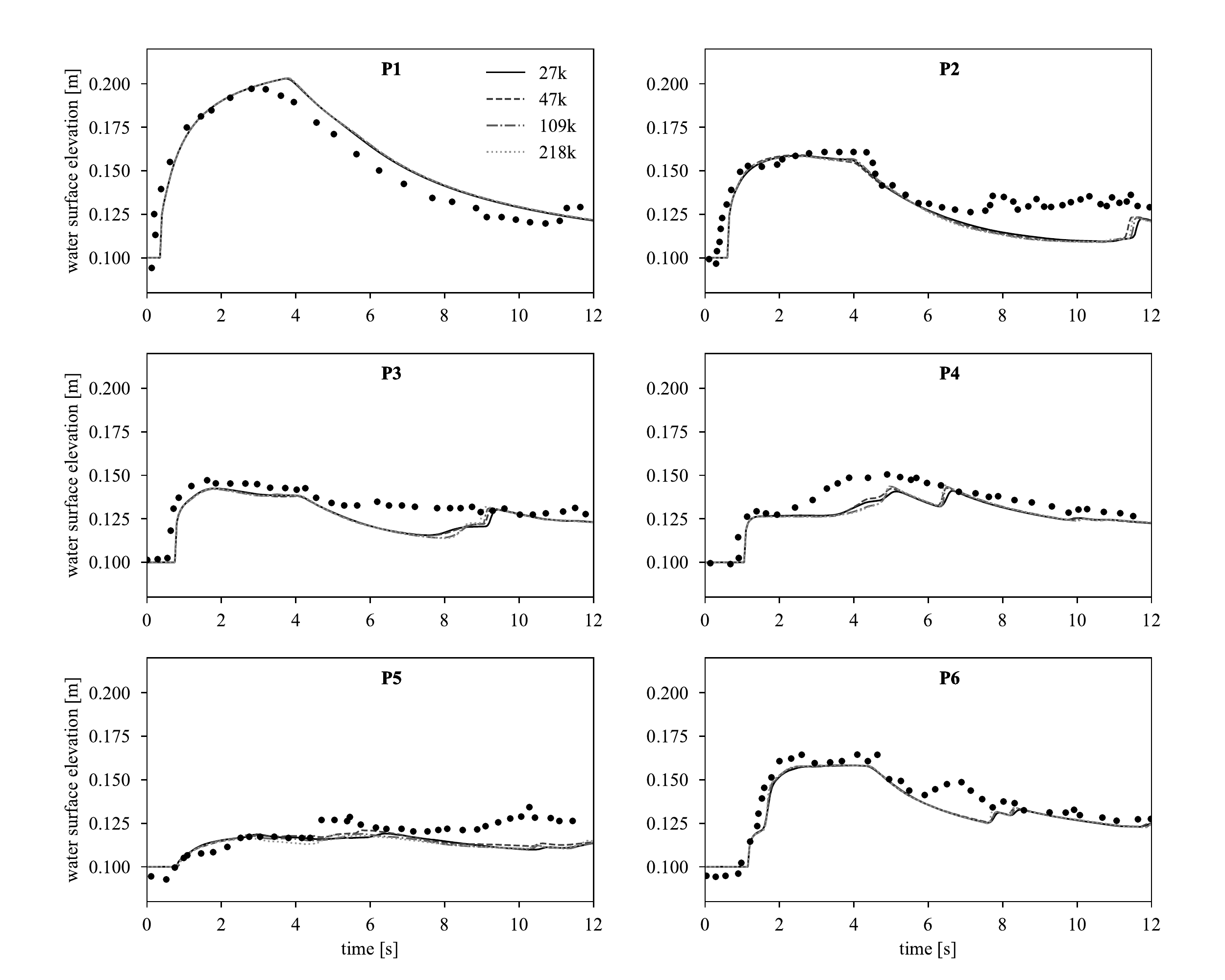}
    \caption{\textbf{T3: experimental and numerical water elevation at six survey points.} Sub-panels represent the survey points P1 to P6 as in Fig.~\ref{fig:XIAflume}: experimental points are given as full circles, whilst numerical results as lines, with four different mesh sizes (27k, 54k, 108k, 216k computational cells).}
    \label{fig:XIApoints}
\end{figure}

The evolution of the water elevation during the simulation is shown in Fig.~\ref{fig:XIApoints} for the six survey points. The simulated series show a fairly good agreement with the experimental values: the dam break wave arrival time is well captured and the maximum elevation values are comparable with the measured ones. Moreover, the simulated series show minor discrepancies with the experimental ones after the arrival of the first wave. As already pointed out by previous works \cite{Siviglia2013,Xia2010}, discrepancies are due to the extremely complex flow pattern generate by multiple wave reflections while simulation proceeds in time, which potentially generate tri-dimensional flow structures. Nevertheless, obtained series are coherent with the ones of \cite{Siviglia2013}, where a second-order accuracy model was employed. 

\begin{figure}[tbp]
    \centering
    \includegraphics[angle=0, width=0.8\columnwidth]{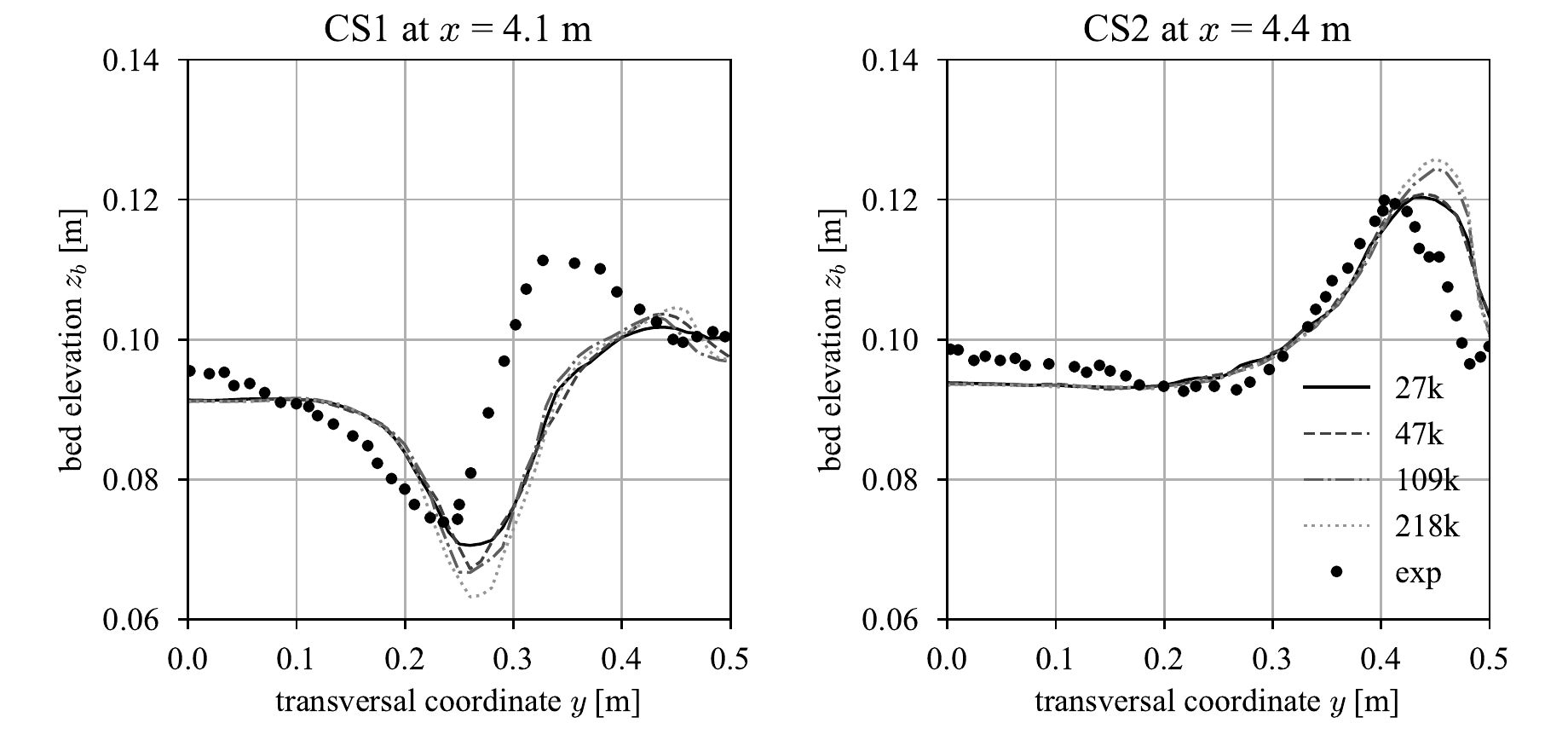}
    \caption{\textbf{T3: experimental and numerical bed elevation at different cross-sections.} Sub-panels represent cross-section CS1 and CS2 as in Fig.~\ref{fig:XIAflume}: experimental points are given as full circles, whilst numerical results as lines, with four different mesh sizes (27k, 54k, 108k, 216k computational cells).}
    \label{fig:XIAcrosssections}
\end{figure}

Numerical bed elevations after 12~s are compared to the experimental results in Fig.~\ref{fig:XIAcrosssections}. The simulated scour and deposition patterns are well reproduced. The magnitude of the scour at cross section CS1 ($y \approx 0.25$~m) matches well, with an underestimation of the deposition pattern ($y \approx 0.35$~m). At cross section CS2, the simulated deposition magnitude matches well with the experimental one, but with a small shift toward the lateral boundary.

\subsection{T4: Scour and deposition on a channel bend}
The scope of the simulation is to test the correct reproduction, both in term of positioning and magnitude, of a river point bar generated by a channel bend. In this test we numerically reproduced one experiment from \cite{yen1995}, already adopted as morphodynamic benchmark test \citep[e.g.][]{kaveh2019}. The flume is U-shaped, with a bend of \ang{180} having a costant radius along the center line of $R_c$=4~m. The cross section is rectangular with width $W$=1~m and slope $S$=0.2\%. The two straight reaches before and after the bend are 11.5~m long. The median diameter of the bed material was $d_{m}$=1~mm. In the experimental run the flume was fed with a simplified (triangular) flood hydrograph, having a base flow of \SI{0.02}{\cubic\meter\per\second}, and peak flow of \SI{0.053}{\cubic\meter\per\second}. The rising and falling limbs last 100 and 200 minutes, respectively. Afterwards a constant baseflow was kept for another 100 minutes. During the experimental run, a steady point bar in the inner side of the bed develops and grows, with a corresponding erosion on the outer side.

In the numerical setup, the domain is discretized with 24,523 computational cells. We set the porosity to 0.4, and used the MPM-like formula with parameters from \cite{wong2006}, as in Table~\ref{tab:sedtransport_closure}. The lateral slope factor $N_l$ is equal to 1.4 and the curvature factor $N*$ is set to 11 (\ref{eq:lateral_slope_correction} and \ref{eq:spiral_flow_correction}). At the numerical domain boundaries uniform flow and equilibrium sediment transport conditions are imposed. The simulation, as the experimental run, lasts 400 minutes.

\begin{figure}[tbp]
    \centering
    \includegraphics[angle=0, width=0.8\columnwidth]{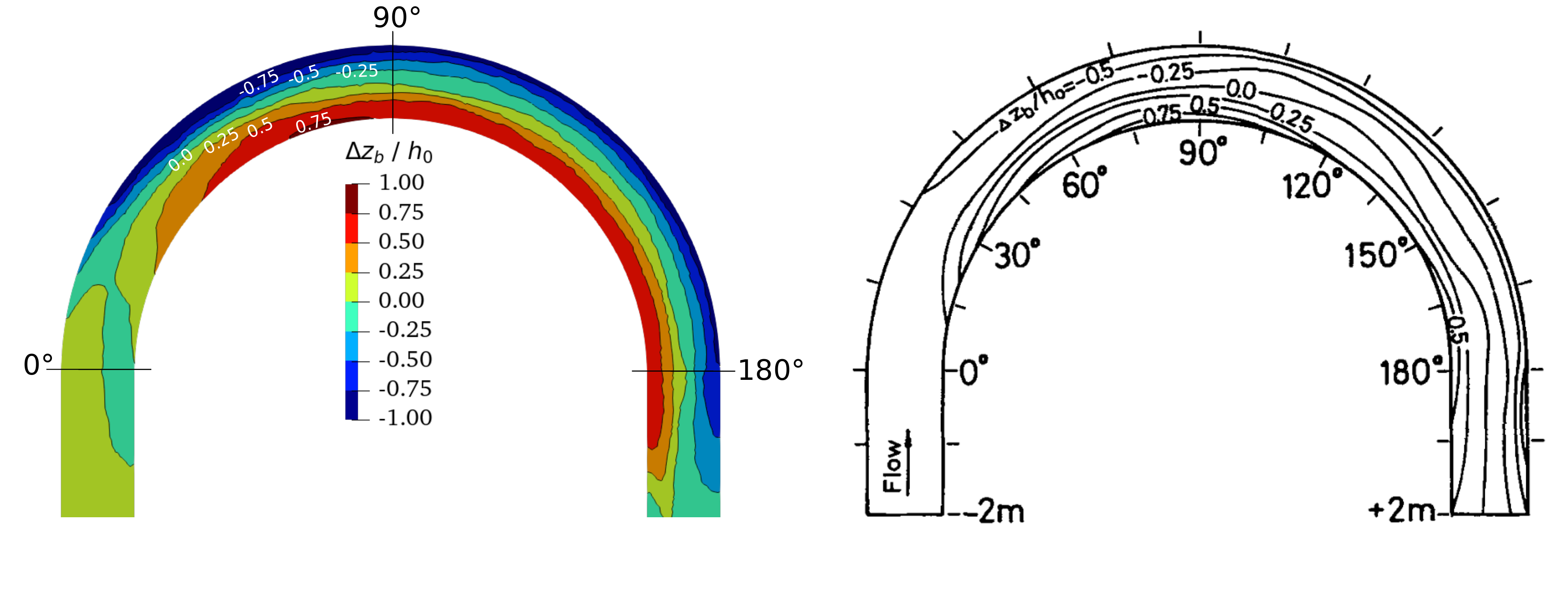}
    \caption{\textbf{T4: scour and deposition on a channel bend.} Planar view of relative bed change $\Delta$ $z_b$/ $h_0$, with $h_0$ approaching flow depth. Numerical results on the left, laboratory results from \citep{yen1995} on the right.}
    \label{fig:UFlumeContour}
\end{figure}

A planar view comparison between numerical and experimental run is depicted in Fig.~\ref{fig:UFlumeContour}. The final bed variation with respect to the initial flat configuration is scaled with the approaching (i.e. upstrem reach) flow depth $h_0$. The magnitude of scours and depositions for the numerical run ranges between -0.75 and 0.75, matching fairly well with the experimental values. Also the positioning of the point bar, with the maximum deposition anticipating the middle of the bend (\ang{90}) is well reproduced numerically.

\begin{figure}[tbp]
    \centering
    \includegraphics[angle=0, width=0.6\columnwidth]{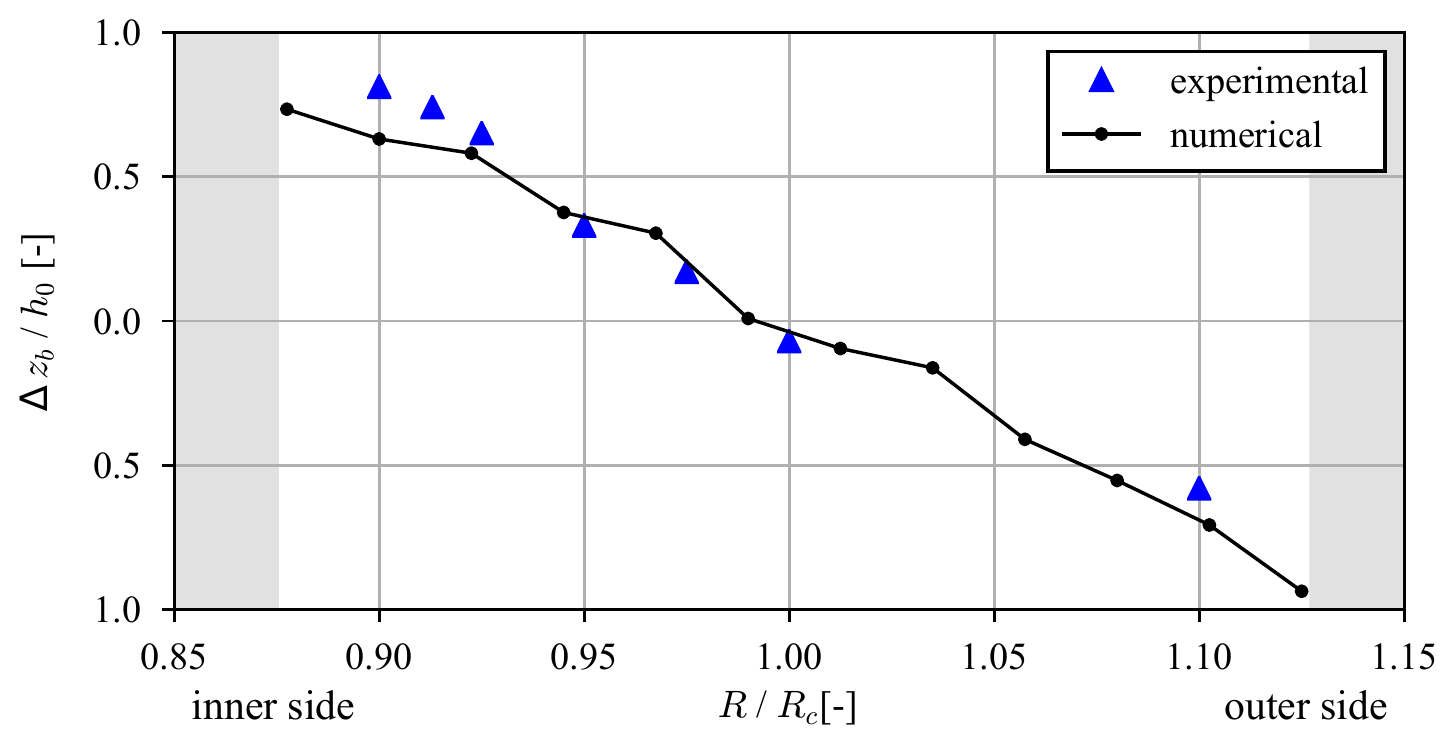}
    \caption{\textbf{T4: scour and deposition at one cross-section.} Cross-sectional view (at \ang{90}) of relative bed change $\Delta$ $z_b$/ $h_0$, with $h_0$ approaching flow depth. In the $x$-axis the radial coordinate $R$ is scaled with the center-line curvature radius $R_c$; black line is the numerical solution, blue triangles are from the laboratory experiment of \citep{yen1995}.}
    \label{fig:UFlume90deg}
\end{figure}

Fig.~\ref{fig:UFlume90deg} shows the cross-sectional profile of the relative bed change for the numerical simulation and the experimental run in the middle of the flume bend (at \ang{90}). The numerical profile reproduces well the experimental trend. This test demonstrates the software capability in simulating an unsteady morphological building process, i.e. a point bar development during a flood. Such process can be well reproduced only by implementing suitable corrections of the sediment transport direction due to gravity and curvature, as presented in \S\ref{sec:morpho_closures}.

\subsection{T5: Steady scalar discontinuity with two diverging hydrodynamic waves}
The test assesses the correct advection of scalar concentration. This is assessed with a challenging test: a steady discontinuity of a scalar concentration subjected to strongly variable flow conditions. The chosen test is an idealized one-dimensional problem, but nevertheless it is particularly challenging for a pletora of numerical scheme \citep{Toro2001}. The domain is a simple, flat-bed, channel 100~m long and 0.1~m wide. The domain is deliberately chosen very narrow, to mimic a 1D setup, given that the exact solution of the problem is available in one-dimension case. As initial conditions, the water depth $h$ is set even in all the domain, whilst the initial longitudinal specific discharge $q_x$  and the concentration of a generic scalar $\phi$ present a discontinuity:
\begin{equation}\label{eq:T5_IC}
\left\{
\begin{aligned}
q_x&=-3.0~\si{m^2/s} & \text{if }x&<50~\si{m}, & q_x&=3.0~\si{m^2/s} &\text{\:otherwise},\\
\phi&=1.0       & \text{if }x&<50~\si{m}, & \phi&=0.0 
&\text{\:otherwise},\\
h & =1.0~\si{m} & \forall x&.
\end{aligned}
\right.
\end{equation}

The domain is discretized with 1362 computational cells, lateral walls are reflective and inviscid, whilst transparent boundary conditions are set at beginning and end of the channel. The CFL is set to 0.95 and the simulation timeout is $t$=2.5~s. As the simulation starts, two strong rarefaction waves start diverging from the center of the domain towards the two extremities, suddenly forming a water depression in the center. Despite the strong unsteadiness of the hydrodynamic quantities during the simulation, a steady contact wave persists in the domain, avoiding the scalar quantity to mix in the domain. 

\begin{figure}
\centering
\subfloat[]{\includegraphics[width=0.45\textwidth]{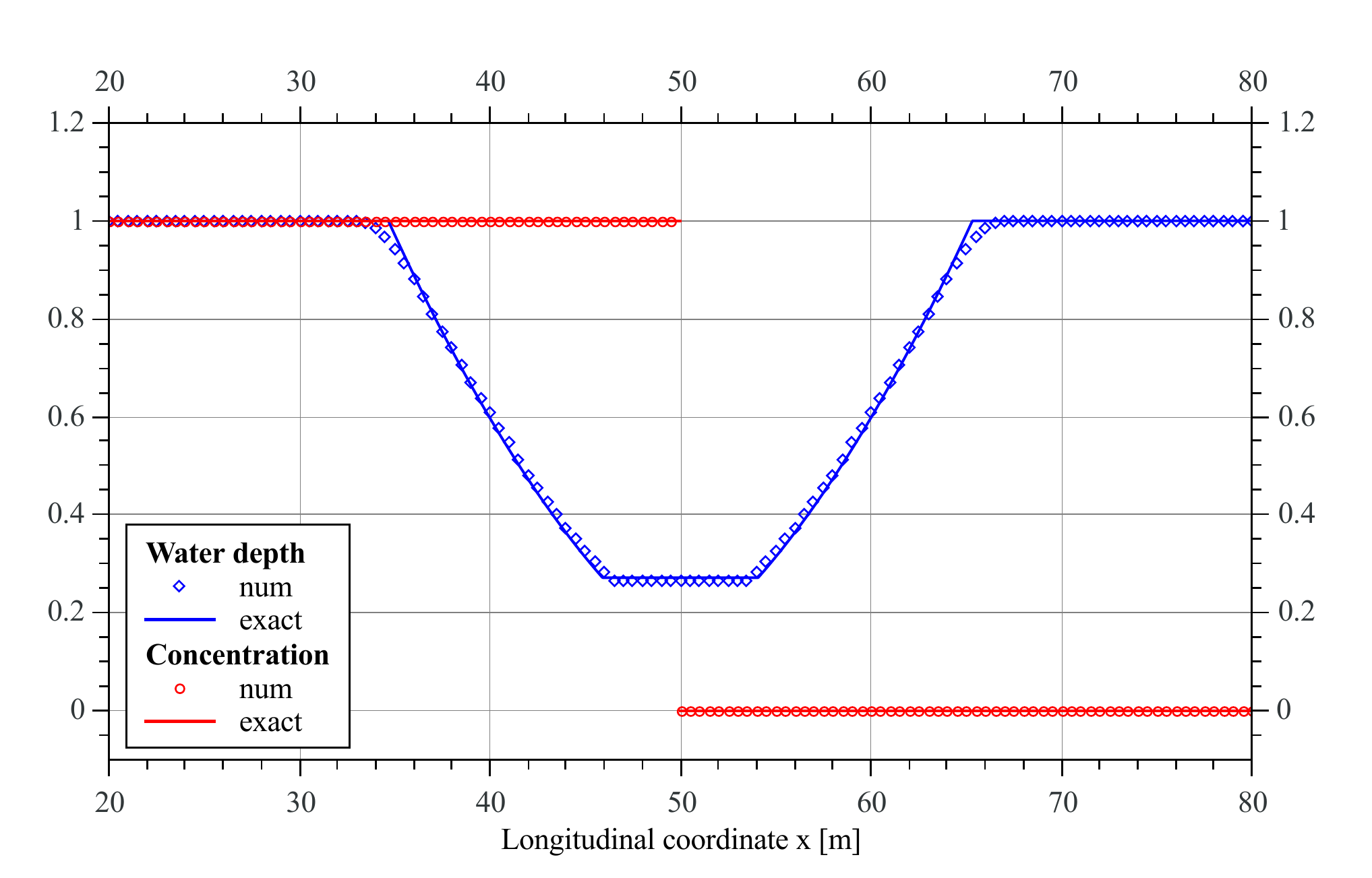}}\qquad
\subfloat[]{\includegraphics[width=0.45\textwidth]{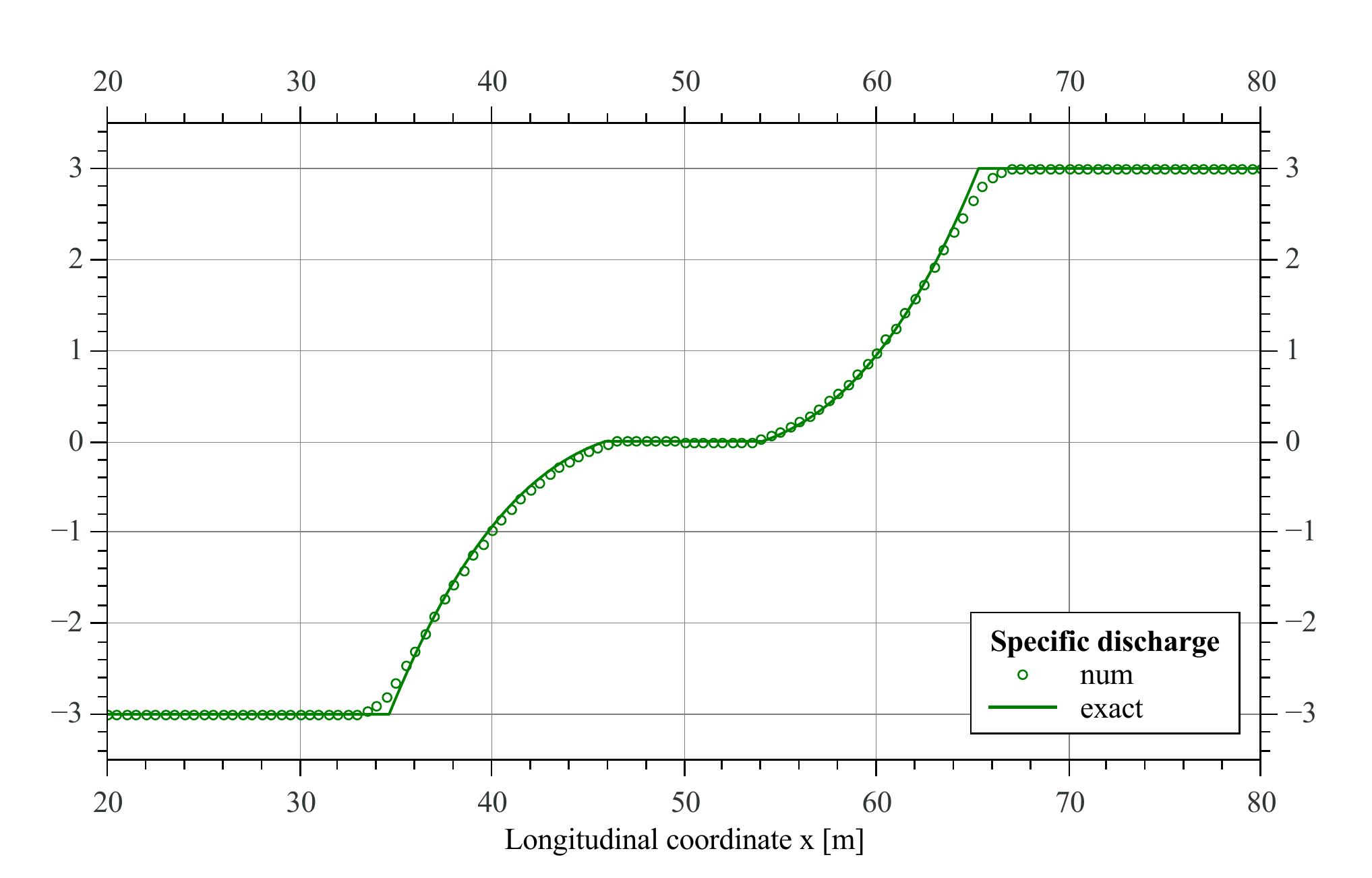}}\\
\caption{\textbf{T5: comparison between numerical and exact solution of a steady contact wave.} Solution is given at timeout $t$=2~s. Exact solutions are depicted with lines, whereas numerical values are symbols, decimated for the sake of visualization. Panel \textbf{(a)} shows the water depth (blue diamonds) and the scalar concentration (red circles), panel \textbf{(b)} shows the specific discharge on the x direction (green circles).}\label{fig:T5_redsea}
\end{figure}

The numerical solution at simulation timeout is compared with the exact solution of the problem in Fig.~\ref{fig:T5_redsea}. The hydrodynamic exact solution is obtained by resolving the two-rarefaction Riemann Problem \citep{Toro2001}, whilst the exact solution for the scalar advection is identical to the given initial conditions. Fig.~\ref{fig:T5_redsea} underlines how the numerical solution correctly approximates the exact solution in all the domain sections. The scalar discontinuity is perfectly maintained throughout the simulation, confirming the accurate resolution of the steady contact wave.

\subsection{T6: Scalar advection and diffusion in a dam-break over a complex domain}\label{sec:3humps_tracer_test}
With this test we assess the solver capability in correctly preserving the liquid and scalar species mass when simulating the advection and diffusion of species during a dam-break phenomena. The test is particularly harsh, due to the presence of fast wetting-drying fronts and multiple discontinuous flow regions.

The test is an \textit{ad-hoc} setup inspired to a common benchmark for hydrodynamic codes \citep[e.g.][]{Vanzo2016,Brufau2002}. The domain is composed by a rectangle $[0;75]\times[-15;15]$~m. The bottom $\eta(x,y)$ is fixed during the simulation, and defined as
\begin{equation}\label{eq:T6_bedelevation}
    \begin{aligned}
        \eta(x,y) &= \mathrm{max} \left( 0, \eta_1(x,y), \eta_2(x,y), \eta_3(x,y) \right)\text{, with} \\
        \eta_1(x,y) &= 1 - \frac{1}{8}  \sqrt{(x-30)^2+(y + 9)^2},\\
        \eta_2(x,y) &= 1 - \frac{1}{8}  \sqrt{(x-30)^2+(y - 9)^2}, \\
        \eta_3(x,y) &= 3 - \frac{3}{10} \sqrt{(x-47.5)^2+y^2}. \\
    \end{aligned}
\end{equation}
The initial conditions are given by:
\begin{equation}\label{IC_3humps}
\left\{
\begin{aligned}
h       &= 1.0~\si{m}   &&\text{if } x < 16~\si{m}, \:\: &h &= 0.125~\si{m}  &\text{\:otherwise},\\
\phi_1  &= 0.5     &&\text{if } x < 16~\si{m},      &\phi_1 &= 0   &\text{\:otherwise},\\
\phi_2  &= 0       &&\text{if } x < 16~\si{m},      &\phi_1 &= 0.1 &\text{\:otherwise},\\
q_{x} &=q_{y}= 0~\si{m^2/s}       &&\text{everywhere},
\end{aligned}
\right.
\end{equation}
presenting a virtual dam at $x$=16~m separating two discontinuous volumes of water and scalar mass. 
Here the domain is discretized with 492,277 triangular cells with a maximum of characteristic length of 0.1~m. The hydrodynamics setup features reflective wall boundaries, a CFL of 0.95 and frictional sources compatible with a Manning coefficient of $n$ = \SI{0.01}{m^{-1/3}s}. The scalar setup features two initially unmixed species, both with a constant and isotropic diffusion coefficient $K_c$=\SI{0.25}{m^2/s}. Fig.~\ref{fig:scl_3humps}a illustrates the hydrodynamic (left) and scalar solutions (right) at the initial condition ($t$=0~s).

At simulation start, the virtual dam collapses instantaneously, with an advancing wave that overtops the two small lateral humps, fully circumvents the larger hump and reaches the opposite wall in about $t$=20~s. At this time the interface between the two species, in what would otherwise be a contact discontinuity on flat topography, is still lagging by approximately 15~m (Fig.~\ref{fig:scl_3humps}b). At this point, the reflected bores propagate upstream and further mix both species, symmetrically around the $x$ axis. By $t$=50~s these bores overcome the two smaller obstacles and propagate upstream on flat ground (Fig.~\ref{fig:scl_3humps}c).

\begin{figure}[tbp]
	\begin{minipage}{0.45\linewidth}
	    \centering
		\def\seriesfile{figures/scalars/series/3humps-left.dat}
        \def\stillfile{figures/scalars/series/3humps-still.dat}
	    \definecolor{blue}{RGB} {0, 158, 227}
\definecolor{gray}{RGB} {75, 86, 92}

\definecolor{darkBlue}{RGB} {9, 101, 140}
\definecolor{darkGray}{RGB} {20, 23, 25}
\definecolor{pastelRed}{RGB} {210, 50, 33}

\begin{tikzpicture}
	
	\pgfplotsset{
	  outer axis line style={line width=1pt, color=darkGray},
	  axis lines=left,
	  grid=major,
	  major grid style={line width=0.3pt, draw=darkGray!5},
	}

	\begin{axis}[
		unbounded coords=jump,
		width = 1.0\linewidth,
		height = 0.67\linewidth,
		xmin=-10, xmax=12100,
		ymin=0, ymax=0.525,
		xtick={0.0,1800.0,...,10800.0},
		ytick={0.0,0.1,...,1.0},
		xticklabels={0.0,0.5,...,3.0},
		x tick label style={
			/pgf/number format/.cd,
			fixed,
			fixed zerofill,
			precision=1,
			/tikz/.cd
		},
		y tick label style={
			/pgf/number format/.cd,
			fixed,
			fixed zerofill,
			precision=1,
			/tikz/.cd
		},
		scaled x ticks=false,
		xlabel= $t$ (h),
		legend pos=north east,
		legend columns=1,
		legend style={draw=none, fill=white, column sep=0.5em},
		legend cell align={left},
		]
				
		\addplot[darkGray, ultra thick, smooth] table [x=t, y=h]{\seriesfile};
		\addplot[darkBlue, ultra thick, smooth] table [x=t, y=qx]{\seriesfile};
		\addplot[red, ultra thick, smooth] table [x=t, y=c1]{\seriesfile};
		\addplot[blue, ultra thick, smooth] table [x=t, y=c2]{\seriesfile};
		
		\addplot[darkGray, thick, dotted] table [x=t, y=h]{\stillfile};
		\addplot[red, thick, dashed] table [x=t, y=c1]{\stillfile};
		\addplot[blue, thick, dashed] table [x=t, y=c2]{\stillfile};
		
	\end{axis}
\end{tikzpicture}

\let\datafile=\relax

				

		

		
		
	\end{minipage}
	\begin{minipage}{0.45\linewidth}
	    \centering
		\def\seriesfile{figures/scalars/series/3humps-right.dat}
        \def\stillfile{figures/scalars/series/3humps-still.dat}
	    \definecolor{blue}{RGB} {0, 158, 227}
\definecolor{gray}{RGB} {75, 86, 92}

\definecolor{darkBlue}{RGB} {9, 101, 140}
\definecolor{darkGray}{RGB} {20, 23, 25}
\definecolor{pastelRed}{RGB} {210, 50, 33}

\begin{tikzpicture}
	
	\pgfplotsset{
	  outer axis line style={line width=1pt, color=darkGray},
	  axis lines=left,
	  grid=major,
	  major grid style={line width=0.3pt, draw=darkGray!5},
	}

	\begin{axis}[
		unbounded coords=jump,
		width = 1.0\linewidth,
		height = 0.67\linewidth,
		xmin=-10, xmax=12100,
		ymin=0, ymax=0.525,
		xtick={0.0,1800.0,...,10800.0},
		ytick={0.0,0.1,...,1.0},
		xticklabels={0.0,0.5,...,3.0},
		x tick label style={
			/pgf/number format/.cd,
			fixed,
			fixed zerofill,
			precision=1,
			/tikz/.cd
		},
		y tick label style={
			/pgf/number format/.cd,
			fixed,
			fixed zerofill,
			precision=1,
			/tikz/.cd
		},
		scaled x ticks=false,
		xlabel= $t$ (h),
		legend pos=north east,
		legend columns=1,
		legend style={draw=none, fill=white, column sep=0.5em},
		legend cell align={left},
		]
				
		\addplot[darkGray, ultra thick, smooth] table [x=t, y=h]{\seriesfile};
		\addplot[darkBlue, ultra thick, smooth] table [x=t, y=qx]{\seriesfile};
		\addplot[red, ultra thick, smooth] table [x=t, y=c1]{\seriesfile};
		\addplot[blue, ultra thick, smooth] table [x=t, y=c2]{\seriesfile};
		
		\addplot[darkGray, thick, dotted] table [x=t, y=h]{\stillfile};
		\addplot[red, thick, dashed] table [x=t, y=c1]{\stillfile};
		\addplot[blue, thick, dashed] table [x=t, y=c2]{\stillfile};
		
	\end{axis}
\end{tikzpicture}

\let\datafile=\relax

				

		

		
		
	\end{minipage}
	\\ \vspace{5pt} \\
	\begin{minipage}{1.0\linewidth}
	    \centering
	    \definecolor{blue}{RGB} {0, 158, 227}
\definecolor{gray}{RGB} {75, 86, 92}

\definecolor{darkBlue}{RGB} {9, 101, 140}
\definecolor{darkGray}{RGB} {20, 23, 25}
\definecolor{pastelRed}{RGB} {210, 50, 33}

\def\datafile{scalar-test-1.dat}

\begin{tikzpicture}
	
	\begin{axis}[
		hide axis,
		legend entries={{111},{222},{333},{444},{555}},
		legend pos=north east,
		legend columns=-1,
				legend style={draw=none,
		            fill=white,
		            /tikz/every even column/.append style={column sep=10pt}
		            },
		legend cell align={center},
		]
				
		\addplot[darkGray, ultra thick] coordinates {(0,0)};
		\addlegendentry{$h$ ($m$)}
		
		\addplot[darkBlue, ultra thick] coordinates {(0,0)};
		\addlegendentry{$q_x$ ($m^2/s$)}

		\addplot[red, ultra thick] coordinates {(0,0)};
		\addlegendentry{$\phi_1$ (-)}
		
		\addplot[blue, ultra thick] coordinates {(0,0)};
		\addlegendentry{$\phi_2$ (-)}
	
	\end{axis}
\end{tikzpicture}

\let\datafile=\relax
	\end{minipage}
	\\ \vspace{-60pt} \\
	\caption{\textbf{T6: time series at two locations.} Evolution of hydrodynamic quantities ($h$ and $q_x$) and scalar concentrations $\phi_1$ (red) and $\phi_2$ (blue) in time at position $(x,y) = (7.5,0.0)$~m (left) and $(x,y) = (67.5,0.0)$~m (right). Dashed lines represent theoretical scalar concentration values, at rest.}
    \label{fig:scl_3humps2}
\end{figure}
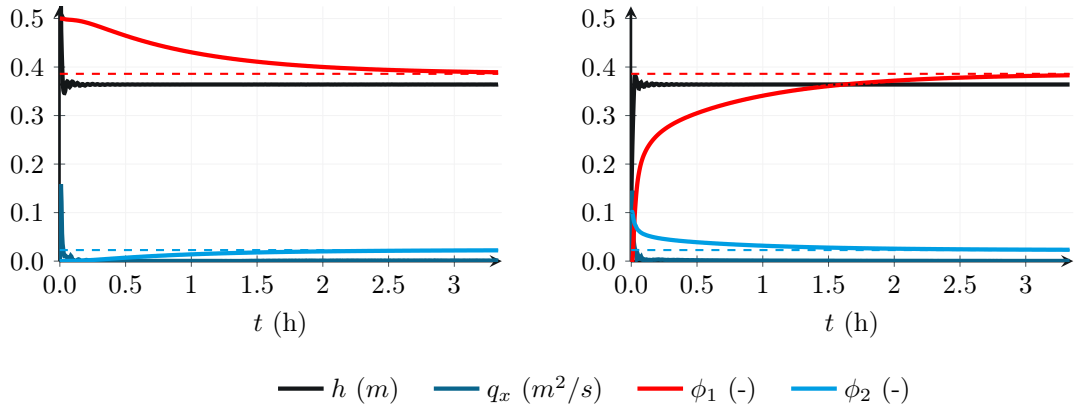

After a continuous sloshing and interaction of reflected waves, topography and lateral walls, the friction sources gain relevance and dissipate most of the kinetic energy in the flow, with a near-static solution being obtained at approximately $t$=20~min. The scalars continue to mix, now due mostly to molecular diffusion, in what is a much slower process, that only vanishes at around 3.5~hours as both scalars become fully homogeneous across the domain (Fig.~\ref{fig:scl_3humps2}).

The model is fully conservative, with the total liquid and scalar mass preserved during the entire simulation. As the simulation approaches the lake-at-rest conditions, the observed quantities correctly converge to their resting values of $h$=0.364~m, $\phi_1$=0.386 and $\phi_2$=0.023 (Fig.~\ref{fig:scl_3humps2}).

\begin{figure}[tbp]
    \centering
    \def\widthHyd{0.375\linewidth}
    \def\widthScl{0.4\linewidth}
    \def\gap{0.01\linewidth}
	\begin{minipage}{\widthHyd}
	    \centering
        \includegraphics[trim=225 100 275 100,clip,
        width=1.0\linewidth]{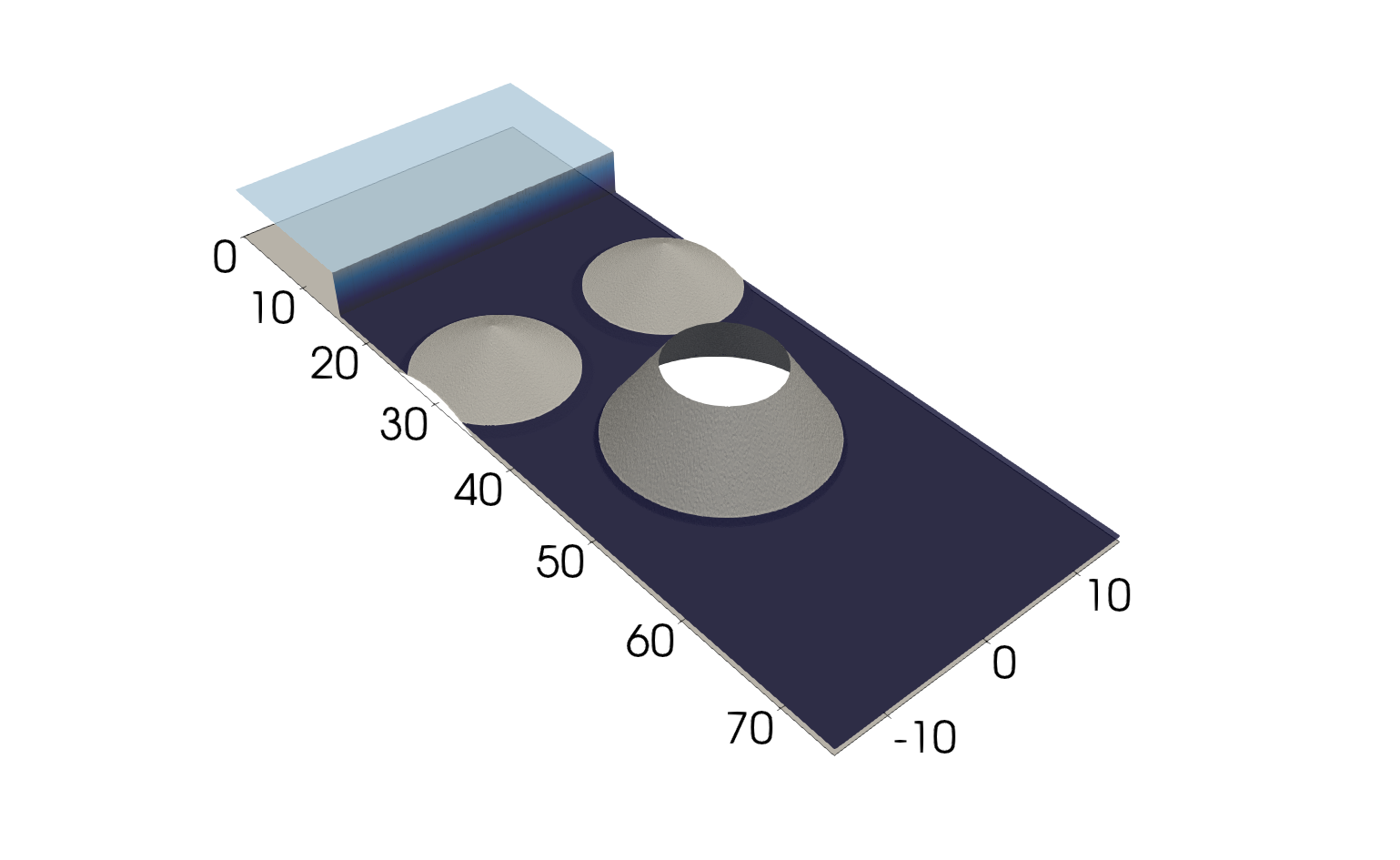}
	\end{minipage}
	\hspace{\gap}
	\begin{minipage}{\widthScl}
	    \centering
	    \vspace{25pt}
		\def\scalarimg{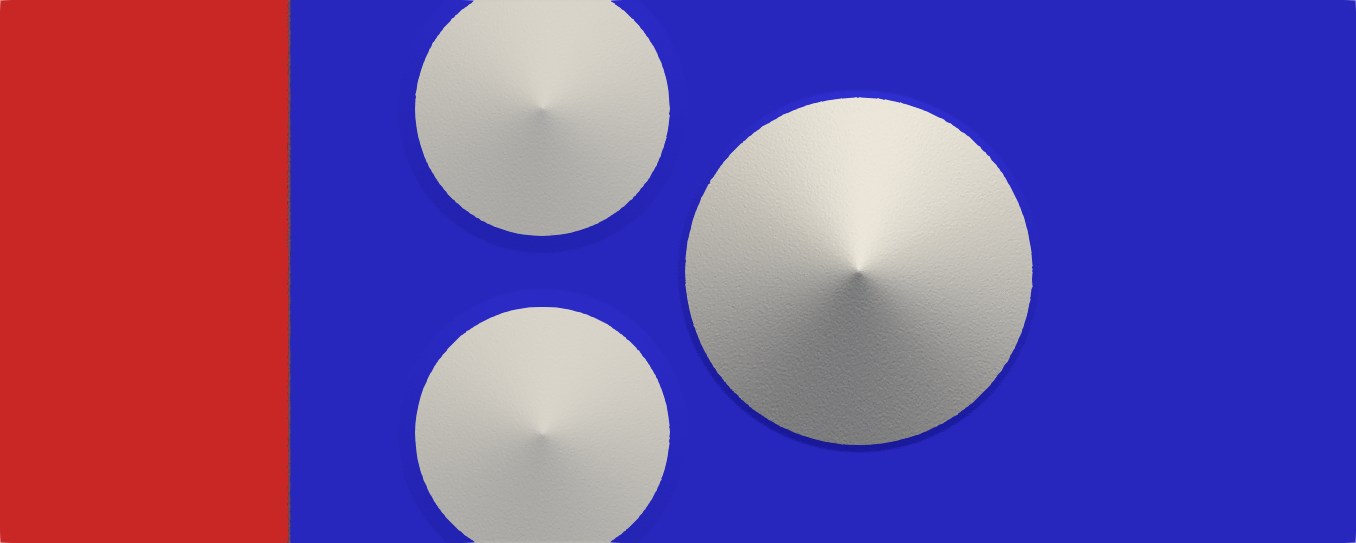}
	    \begin{tikzpicture}
	
	\begin{axis}[
	width=1.0\linewidth,
	height=0.4\linewidth,
	scale only axis,,
	axis lines=left,
	outer axis line style={line width=1pt, color=gray},
	enlargelimits=false,
	xmin=0, xmax=76,
	ymin=-15, ymax=16,
	xtick={0,15,...,75},
	xlabel={$x \ (m)$},
	xlabel near ticks,
	x tick label style={
		/pgf/number format/.cd,
		fixed,
		fixed zerofill,
		precision=0,
		/tikz/.cd
	},
	ylabel={$y \ (m)$},
	ylabel near ticks,
	ytick={-10,0,10},
	y tick label style={
	/pgf/number format/.cd,
		fixed,
		fixed zerofill,
		precision=0,
		/tikz/.cd
	},
	clip=true,
    set layers,
    clip mode=individual
	]
	       
	\addplot[thick, color=blue, on layer=axis background] graphics[xmin=0,ymin=-15,xmax=75,ymax=15] {\scalarimg};
		
	\end{axis}	
	
\end{tikzpicture}

\let\scalarimg=\relax
	\end{minipage}
	\\ a) $t$=0~s \vspace{10pt} \\
	\begin{minipage}{\widthHyd}
	    \centering
        \includegraphics[trim=225 100 275 100,clip,
        width=1.0\linewidth]{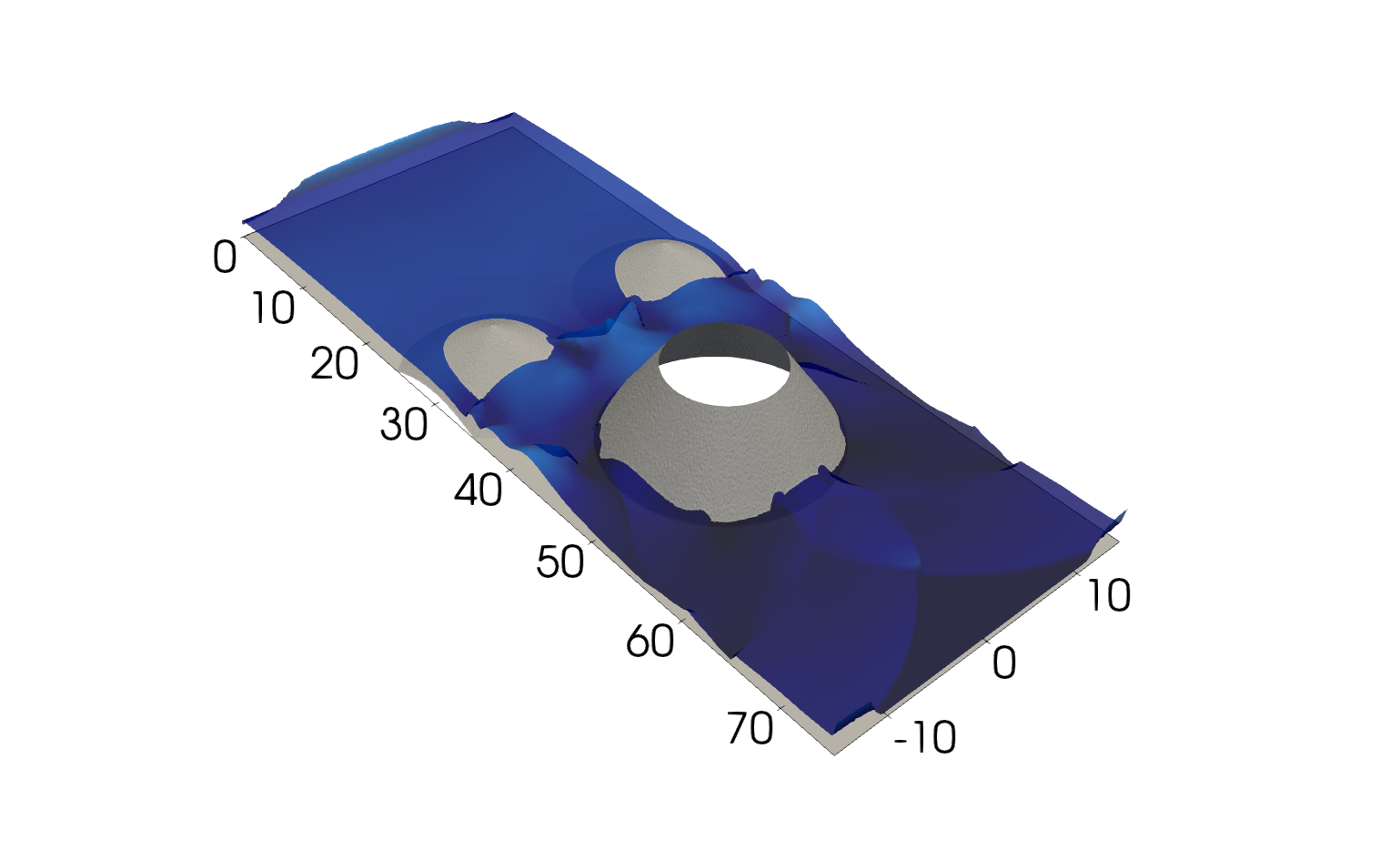}
	\end{minipage}
	\hspace{\gap}
	\begin{minipage}{\widthScl}
	    \centering
	    \vspace{25pt}
		\def\scalarimg{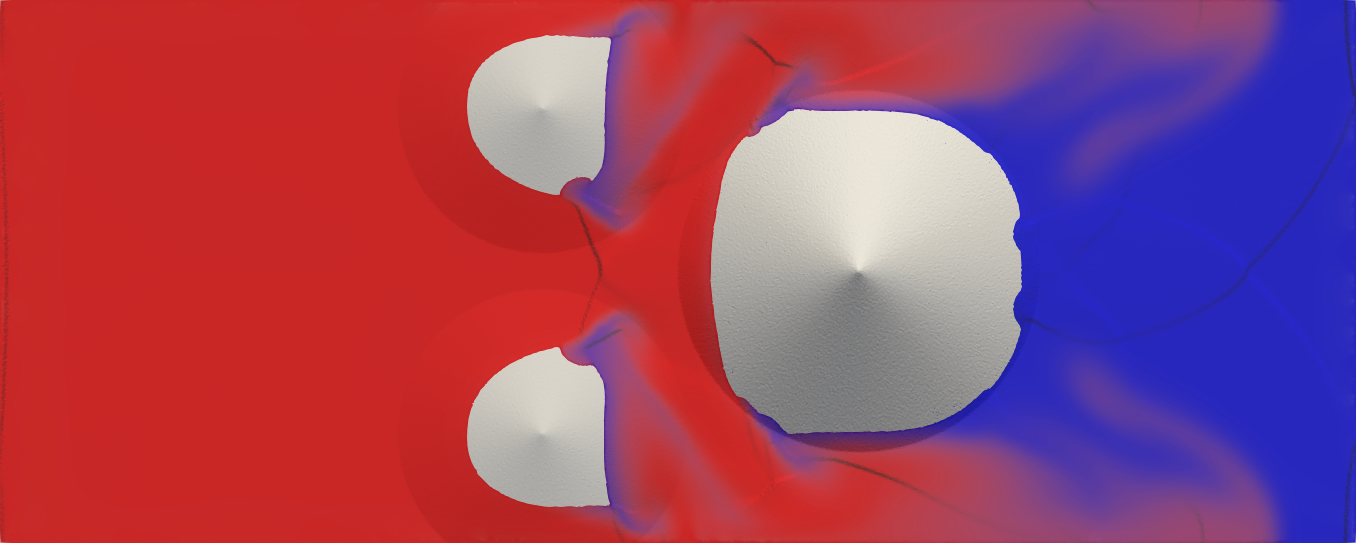}
	    \begin{tikzpicture}
	
	\begin{axis}[
	width=1.0\linewidth,
	height=0.4\linewidth,
	scale only axis,,
	axis lines=left,
	outer axis line style={line width=1pt, color=gray},
	enlargelimits=false,
	xmin=0, xmax=76,
	ymin=-15, ymax=16,
	xtick={0,15,...,75},
	xlabel={$x \ (m)$},
	xlabel near ticks,
	x tick label style={
		/pgf/number format/.cd,
		fixed,
		fixed zerofill,
		precision=0,
		/tikz/.cd
	},
	ylabel={$y \ (m)$},
	ylabel near ticks,
	ytick={-10,0,10},
	y tick label style={
	/pgf/number format/.cd,
		fixed,
		fixed zerofill,
		precision=0,
		/tikz/.cd
	},
	clip=true,
    set layers,
    clip mode=individual
	]
	       
	\addplot[thick, color=blue, on layer=axis background] graphics[xmin=0,ymin=-15,xmax=75,ymax=15] {\scalarimg};
		
	\end{axis}	
	
\end{tikzpicture}

\let\scalarimg=\relax
	\end{minipage}
	\\ b) $t$=20~s \vspace{10pt} \\
	\begin{minipage}{\widthHyd}
	    \centering
        \includegraphics[trim=225 100 275 100,clip,
        width=1.0\linewidth]{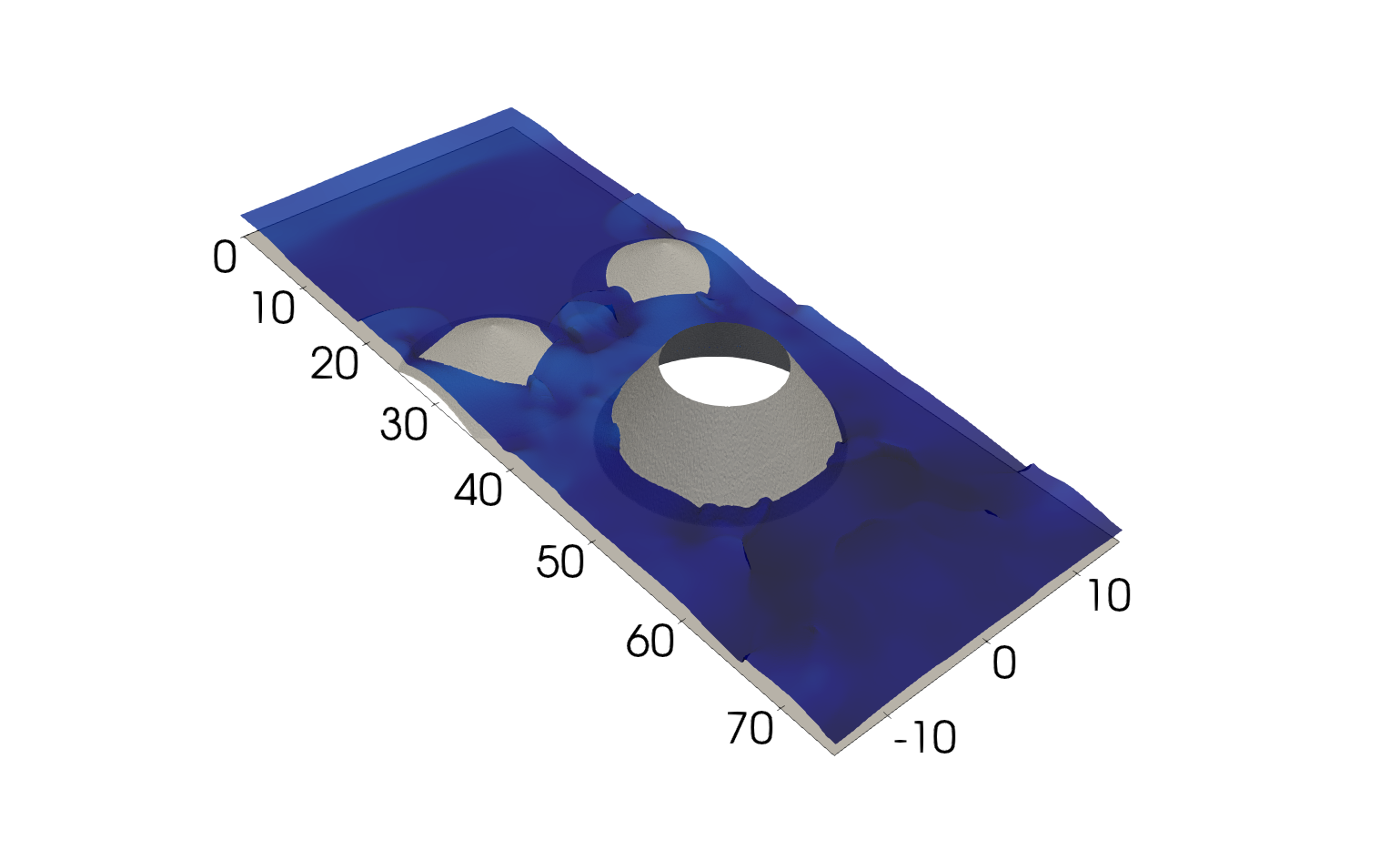}
	\end{minipage}
	\hspace{\gap}
	\begin{minipage}{\widthScl}
	    \centering
	    \vspace{25pt}
		\def\scalarimg{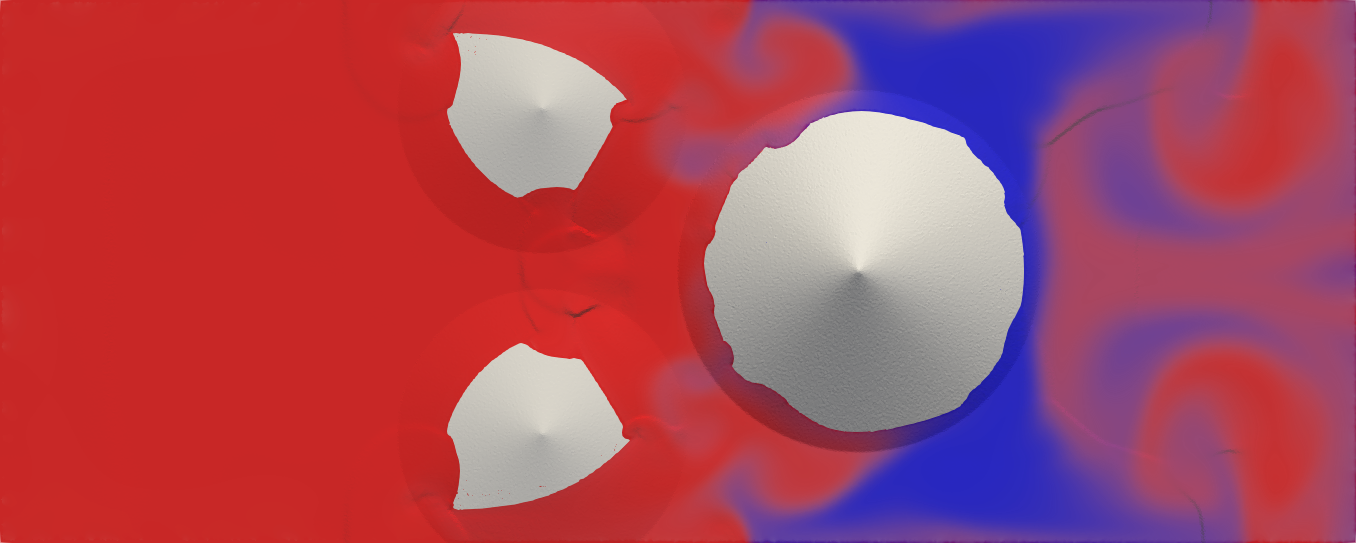}
	    \begin{tikzpicture}
	
	\begin{axis}[
	width=1.0\linewidth,
	height=0.4\linewidth,
	scale only axis,,
	axis lines=left,
	outer axis line style={line width=1pt, color=gray},
	enlargelimits=false,
	xmin=0, xmax=76,
	ymin=-15, ymax=16,
	xtick={0,15,...,75},
	xlabel={$x \ (m)$},
	xlabel near ticks,
	x tick label style={
		/pgf/number format/.cd,
		fixed,
		fixed zerofill,
		precision=0,
		/tikz/.cd
	},
	ylabel={$y \ (m)$},
	ylabel near ticks,
	ytick={-10,0,10},
	y tick label style={
	/pgf/number format/.cd,
		fixed,
		fixed zerofill,
		precision=0,
		/tikz/.cd
	},
	clip=true,
    set layers,
    clip mode=individual
	]
	       
	\addplot[thick, color=blue, on layer=axis background] graphics[xmin=0,ymin=-15,xmax=75,ymax=15] {\scalarimg};
		
	\end{axis}	
	
\end{tikzpicture}

\let\scalarimg=\relax
	\end{minipage}
	\\ c) $t$=50~s \vspace{15pt} \\
	\hspace{0.09\linewidth}
	\begin{minipage}{\widthHyd}
	    \centering
        \pgfplotsset{ colormap={ocean}{ HTML=(21171f) HTML=(187de2) HTML=(a6d7fc) HTML=(ffffff) } }

\begin{tikzpicture}
	
	\begin{axis}[
	    hide axis,
	    scale only axis,
	    height=0pt,
	    width=0pt,
	    colormap name = ocean,
	    colorbar horizontal,
	    point meta min=0,
	    point meta max=1,
	    colorbar style={
	    	height=0.6em,
	        width=0.725\linewidth,
	        xtick={0,0.2,...,1},
	        xlabel={$H$ (m)},
		    x label style={
				text width=10em,
				align=center},
	    }]
	    \addplot [draw=none] coordinates {(0,0)};
	\end{axis}
	
\end{tikzpicture}
	\end{minipage}
	\hspace{0.11\linewidth}
	\begin{minipage}{\widthScl}
	    \centering
        \pgfplotsset{ colormap={red-blue}{ HTML=(FF0000) HTML=(FFFFFF) HTML=(002EFF) } }

\begin{tikzpicture}
	
	\begin{axis}[
	    hide axis,
	    scale only axis,
	    height=0pt,
	    width=0pt,
	    colormap name = red-blue,
	    colorbar horizontal,
	    point meta min=-1,
	    point meta max=1,
	    colorbar style={
	    	height=0.6em,
	        width=0.725\linewidth,
	        xtick={-1,-0.5,...,1},
	        xticklabels={0.5,0.25,0.0,0.05,0.1},
	        xlabel={$\phi_1$ (-) $\hspace{0.425\linewidth}$ $\phi_2$ (-)},
		    x label style={
				text width=10em,
				align=center},
	    }]
	    \addplot [draw=none] coordinates {(0,0)};
	\end{axis}
	
\end{tikzpicture}
	\end{minipage}
	\let\widthHyd=\relax
	\let\widthScl=\relax
	\caption{\textbf{T6: dam break over complex topography}. 3D visualization of the water surface elevation $H$ (left panels) and planar view of concentration distribution ($\phi_1$ and $\phi_2$) for both scalar species 1 and 2 (right panes). Subpanels (a,b,c) report different simulation timeout.}
    \label{fig:scl_3humps}
\end{figure}
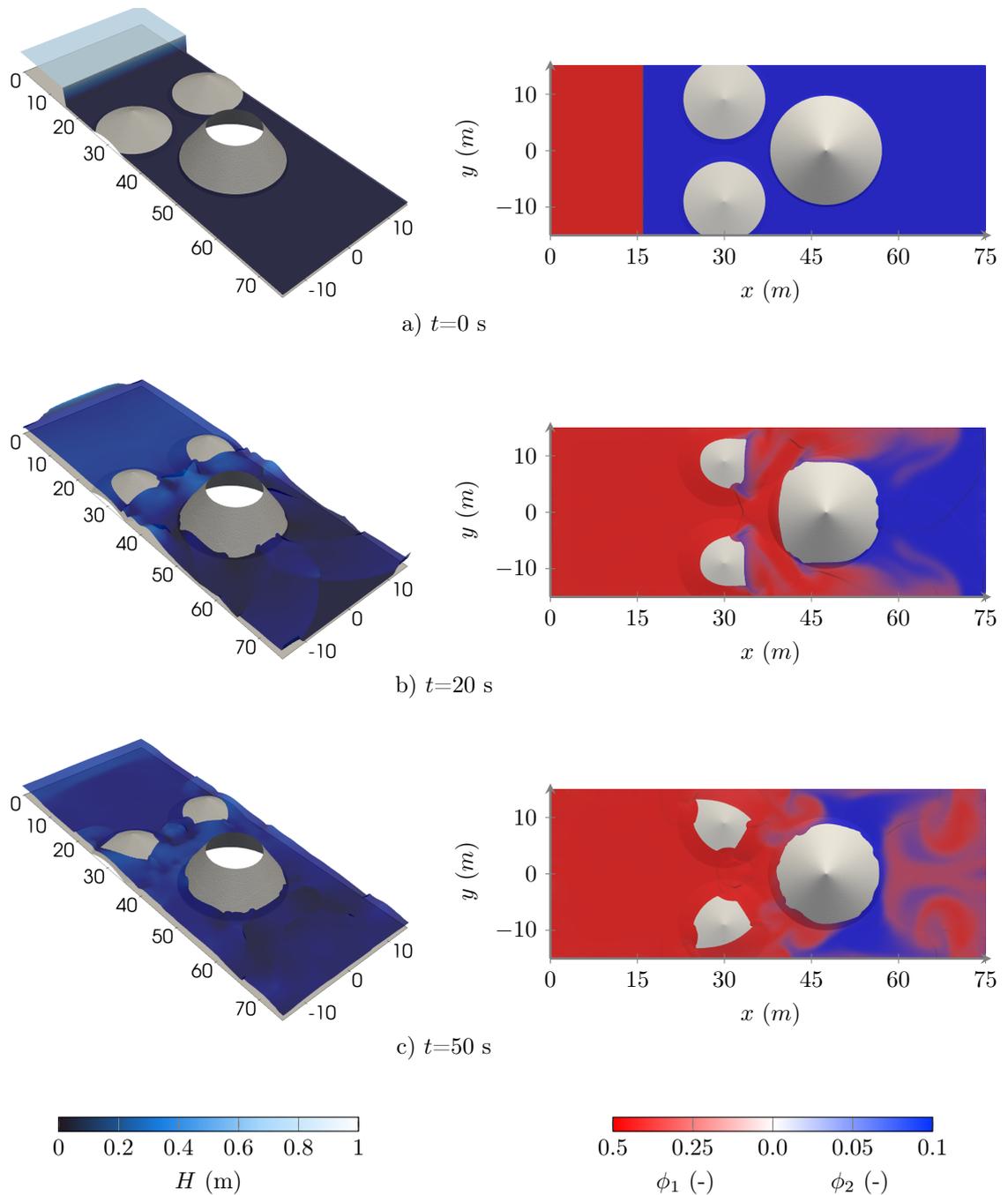

\subsection{Performance and scalability}\label{sec:performance}

The performance and scalability of the software depends not only on the implemented parallelization strategies but also on the physical model to be reproduced. In general, "simpler" models, i.e. only few simulated physical processes, are likely to show higher computational performances. To test \BM's computational performance, we selected the benchmarks T1 (hydrodynamic), T3 (morphodynamic) and T6 (scalar advection-diffusion).

Each of the selected numerical experiments (T1, T3, T6) has been conducted with four different computational meshes. The sizes of these meshes are given in Table~\ref{tab:Ncells_performance} and have been chosen to cover a broad range of spatial resolutions, ranging from thousands to hundred of thousands computational cells. The simulations have been run with a set of computational backends. In particular, CPU-based simulations have been performed on an Intel Xeon Gold 6154 (3.00GHz) workstation equipped with 36 cores (two sockets with 18 physical cores each), whilst GPU-based simulations have been run on three GPUs (GeForce GTX 1050 Ti, GeForce GTX 1080 Ti, and Tesla P100; see Table~\ref{tab:gpu-characteristics} for the main characteristics). Moreover, the simulations have been benchmarked in both single and double precision mode. The GPUs were integrated in a workstation with a 32-core Intel Xeon Gold 5218 (2.30GHz) processor (two sockets with 16 physical cores each).

\begin{table}[tbp]
    \centering
        \caption{\textbf{Number of computational cells for the performance and scalability benchmarks.}}\label{tab:Ncells_performance}
        \begin{tabular}{r|lll}\hline
            \textbf{Mesh ID} & \textbf{T1} & \textbf{T3}  & \textbf{T6}\\
            1 & \SI{24945}{} & \SI{27444}{} & \SI{24388}{}\\
            2 & \SI{52102}{} & \SI{47187}{} & \SI{49155}{}\\
            3 & \SI{101417}{}& \SI{109344}{} & \SI{98163}{}\\
            4 & \SI{499060}{}& \SI{218912}{} & \SI{196829}{}\\\hline
        \end{tabular}
\end{table}

\begin{table}[tbp]
    \centering
        \caption{\textbf{Characteristics of the benchmarked Nvidia GPU cards.} Further specifications available at www.nvidia.com/en-gb/geforce/10-series/ and www.nvidia.com/en-gb/data-center/tesla-p100/.}\label{tab:gpu-characteristics}
        \begin{tabular}{lll}\hline
            \textbf{Type} & \textbf{Architecture} & \textbf{CUDA Cores}\\
            GeForce GTX 1050 Ti & Pascal & 768\\
            GeForce GTX 1080 Ti & Pascal & 3584\\
            Tesla P100 & Pascal & 3584\\\hline
        \end{tabular}
\end{table}

\begin{figure}[htp]
    \subfloat[T1]{ \includegraphics[width=0.6\textwidth]{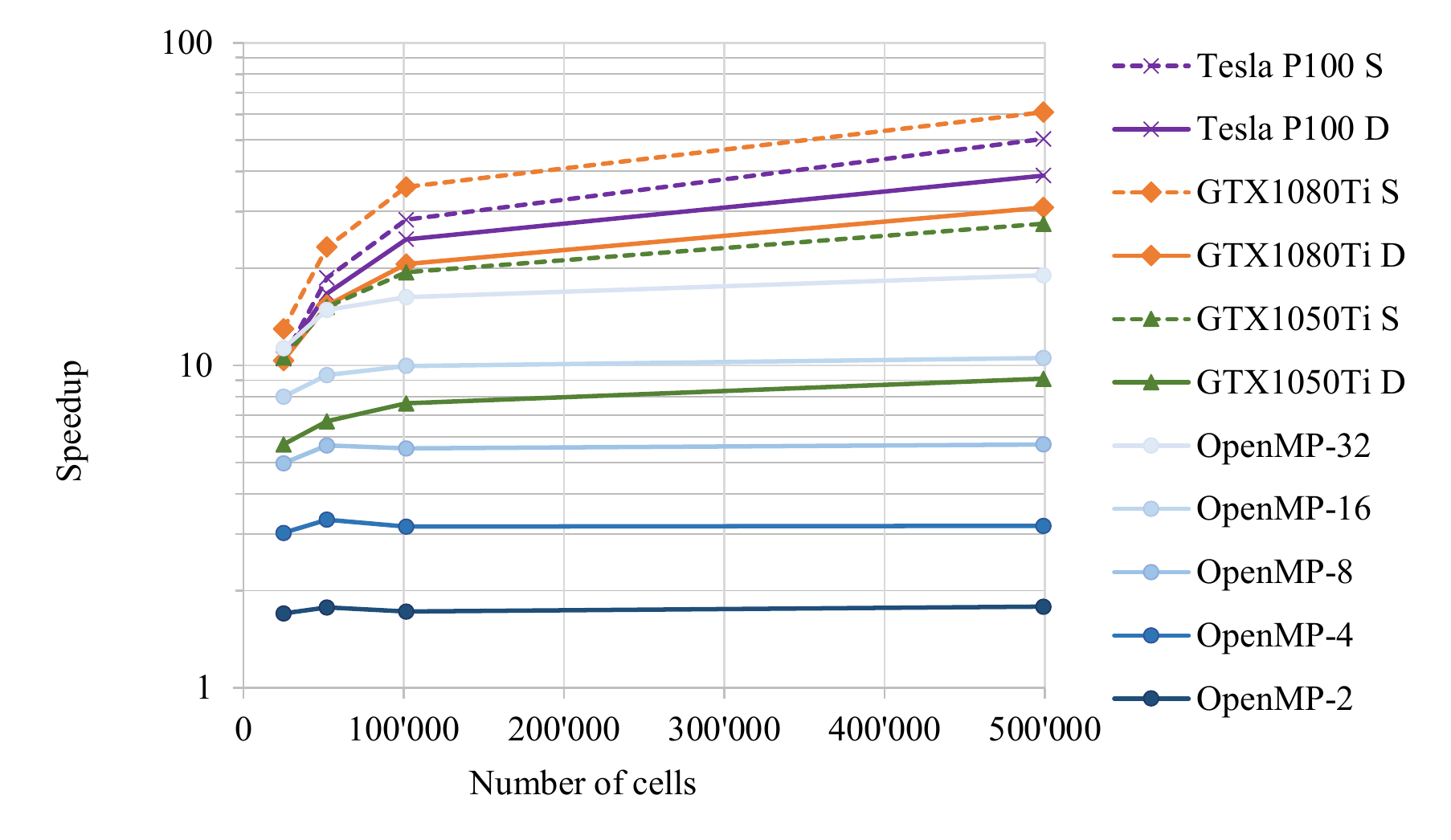}}\\
    \subfloat[T3]{ \includegraphics[width=0.6\textwidth]{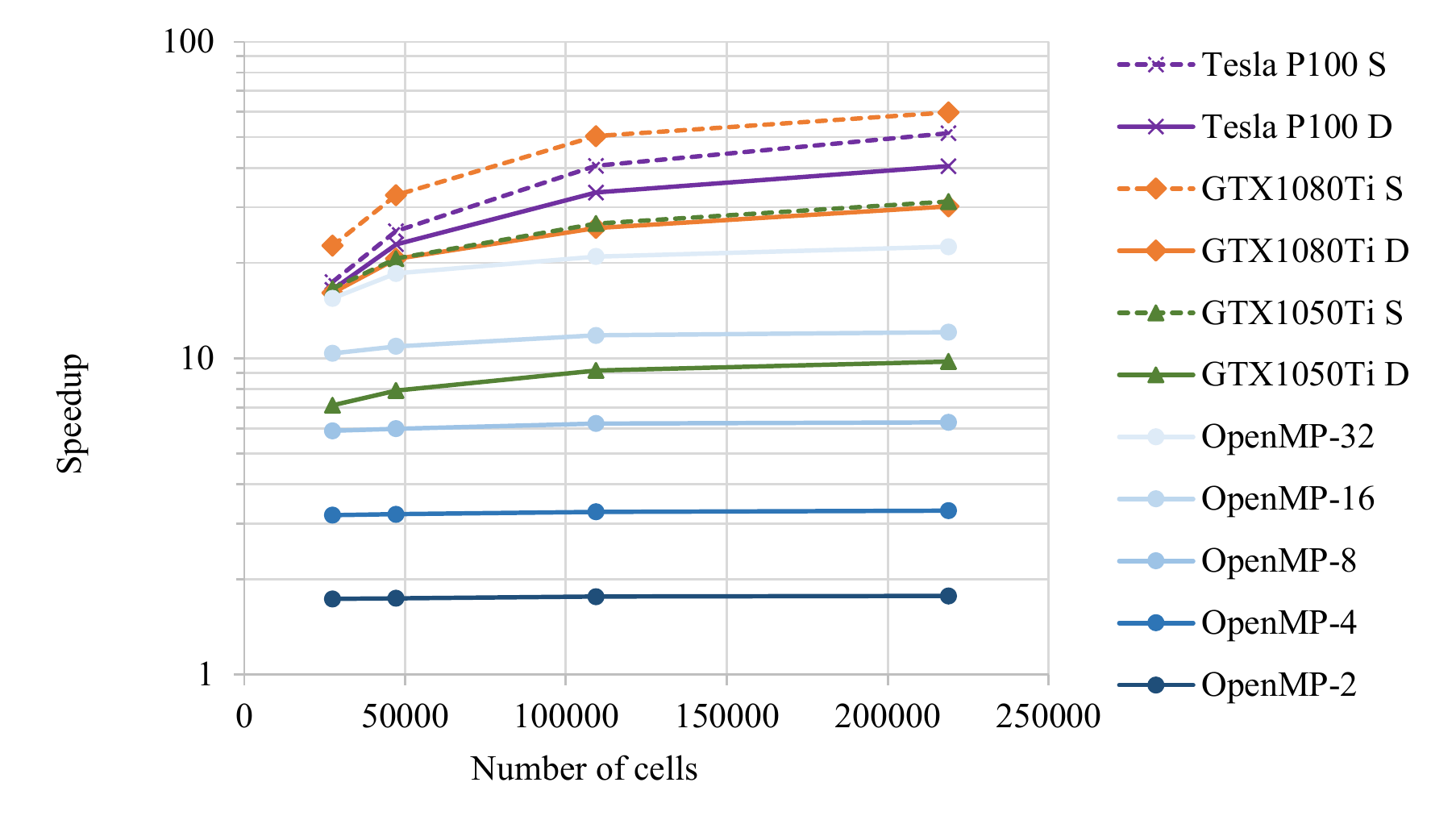}}\\
    \subfloat[T6]{ \includegraphics[width=0.6\textwidth]{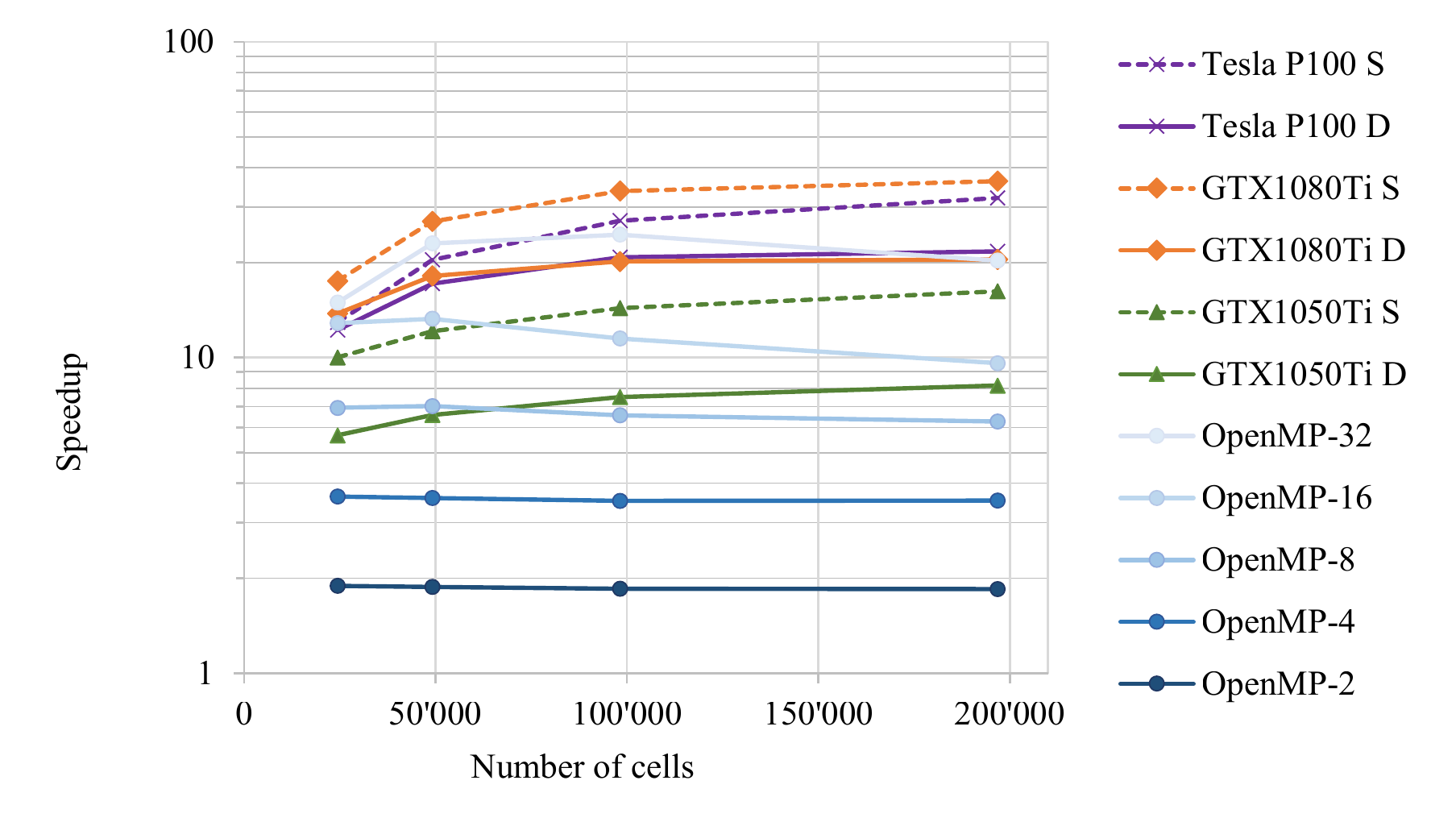}}
    \caption{\textbf{Speedup of backends for varying mesh size, test cases T1, T3 and T6.} The benchmarks were executed with different degrees of parallelism on the CPU (using OpenMP) and the GPU. The final "S" and "D" in GPU series denote single and double precision, respectively.}
    \label{fig:performance_speedup}
\end{figure}

For a given a mesh size, the speedup achieved by a parallelized backend $p$ is computed using the formula $\textrm{speedup} = T_s/T_{p}$, where $T_s$ ($T_{p}$) is the total computational time used by the serial (parallel) backend. The results for all the investigated benchmarks are depicted in Fig.~\ref{fig:performance_speedup}.

As anticipated, the speedup depends on the simulated processes. Comparing the speedup values among different benchmarks in Fig.~\ref{fig:performance_speedup}, cases T1 (hydrodynamics) and T3 (morphodynamics) show higher values on average than T6 (advection-diffusion). Such results are expected, given the increased complexity (number of equations and operations to be solved) of T6.

The performance benefits of \BM's parallelization can be evaluated in more detail by comparing the speedup values along the vertical axis of the plots. In the following we focus on benchmark T3 (morphodynamics) which shows an "intermediate" scalability among the three benchmarks (Fig.~\ref{fig:performance_speedup}b). Looking at the mesh with 47k elements as an example, the CPU-based family (i.e. OpenMP on multiple cores) shows a speedup efficiency (i.e. $\textrm{speedup}/\textrm{ncores}\cdot 100$) of 87\% with 2 CPU cores (Speedup=1.7), and of 58\% with 32 cores (Speedup=18.5), with an average efficiency of 74\%. \BM\ performs even better on some GPU cards. The least performing card (GTX 1050 Ti with double precision) has a speedup of about 8. However, note that speedup jumps to 20 when using the single precision version. Overall, the speedup provided by the tested GPUs ranges between 7 and 60.

The benchmarks in Fig.~\ref{fig:performance_speedup} also show how the maximum speedup changes with mesh size, computational backend and simulated processes. For all three cases, the CPU-based parallelizations show a mild speedup increase with an increasing number of computational cells. The dependency on the mesh size is slightly more pronounced when the number of computational cores is increased. This reflects the fact that CPU-based solutions have shared memory and minimal overhead (for multi-threading handling), thus the domain size (i.e. the data size) does not represent a potential performance bottleneck. Focussing on T3 (Fig.~\ref{fig:performance_speedup}b), the parallelization efficiency for 2-4 CPU cores is almost constant for all mesh sizes and above 80\%. On the other hand, the efficiency for 16-32 CPU cores is larger than 70\% only for mesh 4 (218k).

The speedup of the GPU-accelerated solutions shows not only a more marked dependency on the problem size, but also on the simulated processes. In benchmark T1 (Fig.~\ref{fig:performance_speedup}a), the speedup clearly increases with problem size. This can be explained with the overhead of GPU parallelization, which becomes more and more negligible with increasing domain size. Conversely, benchmark T6 (Fig.~\ref{fig:performance_speedup}c) shows little impact of the domain size on the speedup. In this case the scalability is limited by data transfers (i.e. data bandwidth) given the larger amount of data (compared to test case T1) needed for this simulation.

It is worth remarking that the GPU-accelerated backends show an average speedup difference greater than 10 between the single and double precision versions. Of course, the adequate choice depends on the requirements of the specific application.

The analysis above shows \BM's performance on different computational backends and underlines the differences when simulating different processes. The results summarized in Fig.~\ref{fig:performance_speedup} can also serve as a guideline for the interested reader/user when choosing an appropriate computational configuration for a given application. Finally it is worth highlighting that all the tested hardware configurations can be easily installed in standard office workstations.

\section{Conclusions}\label{sec:conclusions}
In this paper we introduced the main features of \BM\ version 3, a freeware tool for river simulation. \BM\ allows the simulation of a wide variety of hydro-, morphodynamic, and scalar advection-diffusion scenarios. As illustrated with the test cases, the software is able to efficiently capture large scale hydrodynamic processes modelled with several hundreds of thousand elements in good agreement with the measurements. On the opposite end of the spectrum, the morphological solver is able to handle demanding sediment transport scenarios well, albeit with known limitations. With the scalar advection-diffusion module a further set of physical processes such as the fate of river pollutants can be accurately modelled.

The impact of this flexibility on the software performance is minimized by activating feature sets on request in \BM's pre-simulation step. The advantage of this approach is that only the required tasks are scheduled for execution. This, together with OP2's ability to generate executable code for both multi-core CPU's and GPUs, permit \BM\ to scale with both available features and available computational power. Such advantages are reflected in the presented benchmarks. Given a large enough domain, the software shows a good parallel efficiency on the CPU and an even higher speedup when using GPUs. 

The \BM\ project is in continuous advancement to optimize and include further features in the existing basic modules. As an example, the modelling of the sediment transport in presence of non uniform sediment size and the simulation of water temperature dynamics are in implementation phase. On the other hand, efforts are dedicated also to develop novel modelling solutions for river processes such as the bio-morphodynamic feedbacks between vegetation and sediment transport. Table~2 of the Supplementary Material provides an overview of under development features. The modularity of the development framework allows also for further refinement of single specific numerical solvers and the implementation of high-order schemes when needed. 

Overall, the combination of different river processes that can be modelled, the computational efficiency, the flexibility in the backend choice, but also the availability of a light Graphical User Interface, make \BM\ a valuable tool for a broad family of river modelers in both Academia and Practice.

\section*{Acknowledgments}
The Authors greatly thank the many former collaborators and developers within the \BM\ project. Particular thanks to Aurélie Koch, for her valuable contribution in testing and documenting the software.

The design of \BM\ was conducted by DFV, DV, SP, LV. The software prototyping, development and implementation was done by DV, SP, LV, MB, MW, with the coordination and supervision of DFV and AS. Implementation and testing of the reported features was done by MB, DV, MW and DC. 
The manuscript was conceptualized by DV, AS and DFV, and drafted by DV, with support of MB and MW. All Authors contributed to the manuscript review.

\section*{Conflicts of interest}
The Authors declare that there are no conflicts of interest.

\section*{Funding}
The development of the software \BM\ is financially supported by the Swiss Federal Office for the Environment (BAFU).

\clearpage
\section*{References}
\bibliographystyle{elsarticle-num-names}
\bibliography{bib_BM3}

\begin{thebibliography}{81}
\providecommand{\natexlab}[1]{#1}
\providecommand{\url}[1]{\texttt{#1}}
\providecommand{\urlprefix}{URL }
\expandafter\ifx\csname urlstyle\endcsname\relax
  \providecommand{\doi}[1]{doi:\discretionary{}{}{}#1}\else
  \providecommand{\doi}[1]{doi:\discretionary{}{}{}\begingroup
  \urlstyle{rm}\url{#1}\endgroup}\fi
\providecommand{\bibinfo}[2]{#2}

\bibitem[{Gilvear et~al.(2016)Gilvear, Greenwood, Thoms, and
  Wood}]{Gilvear2016}
\bibinfo{editor}{D.~J. Gilvear}, \bibinfo{editor}{M.~T. Greenwood},
  \bibinfo{editor}{M.~C. Thoms}, \bibinfo{editor}{P.~J. Wood} (Eds.),
  \bibinfo{title}{River Science}, \bibinfo{publisher}{John Wiley {\&} Sons,
  Ltd}, \doi{\bibinfo{doi}{10.1002/9781118643525}}, \bibinfo{year}{2016}.

\bibitem[{Brewer et~al.(2018)Brewer, Worthington, Mollenhauer, Stewart,
  McManamay, Guertault, and Moore}]{Brewer_2018}
\bibinfo{author}{S.~K. Brewer}, \bibinfo{author}{T.~A. Worthington},
  \bibinfo{author}{R.~Mollenhauer}, \bibinfo{author}{D.~R. Stewart},
  \bibinfo{author}{R.~A. McManamay}, \bibinfo{author}{L.~Guertault},
  \bibinfo{author}{D.~Moore}, \bibinfo{title}{Synthesizing models useful for
  ecohydrology and ecohydraulic approaches: An emphasis on integrating models
  to address complex research questions}, \bibinfo{journal}{Ecohydrology}
  \bibinfo{volume}{11}~(\bibinfo{number}{7}) (\bibinfo{year}{2018})
  \bibinfo{pages}{e1966}, ISSN \bibinfo{issn}{19360584},
  \doi{\bibinfo{doi}{10.1002/eco.1966}}.

\bibitem[{Shimizu et~al.(2019)Shimizu, Nelson, Ferrel, Asahi, Giri, Inoue,
  Iwasaki, Jang, Kang, Kimura, Kyuka, Mishra, Nabi, Patsinghasanee, and
  Yamaguchi}]{Shimizu2019}
\bibinfo{author}{Y.~Shimizu}, \bibinfo{author}{J.~Nelson},
  \bibinfo{author}{K.~A. Ferrel}, \bibinfo{author}{K.~Asahi},
  \bibinfo{author}{S.~Giri}, \bibinfo{author}{T.~Inoue},
  \bibinfo{author}{T.~Iwasaki}, \bibinfo{author}{C.-L. Jang},
  \bibinfo{author}{T.~Kang}, \bibinfo{author}{I.~Kimura},
  \bibinfo{author}{T.~Kyuka}, \bibinfo{author}{J.~Mishra},
  \bibinfo{author}{M.~Nabi}, \bibinfo{author}{S.~Patsinghasanee},
  \bibinfo{author}{S.~Yamaguchi}, \bibinfo{title}{Advances in computational
  morphodynamics using the International River Interface Cooperative ({iRIC})
  software}, \bibinfo{journal}{Earth Surface Processes and Landforms}
  \bibinfo{volume}{45}~(\bibinfo{number}{1}) (\bibinfo{year}{2019})
  \bibinfo{pages}{11--37}, \doi{\bibinfo{doi}{10.1002/esp.4653}}.

\bibitem[{Crosato and Saleh(2011)}]{Crosato2011}
\bibinfo{author}{A.~Crosato}, \bibinfo{author}{M.~S. Saleh},
  \bibinfo{title}{Numerical study on the effects of floodplain vegetation on
  river planform style}, \bibinfo{journal}{Earth Surface Processes and
  Landforms} \bibinfo{volume}{36}~(\bibinfo{number}{6}) (\bibinfo{year}{2011})
  \bibinfo{pages}{711--720}.

\bibitem[{Sharma and Kansal(2012)}]{Sharma2012}
\bibinfo{author}{D.~Sharma}, \bibinfo{author}{A.~Kansal},
  \bibinfo{title}{Assessment of river quality models: a review},
  \bibinfo{journal}{Reviews in Environmental Science and Bio/Technology}
  \bibinfo{volume}{12}~(\bibinfo{number}{3}) (\bibinfo{year}{2012})
  \bibinfo{pages}{285--311}, \doi{\bibinfo{doi}{10.1007/s11157-012-9285-8}}.

\bibitem[{Williams et~al.(2016)Williams, Brasington, and Hicks}]{Williams2016}
\bibinfo{author}{R.~D. Williams}, \bibinfo{author}{J.~Brasington},
  \bibinfo{author}{D.~M. Hicks}, \bibinfo{title}{Numerical Modelling of Braided
  River Morphodynamics: Review and Future Challenges},
  \bibinfo{journal}{Geography Compass}
  \bibinfo{volume}{10}~(\bibinfo{number}{3}) (\bibinfo{year}{2016})
  \bibinfo{pages}{102--127}, \doi{\bibinfo{doi}{10.1111/gec3.12260}}.

\bibitem[{Dugdale et~al.(2017)Dugdale, Hannah, and Malcolm}]{Dugdale2017}
\bibinfo{author}{S.~J. Dugdale}, \bibinfo{author}{D.~M. Hannah},
  \bibinfo{author}{I.~A. Malcolm}, \bibinfo{title}{River temperature modelling:
  A review of process-based approaches and future directions},
  \bibinfo{journal}{Earth-Science Reviews} \bibinfo{volume}{175}
  (\bibinfo{year}{2017}) \bibinfo{pages}{97--113},
  \doi{\bibinfo{doi}{10.1016/j.earscirev.2017.10.009}}.

\bibitem[{Teng et~al.(2017)Teng, Jakeman, Vaze, Croke, Dutta, and
  Kim}]{Teng2017}
\bibinfo{author}{J.~Teng}, \bibinfo{author}{A.~J. Jakeman},
  \bibinfo{author}{J.~Vaze}, \bibinfo{author}{B.~F.~W. Croke},
  \bibinfo{author}{D.~Dutta}, \bibinfo{author}{S.~Kim}, \bibinfo{title}{{Flood
  inundation modelling: A review of methods, recent advances and uncertainty
  analysis}}, \bibinfo{journal}{Environmental Modelling and Software}
  \bibinfo{volume}{90} (\bibinfo{year}{2017}) \bibinfo{pages}{201--216}, ISSN
  \bibinfo{issn}{13648152}, \doi{\bibinfo{doi}{10.1016/j.envsoft.2017.01.006}}.

\bibitem[{Zischg et~al.(2018)Zischg, Mosimann, Bernet, and
  R{\"o}thlisberger}]{zischg2018}
\bibinfo{author}{A.~P. Zischg}, \bibinfo{author}{M.~Mosimann},
  \bibinfo{author}{D.~B. Bernet}, \bibinfo{author}{V.~R{\"o}thlisberger},
  \bibinfo{title}{Validation of 2D flood models with insurance claims},
  \bibinfo{journal}{Journal of hydrology} \bibinfo{volume}{557}
  (\bibinfo{year}{2018}) \bibinfo{pages}{350--361}.

\bibitem[{Marcus and Fonstad(2010)}]{Marcus_2010}
\bibinfo{author}{W.~A. Marcus}, \bibinfo{author}{M.~A. Fonstad},
  \bibinfo{title}{Remote sensing of rivers: the emergence of a subdiscipline in
  the river sciences}, \bibinfo{journal}{Earth Surface Processes and Landforms}
  \bibinfo{volume}{35}~(\bibinfo{number}{15}) (\bibinfo{year}{2010})
  \bibinfo{pages}{1867--1872}, \doi{\bibinfo{doi}{10.1002/esp.2094}}.

\bibitem[{Thomas Steven~Savage et~al.(2016)Thomas Steven~Savage, Pianosi,
  Bates, Freer, and Wagener}]{thomas2016}
\bibinfo{author}{J.~Thomas Steven~Savage}, \bibinfo{author}{F.~Pianosi},
  \bibinfo{author}{P.~Bates}, \bibinfo{author}{J.~Freer},
  \bibinfo{author}{T.~Wagener}, \bibinfo{title}{Quantifying the importance of
  spatial resolution and other factors through global sensitivity analysis of a
  flood inundation model}, \bibinfo{journal}{Water Resources Research}
  \bibinfo{volume}{52}~(\bibinfo{number}{11}) (\bibinfo{year}{2016})
  \bibinfo{pages}{9146--9163}.

\bibitem[{Pasternack(2011)}]{Pasternack2011}
\bibinfo{author}{G.~Pasternack}, \bibinfo{title}{2D modeling and ecohydraulic
  analysis}, \bibinfo{publisher}{University of California at Davis},
  \bibinfo{address}{California}, ISBN \bibinfo{isbn}{9781466320093},
  \bibinfo{year}{2011}.

\bibitem[{Maddock(1999)}]{Maddock_1999}
\bibinfo{author}{I.~Maddock}, \bibinfo{title}{The importance of physical
  habitat assessment for evaluating river health}, \bibinfo{journal}{Freshwater
  Biology} \bibinfo{volume}{41}~(\bibinfo{number}{2}) (\bibinfo{year}{1999})
  \bibinfo{pages}{373--391},
  \doi{\bibinfo{doi}{10.1046/j.1365-2427.1999.00437.x}}.

\bibitem[{Siviglia et~al.(2013)Siviglia, Stecca, Vanzo, Zolezzi, Toro, and
  Tubino}]{Siviglia2013}
\bibinfo{author}{A.~Siviglia}, \bibinfo{author}{G.~Stecca},
  \bibinfo{author}{D.~Vanzo}, \bibinfo{author}{G.~Zolezzi},
  \bibinfo{author}{E.~F. Toro}, \bibinfo{author}{M.~Tubino},
  \bibinfo{title}{Numerical modelling of two-dimensional morphodynamics with
  applications to river bars and bifurcations}, \bibinfo{journal}{Advances in
  Water Resources} \bibinfo{volume}{52} (\bibinfo{year}{2013})
  \bibinfo{pages}{243--260}, ISSN \bibinfo{issn}{03091708},
  \doi{\bibinfo{doi}{10.1016/j.advwatres.2012.11.010}}.

\bibitem[{Wyrick et~al.(2014)Wyrick, Senter, and Pasternack}]{Wyrick2014}
\bibinfo{author}{J.~Wyrick}, \bibinfo{author}{A.~Senter},
  \bibinfo{author}{G.~Pasternack}, \bibinfo{title}{Revealing the natural
  complexity of fluvial morphology through 2D hydrodynamic delineation of river
  landforms}, \bibinfo{journal}{Geomorphology} \bibinfo{volume}{210}
  (\bibinfo{year}{2014}) \bibinfo{pages}{14--22},
  \doi{\bibinfo{doi}{10.1016/j.geomorph.2013.12.013}}.

\bibitem[{Guan et~al.(2016)Guan, Wright, Sleigh, Ahilan, and Lamb}]{guan2016}
\bibinfo{author}{M.~Guan}, \bibinfo{author}{N.~Wright},
  \bibinfo{author}{P.~Sleigh}, \bibinfo{author}{S.~Ahilan},
  \bibinfo{author}{R.~Lamb}, \bibinfo{title}{Physical complexity to model
  morphological changes at a natural channel bend}, \bibinfo{journal}{Water
  resources research} \bibinfo{volume}{52}~(\bibinfo{number}{8})
  (\bibinfo{year}{2016}) \bibinfo{pages}{6348--6364},
  \doi{\bibinfo{doi}{10.1002/2015WR017917}}.

\bibitem[{Siviglia and Crosato(2016)}]{Siviglia2016}
\bibinfo{author}{A.~Siviglia}, \bibinfo{author}{A.~Crosato},
  \bibinfo{title}{Numerical modelling of river morphodynamics: Latest
  developments and remaining challenges}, \bibinfo{journal}{Advances in Water
  Resources} \bibinfo{volume}{93} (\bibinfo{year}{2016}) \bibinfo{pages}{1--3},
  \doi{\bibinfo{doi}{10.1016/j.advwatres.2016.01.005}}.

\bibitem[{Toro(2001)}]{Toro2001}
\bibinfo{author}{E.~Toro}, \bibinfo{title}{{Shock-capturing methods for
  free-surface shallow flows}}, \bibinfo{publisher}{Wiley and Sons Ltd}, ISBN
  \bibinfo{isbn}{0471987662}, \doi{\bibinfo{doi}{10.1080/00221680309499935}},
  \bibinfo{year}{2001}.

\bibitem[{Nahorniak et~al.(2018)Nahorniak, Wheaton, Volk, Bailey, Reimer, Wall,
  Whitehead, and Jordan}]{Nahorniak2018}
\bibinfo{author}{M.~Nahorniak}, \bibinfo{author}{J.~Wheaton},
  \bibinfo{author}{C.~Volk}, \bibinfo{author}{P.~Bailey},
  \bibinfo{author}{M.~Reimer}, \bibinfo{author}{E.~Wall},
  \bibinfo{author}{K.~Whitehead}, \bibinfo{author}{C.~Jordan},
  \bibinfo{title}{{How do we efficiently generate high-resolution hydraulic
  models at large numbers of riverine reaches?}}, \bibinfo{journal}{Computers
  {\&} Geosciences} \bibinfo{volume}{119}~(\bibinfo{number}{November 2017})
  (\bibinfo{year}{2018}) \bibinfo{pages}{80--91}, ISSN
  \bibinfo{issn}{00983004}, \doi{\bibinfo{doi}{10.1016/j.cageo.2018.07.001}}.

\bibitem[{Costabile and Macchione(2015)}]{Costabile2015b}
\bibinfo{author}{P.~Costabile}, \bibinfo{author}{F.~Macchione},
  \bibinfo{title}{{Enhancing river model set-up for 2-D dynamic flood
  modelling}}, \bibinfo{journal}{Environmental Modelling {\&} Software}
  \bibinfo{volume}{67} (\bibinfo{year}{2015}) \bibinfo{pages}{89--107}, ISSN
  \bibinfo{issn}{13648152}, \doi{\bibinfo{doi}{10.1016/j.envsoft.2015.01.009}}.

\bibitem[{Dazzi et~al.(2020)Dazzi, Vacondio, and Mignosa}]{dazzi2020}
\bibinfo{author}{S.~Dazzi}, \bibinfo{author}{R.~Vacondio},
  \bibinfo{author}{P.~Mignosa}, \bibinfo{title}{Internal boundary conditions
  for a GPU-accelerated 2D shallow water model: Implementation and
  applications}, \bibinfo{journal}{Advances in Water Resources}
  \bibinfo{volume}{137} (\bibinfo{year}{2020}) \bibinfo{pages}{103525}.

\bibitem[{Sanders(2008)}]{Sanders2008}
\bibinfo{author}{B.~F. Sanders}, \bibinfo{title}{Integration of a shallow water
  model with a local time step}, \bibinfo{journal}{Journal of Hydraulic
  Research} \bibinfo{volume}{46}~(\bibinfo{number}{4}) (\bibinfo{year}{2008})
  \bibinfo{pages}{466--475}, \doi{\bibinfo{doi}{10.3826/jhr.2008.3243}}.

\bibitem[{Dazzi et~al.(2018)Dazzi, Vacondio, Pal{\`{u}}, and
  Mignosa}]{Dazzi2018}
\bibinfo{author}{S.~Dazzi}, \bibinfo{author}{R.~Vacondio},
  \bibinfo{author}{A.~D. Pal{\`{u}}}, \bibinfo{author}{P.~Mignosa},
  \bibinfo{title}{A local time stepping algorithm for {GPU}-accelerated 2D
  shallow water models}, \bibinfo{journal}{Advances in Water Resources}
  \bibinfo{volume}{111} (\bibinfo{year}{2018}) \bibinfo{pages}{274--288},
  \doi{\bibinfo{doi}{10.1016/j.advwatres.2017.11.023}}.

\bibitem[{Powell et~al.(1993)Powell, Roe, and Quirk}]{Powell1993}
\bibinfo{author}{K.~G. Powell}, \bibinfo{author}{P.~L. Roe},
  \bibinfo{author}{J.~Quirk}, \bibinfo{title}{Adaptive-Mesh Algorithms for
  Computational Fluid Dynamics}, in: \bibinfo{booktitle}{Algorithmic Trends in
  Computational Fluid Dynamics}, \bibinfo{publisher}{Springer New York},
  \bibinfo{pages}{303--337}, \doi{\bibinfo{doi}{10.1007/978-1-4612-2708-3_18}},
  \bibinfo{year}{1993}.

\bibitem[{Carraro et~al.(2018)Carraro, Vanzo, Caleffi, Valiani, and
  Siviglia}]{Carraro2018}
\bibinfo{author}{F.~Carraro}, \bibinfo{author}{D.~Vanzo},
  \bibinfo{author}{V.~Caleffi}, \bibinfo{author}{A.~Valiani},
  \bibinfo{author}{A.~Siviglia}, \bibinfo{title}{Mathematical study of linear
  morphodynamic acceleration and derivation of the MASSPEED approach},
  \bibinfo{journal}{Advances in Water Resources} \bibinfo{volume}{117}
  (\bibinfo{year}{2018}) \bibinfo{pages}{40--52}.

\bibitem[{Morgan et~al.(2020)Morgan, Kumar, Horner-Devine, Ahrendt,
  Istanbullouglu, and Bandaragoda}]{Morgan2020}
\bibinfo{author}{J.~A. Morgan}, \bibinfo{author}{N.~Kumar},
  \bibinfo{author}{A.~R. Horner-Devine}, \bibinfo{author}{S.~Ahrendt},
  \bibinfo{author}{E.~Istanbullouglu}, \bibinfo{author}{C.~Bandaragoda},
  \bibinfo{title}{The use of a morphological acceleration factor in the
  simulation of large-scale fluvial morphodynamics},
  \bibinfo{journal}{Geomorphology} \bibinfo{volume}{356} (\bibinfo{year}{2020})
  \bibinfo{pages}{107088}, \doi{\bibinfo{doi}{10.1016/j.geomorph.2020.107088}}.

\bibitem[{Vanzo et~al.(2016)Vanzo, Siviglia, and Toro}]{Vanzo2016}
\bibinfo{author}{D.~Vanzo}, \bibinfo{author}{A.~Siviglia},
  \bibinfo{author}{E.~F. Toro}, \bibinfo{title}{{Pollutant transport by shallow
  water equations on unstructured meshes: Hyperbolization of the model and
  numerical solution via a novel flux splitting scheme}},
  \bibinfo{journal}{Journal of Computational Physics} \bibinfo{volume}{321}
  (\bibinfo{year}{2016}) \bibinfo{pages}{1--20}.

\bibitem[{Afzal et~al.(2016)Afzal, Ansari, Faizabadi, and Ramis}]{Afzal2016}
\bibinfo{author}{A.~Afzal}, \bibinfo{author}{Z.~Ansari}, \bibinfo{author}{A.~R.
  Faizabadi}, \bibinfo{author}{M.~K. Ramis}, \bibinfo{title}{Parallelization
  Strategies for Computational Fluid Dynamics Software: State of the Art
  Review}, \bibinfo{journal}{Archives of Computational Methods in Engineering}
  \bibinfo{volume}{24}~(\bibinfo{number}{2}) (\bibinfo{year}{2016})
  \bibinfo{pages}{337--363}, \doi{\bibinfo{doi}{10.1007/s11831-016-9165-4}}.

\bibitem[{Owens et~al.(2008)Owens, Houston, Luebke, Green, Stone, and
  Phillips}]{Owens2008}
\bibinfo{author}{J.~Owens}, \bibinfo{author}{M.~Houston},
  \bibinfo{author}{D.~Luebke}, \bibinfo{author}{S.~Green},
  \bibinfo{author}{J.~Stone}, \bibinfo{author}{J.~Phillips},
  \bibinfo{title}{{GPU} Computing}, \bibinfo{journal}{Proceedings of the
  {IEEE}} \bibinfo{volume}{96}~(\bibinfo{number}{5}) (\bibinfo{year}{2008})
  \bibinfo{pages}{879--899}, \doi{\bibinfo{doi}{10.1109/jproc.2008.917757}}.

\bibitem[{Mudalige et~al.(2012)Mudalige, Giles, Reguly, Bertolli, and
  Kelly}]{Mudalige2012a}
\bibinfo{author}{G.~Mudalige}, \bibinfo{author}{M.~Giles},
  \bibinfo{author}{I.~Reguly}, \bibinfo{author}{C.~Bertolli},
  \bibinfo{author}{P.~Kelly}, \bibinfo{title}{{OP2: An active library framework
  for solving unstructured mesh-based applications on multi-core and many-core
  architectures}}, in: \bibinfo{booktitle}{2012 Innovative Parallel Computing
  (InPar)}, \bibinfo{publisher}{IEEE}, ISBN \bibinfo{isbn}{978-1-4673-2633-9},
  \bibinfo{pages}{1--12}, \doi{\bibinfo{doi}{10.1109/InPar.2012.6339594}},
  \bibinfo{year}{2012}.

\bibitem[{Brodtkorb et~al.(2012)Brodtkorb, S{\ae}tra, and
  Altinakar}]{Brodtkorb2012}
\bibinfo{author}{A.~R. Brodtkorb}, \bibinfo{author}{M.~L. S{\ae}tra},
  \bibinfo{author}{M.~Altinakar}, \bibinfo{title}{{Efficient shallow water
  simulations on GPUs: Implementation, visualization, verification, and
  validation}}, \bibinfo{journal}{Computers and Fluids} \bibinfo{volume}{55}
  (\bibinfo{year}{2012}) \bibinfo{pages}{1--12}, ISSN \bibinfo{issn}{00457930},
  \doi{\bibinfo{doi}{10.1016/j.compfluid.2011.10.012}}.

\bibitem[{Smith and Liang(2013)}]{Smith2013}
\bibinfo{author}{L.~S. Smith}, \bibinfo{author}{Q.~Liang},
  \bibinfo{title}{{Towards a generalised GPU/CPU shallow-flow modelling tool}},
  \bibinfo{journal}{Computers {\&} Fluids} \bibinfo{volume}{88}
  (\bibinfo{year}{2013}) \bibinfo{pages}{334--343}, ISSN
  \bibinfo{issn}{00457930},
  \doi{\bibinfo{doi}{10.1016/j.compfluid.2013.09.018}}.

\bibitem[{Vacondio et~al.(2014)Vacondio, Pal{\`{u}}, and
  Mignosa}]{Vacondio_2014}
\bibinfo{author}{R.~Vacondio}, \bibinfo{author}{A.~D. Pal{\`{u}}},
  \bibinfo{author}{P.~Mignosa}, \bibinfo{title}{{GPU}-enhanced Finite Volume
  Shallow Water solver for fast flood simulations},
  \bibinfo{journal}{Environmental Modelling {\&} Software} \bibinfo{volume}{57}
  (\bibinfo{year}{2014}) \bibinfo{pages}{60--75}, ISSN
  \bibinfo{issn}{1364-8152},
  \doi{\bibinfo{doi}{10.1016/j.envsoft.2014.02.003}}.

\bibitem[{Vacondio et~al.(2017)Vacondio, {Dal Pal{\`{u}}}, Ferrari, Mignosa,
  Aureli, and Dazzi}]{Vacondio2017}
\bibinfo{author}{R.~Vacondio}, \bibinfo{author}{A.~{Dal Pal{\`{u}}}},
  \bibinfo{author}{A.~Ferrari}, \bibinfo{author}{P.~Mignosa},
  \bibinfo{author}{F.~Aureli}, \bibinfo{author}{S.~Dazzi}, \bibinfo{title}{{A
  non-uniform efficient grid type for GPU-parallel Shallow Water Equations
  models}}, \bibinfo{journal}{Environmental Modelling {\&} Software}
  \bibinfo{volume}{88} (\bibinfo{year}{2017}) \bibinfo{pages}{119--137}, ISSN
  \bibinfo{issn}{13648152}, \doi{\bibinfo{doi}{10.1016/j.envsoft.2016.11.012}}.

\bibitem[{Hou et~al.(2020)Hou, Kang, Hu, Tong, Pan, and Xia}]{Hou2020}
\bibinfo{author}{J.~Hou}, \bibinfo{author}{Y.~Kang}, \bibinfo{author}{C.~Hu},
  \bibinfo{author}{Y.~Tong}, \bibinfo{author}{B.~Pan},
  \bibinfo{author}{J.~Xia}, \bibinfo{title}{A {GPU}-based numerical model
  coupling hydrodynamical and morphological processes},
  \bibinfo{journal}{International Journal of Sediment Research}
  \bibinfo{volume}{35}~(\bibinfo{number}{4}) (\bibinfo{year}{2020})
  \bibinfo{pages}{386--394}, \doi{\bibinfo{doi}{10.1016/j.ijsrc.2020.02.005}}.

\bibitem[{Castro et~al.(2011)Castro, Ortega, de~la Asunci{\'{o}}n, Mantas, and
  Gallardo}]{Castro2011}
\bibinfo{author}{M.~J. Castro}, \bibinfo{author}{S.~Ortega},
  \bibinfo{author}{M.~de~la Asunci{\'{o}}n}, \bibinfo{author}{J.~M. Mantas},
  \bibinfo{author}{J.~M. Gallardo}, \bibinfo{title}{{GPU computing for shallow
  water flow simulation based on finite volume schemes}},
  \bibinfo{journal}{Comptes Rendus M{\'{e}}canique}
  \bibinfo{volume}{339}~(\bibinfo{number}{2-3}) (\bibinfo{year}{2011})
  \bibinfo{pages}{165--184}, ISSN \bibinfo{issn}{16310721},
  \doi{\bibinfo{doi}{10.1016/j.crme.2010.12.004}}.

\bibitem[{Lacasta et~al.(2014)Lacasta, Morales-Hern{\'{a}}ndez, Murillo, and
  Garc{\'{i}}a-Navarro}]{Lacasta2014b}
\bibinfo{author}{A.~Lacasta}, \bibinfo{author}{M.~Morales-Hern{\'{a}}ndez},
  \bibinfo{author}{J.~Murillo}, \bibinfo{author}{P.~Garc{\'{i}}a-Navarro},
  \bibinfo{title}{{An optimized GPU implementation of a 2D free surface
  simulation model on unstructured meshes}}, \bibinfo{journal}{Advances in
  Engineering Software} \bibinfo{volume}{78} (\bibinfo{year}{2014})
  \bibinfo{pages}{1--15}, ISSN \bibinfo{issn}{09659978},
  \doi{\bibinfo{doi}{10.1016/j.advengsoft.2014.08.007}}.

\bibitem[{Lacasta et~al.(2015)Lacasta, Juez, Murillo, and
  Garc{\'\i}a-Navarro}]{Lacasta2015b}
\bibinfo{author}{A.~Lacasta}, \bibinfo{author}{C.~Juez},
  \bibinfo{author}{J.~Murillo}, \bibinfo{author}{P.~Garc{\'\i}a-Navarro},
  \bibinfo{title}{An efficient solution for hazardous geophysical flows
  simulation using GPUs}, \bibinfo{journal}{Computers \& Geosciences}
  \bibinfo{volume}{78} (\bibinfo{year}{2015}) \bibinfo{pages}{63--72},
  \doi{\bibinfo{doi}{10.1016/j.cageo.2015.02.010}}.

\bibitem[{Petaccia et~al.(2016)Petaccia, Leporati, and Torti}]{Petaccia2016}
\bibinfo{author}{G.~Petaccia}, \bibinfo{author}{F.~Leporati},
  \bibinfo{author}{E.~Torti}, \bibinfo{title}{{OpenMP} and {CUDA} simulations
  of Sella Zerbino Dam break on unstructured grids},
  \bibinfo{journal}{Computational Geosciences}
  \bibinfo{volume}{20}~(\bibinfo{number}{5}) (\bibinfo{year}{2016})
  \bibinfo{pages}{1123--1132}, \doi{\bibinfo{doi}{10.1007/s10596-016-9580-5}}.

\bibitem[{Juez et~al.(2016)Juez, Lacasta, Murillo, and
  Garc{\'{i}}a-Navarro}]{Juez2016}
\bibinfo{author}{C.~Juez}, \bibinfo{author}{A.~Lacasta},
  \bibinfo{author}{J.~Murillo}, \bibinfo{author}{P.~Garc{\'{i}}a-Navarro},
  \bibinfo{title}{{An efficient GPU implementation for a faster simulation of
  unsteady bed-load transport}}, \bibinfo{journal}{Journal of Hydraulic
  Research} \bibinfo{volume}{54}~(\bibinfo{number}{3}) (\bibinfo{year}{2016})
  \bibinfo{pages}{275--288}, ISSN \bibinfo{issn}{0022-1686},
  \doi{\bibinfo{doi}{10.1080/00221686.2016.1143042}}.

\bibitem[{Garc{\'{\i}}a-Feal et~al.(2018)Garc{\'{\i}}a-Feal,
  Gonz{\'{a}}lez-Cao, G{\'{o}}mez-Gesteira, Cea, Dom{\'{\i}}nguez, and
  Formella}]{GarciaFeal2018}
\bibinfo{author}{O.~Garc{\'{\i}}a-Feal},
  \bibinfo{author}{J.~Gonz{\'{a}}lez-Cao},
  \bibinfo{author}{M.~G{\'{o}}mez-Gesteira}, \bibinfo{author}{L.~Cea},
  \bibinfo{author}{J.~Dom{\'{\i}}nguez}, \bibinfo{author}{A.~Formella},
  \bibinfo{title}{An Accelerated Tool for Flood Modelling Based on Iber},
  \bibinfo{journal}{Water} \bibinfo{volume}{10}~(\bibinfo{number}{10})
  (\bibinfo{year}{2018}) \bibinfo{pages}{1459},
  \doi{\bibinfo{doi}{10.3390/w10101459}}.

\bibitem[{Reguly et~al.(2016)Reguly, Mudalige, Bertolli, Giles, Betts, Kelly,
  and Radford}]{Regulyb}
\bibinfo{author}{I.~Z. Reguly}, \bibinfo{author}{G.~R. Mudalige},
  \bibinfo{author}{C.~Bertolli}, \bibinfo{author}{M.~B. Giles},
  \bibinfo{author}{A.~Betts}, \bibinfo{author}{P.~H. Kelly},
  \bibinfo{author}{D.~Radford}, \bibinfo{title}{{Acceleration of a Full-Scale
  Industrial CFD Application with OP2}}, \bibinfo{journal}{IEEE Transactions on
  Parallel and Distributed Systems} \bibinfo{volume}{27}~(\bibinfo{number}{5})
  (\bibinfo{year}{2016}) \bibinfo{pages}{1265--1278}, ISSN
  \bibinfo{issn}{1045-9219}, \doi{\bibinfo{doi}{10.1109/TPDS.2015.2453972}}.

\bibitem[{Giles et~al.(2012)Giles, Mudalige, Sharif, Markall, and
  Kelly}]{Giles2012}
\bibinfo{author}{M.~B. Giles}, \bibinfo{author}{G.~R. Mudalige},
  \bibinfo{author}{Z.~Sharif}, \bibinfo{author}{G.~Markall},
  \bibinfo{author}{P.~H.~J. Kelly}, \bibinfo{title}{{Performance Analysis and
  Optimization of the OP2 Framework on Many-Core Architectures}},
  \bibinfo{journal}{The Computer Journal}
  \bibinfo{volume}{55}~(\bibinfo{number}{2}) (\bibinfo{year}{2012})
  \bibinfo{pages}{168--180}, ISSN \bibinfo{issn}{0010-4620},
  \doi{\bibinfo{doi}{10.1093/comjnl/bxr062}}.

\bibitem[{Reguly et~al.(2018)Reguly, Giles, Gopinathan, Quivy, Beck, Giles,
  Guillas, and Dias}]{reguly2018}
\bibinfo{author}{I.~Z. Reguly}, \bibinfo{author}{D.~Giles},
  \bibinfo{author}{D.~Gopinathan}, \bibinfo{author}{L.~Quivy},
  \bibinfo{author}{J.~H. Beck}, \bibinfo{author}{M.~B. Giles},
  \bibinfo{author}{S.~Guillas}, \bibinfo{author}{F.~Dias}, \bibinfo{title}{The
  VOLNA-OP2 tsunami code (version 1.5)}, \bibinfo{journal}{Geoscientific Model
  Development} \bibinfo{volume}{11}~(\bibinfo{number}{11})
  (\bibinfo{year}{2018}) \bibinfo{pages}{4621--4635}.

\bibitem[{Giles et~al.(2020)Giles, Kashdan, Salmanidou, Guillas, and
  Dias}]{giles2020}
\bibinfo{author}{D.~Giles}, \bibinfo{author}{E.~Kashdan},
  \bibinfo{author}{D.~M. Salmanidou}, \bibinfo{author}{S.~Guillas},
  \bibinfo{author}{F.~Dias}, \bibinfo{title}{Performance analysis of
  Volna-OP2--massively parallel code for tsunami modelling},
  \bibinfo{journal}{Computers \& Fluids} \bibinfo{volume}{209}
  (\bibinfo{year}{2020}) \bibinfo{pages}{104649}.

\bibitem[{Beckers et~al.(2020)Beckers, Heredia, Noack, Nowak, Wieprecht, and
  Oladyshkin}]{beckers2020}
\bibinfo{author}{F.~Beckers}, \bibinfo{author}{A.~Heredia},
  \bibinfo{author}{M.~Noack}, \bibinfo{author}{W.~Nowak},
  \bibinfo{author}{S.~Wieprecht}, \bibinfo{author}{S.~Oladyshkin},
  \bibinfo{title}{Bayesian Calibration and Validation of a Large-Scale and
  Time-Demanding Sediment Transport Model}, \bibinfo{journal}{Water Resources
  Research} \bibinfo{volume}{56}~(\bibinfo{number}{7}) (\bibinfo{year}{2020})
  \bibinfo{pages}{e2019WR026966}.

\bibitem[{Jung and Merwade(2015)}]{jung2015}
\bibinfo{author}{Y.~Jung}, \bibinfo{author}{V.~Merwade},
  \bibinfo{title}{Estimation of uncertainty propagation in flood inundation
  mapping using a 1-D hydraulic model}, \bibinfo{journal}{Hydrological
  Processes} \bibinfo{volume}{29}~(\bibinfo{number}{4}) (\bibinfo{year}{2015})
  \bibinfo{pages}{624--640}.

\bibitem[{Peter(2017)}]{Peter2017}
\bibinfo{author}{S.~J. Peter}, \bibinfo{title}{Dam Break Analysis under
  Uncertainty}, Ph.D. thesis, \bibinfo{school}{ETH Zurich},
  \bibinfo{address}{Zurich, Switzerland},
  \doi{\bibinfo{doi}{10.3929/ethz-b-000209879}}, \bibinfo{year}{2017}.

\bibitem[{Armanini(2018)}]{armanini2018}
\bibinfo{author}{A.~Armanini}, \bibinfo{title}{Principles of river hydraulics},
  \bibinfo{publisher}{Springer}, \bibinfo{year}{2018}.

\bibitem[{Graf(1966)}]{graf1966}
\bibinfo{author}{W.~H. Graf}, \bibinfo{title}{On the determination of the
  roughness coefficient in natural and artificial waterways},
  \bibinfo{journal}{Hydrological Sciences Journal}
  \bibinfo{volume}{11}~(\bibinfo{number}{1}) (\bibinfo{year}{1966})
  \bibinfo{pages}{59--68}.

\bibitem[{Bezzola(2002)}]{Bezzola2002}
\bibinfo{author}{G.~R. Bezzola}, \bibinfo{title}{{Fliesswiederstand und
  Sohlenstabilit{\"{a}}t nat{\"{u}}rlicher Gerinne}}, Ph.D. thesis,
  \bibinfo{school}{Eidgen{\"{o}}ssische Technische Hochschule Z{\"{u}}rich},
  \bibinfo{year}{2002}.

\bibitem[{Exner(1925)}]{Exner1925}
\bibinfo{author}{F.~M. Exner}, \bibinfo{title}{{Ueber die Wechselwirkung
  zwischen Wasser und Geschiebe in Fluessen}}, \bibinfo{type}{Tech. Rep.},
  \bibinfo{institution}{Akademie der Wissenschaften, Mathematische
  Naturwissenschaft Abt. IIa, Wien, Austria}, \bibinfo{year}{1925}.

\bibitem[{Parker(2004)}]{parkerEbook}
\bibinfo{author}{G.~Parker}, \bibinfo{title}{1D sediment transport
  morphodynamics with applications to rivers and turbidity currents: E-book},
  \bibinfo{journal}{Minneapolis, MN, available from:
  {http://hydrolab.illinois.edu/people/parkerg/morphodynamics\_e-book.htm}} .

\bibitem[{Engelund and Hansen(1972)}]{Engelund1972}
\bibinfo{author}{F.~Engelund}, \bibinfo{author}{E.~Hansen}, \bibinfo{title}{A
  monograph on sediment transport in alluvial streams},
  \bibinfo{publisher}{Teknisk Forlag, Copenhagen}, \bibinfo{year}{1972}.

\bibitem[{Meyer-Peter and M{\"u}ller(1948)}]{MPM1948}
\bibinfo{author}{E.~Meyer-Peter}, \bibinfo{author}{R.~M{\"u}ller},
  \bibinfo{title}{Formulas for bed-load transport}, in:
  \bibinfo{booktitle}{IAHSR 2nd meeting, Stockholm, appendix 2},
  \bibinfo{organization}{IAHR}, \bibinfo{pages}{--}, \bibinfo{year}{1948}.

\bibitem[{Wong and Parker(2006)}]{wong2006}
\bibinfo{author}{M.~Wong}, \bibinfo{author}{G.~Parker},
  \bibinfo{title}{Reanalysis and correction of bed-load relation of Meyer-Peter
  and M{\"u}ller using their own database}, \bibinfo{journal}{Journal of
  Hydraulic Engineering} \bibinfo{volume}{132}~(\bibinfo{number}{11})
  (\bibinfo{year}{2006}) \bibinfo{pages}{1159--1168}.

\bibitem[{Grass(1981)}]{Grass1981}
\bibinfo{author}{A.~J. Grass}, \bibinfo{title}{{Sediment transport by waves and
  currents}}, \bibinfo{publisher}{University College, London, Dept. of Civil
  Engineering}, \bibinfo{year}{1981}.

\bibitem[{Smart and Jaeggi(1983)}]{Smart1983}
\bibinfo{author}{G.~M. Smart}, \bibinfo{author}{M.~N.~R. Jaeggi},
  \bibinfo{title}{{Sediment Transport on Steep Slopes}},
  \bibinfo{type}{VAW-Mitteilung} \bibinfo{number}{64},
  \bibinfo{institution}{Versuchsanstalt f{\"{u}}r Wasserbau,Hydrologie und
  Glaziologie (VAW). Z{\"{u}}rich, ETH Z{\"{u}}rich.}, \bibinfo{year}{1983}.

\bibitem[{Shields(1936)}]{Shields1936}
\bibinfo{author}{A.~Shields}, \bibinfo{title}{{Anwendungen der
  {\"{A}}hnlichkeitsmechanik und der Turbulenzforschung auf die
  Geschiebebewegungen}}, \bibinfo{type}{Tech. Rep.},
  \bibinfo{institution}{Mitteilung der Preussischen Versuchsanstalt f{\"{u}}r
  Wasserbau und Schiffbau. Berlin, Deutschland}, \bibinfo{year}{1936}.

\bibitem[{van Rijn(1989)}]{VanRijn1989}
\bibinfo{author}{L.~C. van Rijn}, \bibinfo{title}{{Handbook Sediment Transport
  by Current and Waves}}, \bibinfo{publisher}{Delft Hydraulics Laboratory},
  \bibinfo{address}{Delft, The Netherlands}, \bibinfo{year}{1989}.

\bibitem[{Chen et~al.(2010)Chen, Ma, and Dey}]{Chen2010}
\bibinfo{author}{X.~Chen}, \bibinfo{author}{J.~Ma}, \bibinfo{author}{S.~Dey},
  \bibinfo{title}{{Sediment Transport on Arbitrary Slopes: Simplified Model}},
  \bibinfo{journal}{Journal of Hydraulic Engineering}
  \bibinfo{volume}{136}~(\bibinfo{number}{5}) (\bibinfo{year}{2010})
  \bibinfo{pages}{311--317}, ISSN \bibinfo{issn}{0733-9429},
  \doi{\bibinfo{doi}{10.1061/(ASCE)HY.1943-7900.0000175}}.

\bibitem[{Ikeda(1982)}]{Ikeda82r}
\bibinfo{author}{S.~Ikeda}, \bibinfo{title}{{Lateral bed-load transport on side
  slopes}}, \bibinfo{journal}{Journal of the Hydraulics Division, ASCE}
  \bibinfo{volume}{108}~(\bibinfo{number}{11}) (\bibinfo{year}{1982})
  \bibinfo{pages}{1369--1373}.

\bibitem[{Talmon et~al.(1995)Talmon, Struiksma, and {Van Mierlo}}]{Talmon1995}
\bibinfo{author}{A.~M. Talmon}, \bibinfo{author}{N.~Struiksma},
  \bibinfo{author}{M.~{Van Mierlo}}, \bibinfo{title}{{Laboratory measurements
  of the direction of sediment transport on transverse alluvial-bed slopes}},
  \bibinfo{journal}{Journal of Hydraulic Research}
  \bibinfo{volume}{33}~(\bibinfo{number}{4}) (\bibinfo{year}{1995})
  \bibinfo{pages}{495--517}, ISSN \bibinfo{issn}{0022-1686},
  \doi{\bibinfo{doi}{10.1080/00221689509498657}}.

\bibitem[{Engelund(1974)}]{Engelund1974}
\bibinfo{author}{F.~Engelund}, \bibinfo{title}{{Flow and bed topography in
  channel bends.}}, \bibinfo{journal}{Journal of the Hydraulics Division ASCE}
  \bibinfo{volume}{100}~(\bibinfo{number}{11}) (\bibinfo{year}{1974})
  \bibinfo{pages}{1631--1648}.

\bibitem[{Rozovskii(1961)}]{Rozovskii1961}
\bibinfo{author}{I.~L. Rozovskii}, \bibinfo{title}{{Flow of Water in Bends of
  Open Channels}}, \bibinfo{publisher}{Academy of Science of the Ukrainian
  S.S.R, Institute of Hydrology and Hydraulic Engineering},
  \bibinfo{year}{1961}.

\bibitem[{Toro et~al.(1994)Toro, Spruce, and Speares}]{Toro1994}
\bibinfo{author}{E.~F. Toro}, \bibinfo{author}{M.~Spruce},
  \bibinfo{author}{W.~Speares}, \bibinfo{title}{Restoration of the contact
  surface in the HLL-Riemann solver}, \bibinfo{journal}{Shock waves}
  \bibinfo{volume}{4}~(\bibinfo{number}{1}) (\bibinfo{year}{1994})
  \bibinfo{pages}{25--34}.

\bibitem[{Toro(2009)}]{Toro2009a}
\bibinfo{author}{E.~F. Toro}, \bibinfo{title}{{Riemann solvers and numerical
  methods for fluid dynamics.}}, \bibinfo{publisher}{Springer-Verlag GmbH},
  ISBN \bibinfo{isbn}{978-3-540-49834-6},
  \doi{\bibinfo{doi}{10.1007/978-3-540-49834-6}}, \bibinfo{year}{2009}.

\bibitem[{Vanzo(2015)}]{Vanzo2015}
\bibinfo{author}{D.~Vanzo}, \bibinfo{title}{{Eco-hydraulic quantification of
  hydropeaking and thermopeaking: development of modeling and assessment
  tools}}, Ph.D. thesis, \bibinfo{school}{University of Trento},
  \bibinfo{address}{Trento, Italy}, \bibinfo{year}{2015}.

\bibitem[{Duran et~al.(2013)Duran, Liang, and Marche}]{Duran2013}
\bibinfo{author}{A.~Duran}, \bibinfo{author}{Q.~Liang},
  \bibinfo{author}{F.~Marche}, \bibinfo{title}{{On the well-balanced numerical
  discretization of shallow water equations on unstructured meshes}},
  \bibinfo{journal}{Journal of Computational Physics} \bibinfo{volume}{235}
  (\bibinfo{year}{2013}) \bibinfo{pages}{565--586}, ISSN
  \bibinfo{issn}{00219991}, \doi{\bibinfo{doi}{10.1016/j.jcp.2012.10.033}}.

\bibitem[{Soares-Fraz{\~{a}}o and Zech(2011)}]{Soares2011}
\bibinfo{author}{S.~Soares-Fraz{\~{a}}o}, \bibinfo{author}{Y.~Zech},
  \bibinfo{title}{{HLLC scheme with novel wave-speed estimators appropriate for
  two-dimensional shallow-water flow on erodible bed}},
  \bibinfo{journal}{International Journal for Numerical Methods in Fluids}
  \bibinfo{volume}{66}~(\bibinfo{number}{8}) (\bibinfo{year}{2011})
  \bibinfo{pages}{1019--1036}, ISSN \bibinfo{issn}{02712091},
  \doi{\bibinfo{doi}{10.1002/fld.2300}}.

\bibitem[{Hervouet and Petitjean(1999)}]{hervouet1999}
\bibinfo{author}{J.-M. Hervouet}, \bibinfo{author}{A.~Petitjean},
  \bibinfo{title}{Malpasset dam-break revisited with two-dimensional
  computations}, \bibinfo{journal}{Journal of hydraulic research}
  \bibinfo{volume}{37}~(\bibinfo{number}{6}) (\bibinfo{year}{1999})
  \bibinfo{pages}{777--788}.

\bibitem[{Valiani et~al.(2002)Valiani, Caleffi, and Zanni}]{valiani2002}
\bibinfo{author}{A.~Valiani}, \bibinfo{author}{V.~Caleffi},
  \bibinfo{author}{A.~Zanni}, \bibinfo{title}{Case study: Malpasset dam-break
  simulation using a two-dimensional finite volume method},
  \bibinfo{journal}{Journal of Hydraulic Engineering}
  \bibinfo{volume}{128}~(\bibinfo{number}{5}) (\bibinfo{year}{2002})
  \bibinfo{pages}{460--472}.

\bibitem[{Singh et~al.(2011)Singh, Altinakar, and Ding}]{singh2011}
\bibinfo{author}{J.~Singh}, \bibinfo{author}{M.~S. Altinakar},
  \bibinfo{author}{Y.~Ding}, \bibinfo{title}{Two-dimensional numerical modeling
  of dam-break flows over natural terrain using a central explicit scheme},
  \bibinfo{journal}{Advances in Water Resources}
  \bibinfo{volume}{34}~(\bibinfo{number}{10}) (\bibinfo{year}{2011})
  \bibinfo{pages}{1366--1375}.

\bibitem[{Cordier et~al.(2011)Cordier, Le, and De~Luna}]{cordier2011}
\bibinfo{author}{S.~Cordier}, \bibinfo{author}{M.~H. Le},
  \bibinfo{author}{T.~M. De~Luna}, \bibinfo{title}{Bedload transport in shallow
  water models: Why splitting (may) fail, how hyperbolicity (can) help},
  \bibinfo{journal}{Advances in Water Resources}
  \bibinfo{volume}{34}~(\bibinfo{number}{8}) (\bibinfo{year}{2011})
  \bibinfo{pages}{980--989}.

\bibitem[{Bellal et~al.(2003)Bellal, Spinewine, Savary, and Zech}]{bellal2003}
\bibinfo{author}{M.~Bellal}, \bibinfo{author}{B.~Spinewine},
  \bibinfo{author}{C.~Savary}, \bibinfo{author}{Y.~Zech},
  \bibinfo{title}{Morphological evolution of steep-sloped river beds in the
  presence of a hydraulic jump: Experimental study}, in:
  \bibinfo{booktitle}{XXX IAHR Congress}, \bibinfo{organization}{Citeseer},
  \bibinfo{pages}{133--140}, \bibinfo{year}{2003}.

\bibitem[{Goutiere et~al.(2011)Goutiere, Soares-Fraz{\~{a}}o, and
  Zech}]{Goutiere2011}
\bibinfo{author}{L.~Goutiere}, \bibinfo{author}{S.~Soares-Fraz{\~{a}}o},
  \bibinfo{author}{Y.~Zech}, \bibinfo{title}{Dam-break flow on mobile bed in
  abruptly widening channel: experimental data}, \bibinfo{journal}{Journal of
  Hydraulic Research} \bibinfo{volume}{49}~(\bibinfo{number}{3})
  (\bibinfo{year}{2011}) \bibinfo{pages}{367--371},
  \doi{\bibinfo{doi}{10.1080/00221686.2010.548969}}.

\bibitem[{Juez et~al.(2014)Juez, Murillo, and Garc{\'\i}a-Navarro}]{juez2014}
\bibinfo{author}{C.~Juez}, \bibinfo{author}{J.~Murillo},
  \bibinfo{author}{P.~Garc{\'\i}a-Navarro}, \bibinfo{title}{A 2D weakly-coupled
  and efficient numerical model for transient shallow flow and movable bed},
  \bibinfo{journal}{Advances in Water Resources} \bibinfo{volume}{71}
  (\bibinfo{year}{2014}) \bibinfo{pages}{93--109}.

\bibitem[{Xia et~al.(2010)Xia, Lin, Falconer, and Wang}]{Xia2010}
\bibinfo{author}{J.~Xia}, \bibinfo{author}{B.~Lin}, \bibinfo{author}{R.~A.
  Falconer}, \bibinfo{author}{G.~Wang}, \bibinfo{title}{Modelling dam-break
  flows over mobile beds using a 2D coupled approach},
  \bibinfo{journal}{Advances in Water Resources}
  \bibinfo{volume}{33}~(\bibinfo{number}{2}) (\bibinfo{year}{2010})
  \bibinfo{pages}{171--183},
  \doi{\bibinfo{doi}{10.1016/j.advwatres.2009.11.004}}.

\bibitem[{Yen and Lee(1995)}]{yen1995}
\bibinfo{author}{C.-l. Yen}, \bibinfo{author}{K.~T. Lee}, \bibinfo{title}{Bed
  topography and sediment sorting in channel bend with unsteady flow},
  \bibinfo{journal}{Journal of Hydraulic Engineering}
  \bibinfo{volume}{121}~(\bibinfo{number}{8}) (\bibinfo{year}{1995})
  \bibinfo{pages}{591--599}.

\bibitem[{Kaveh et~al.(2019)Kaveh, Reisenb{\"u}chler, Lamichhane, Liepert,
  Nguyen, Bui, and Rutschmann}]{kaveh2019}
\bibinfo{author}{K.~Kaveh}, \bibinfo{author}{M.~Reisenb{\"u}chler},
  \bibinfo{author}{S.~Lamichhane}, \bibinfo{author}{T.~Liepert},
  \bibinfo{author}{N.~D. Nguyen}, \bibinfo{author}{M.~D. Bui},
  \bibinfo{author}{P.~Rutschmann}, \bibinfo{title}{A Comparative Study of
  Comprehensive Modeling Systems for Sediment Transport in a Curved Open
  Channel}, \bibinfo{journal}{Water} \bibinfo{volume}{11}~(\bibinfo{number}{9})
  (\bibinfo{year}{2019}) \bibinfo{pages}{1779}.

\bibitem[{Brufau et~al.(2002)Brufau, Vázquez-Cendón, and
  García-Navarro}]{Brufau2002}
\bibinfo{author}{P.~Brufau}, \bibinfo{author}{M.~E. Vázquez-Cendón},
  \bibinfo{author}{P.~García-Navarro}, \bibinfo{title}{A numerical model for
  the flooding and drying of irregular domains},
  \bibinfo{journal}{International Journal for Numerical Methods in Fluids}
  \bibinfo{volume}{39}~(\bibinfo{number}{3}) (\bibinfo{year}{2002})
  \bibinfo{pages}{247--275}, \doi{\bibinfo{doi}{10.1002/fld.285}}.

\end{thebibliography}


\begin{thebibliography}{15}
\providecommand{\natexlab}[1]{#1}
\providecommand{\url}[1]{\texttt{#1}}
\providecommand{\urlprefix}{URL }
\expandafter\ifx\csname urlstyle\endcsname\relax
  \providecommand{\doi}[1]{doi:\discretionary{}{}{}#1}\else
  \providecommand{\doi}[1]{doi:\discretionary{}{}{}\begingroup
  \urlstyle{rm}\url{#1}\endgroup}\fi
\providecommand{\bibinfo}[2]{#2}

\bibitem[{Soares-Fraz{\~{a}}o and Zech(2011)}]{Soares2011}
\bibinfo{author}{S.~Soares-Fraz{\~{a}}o}, \bibinfo{author}{Y.~Zech},
  \bibinfo{title}{{HLLC scheme with novel wave-speed estimators appropriate for
  two-dimensional shallow-water flow on erodible bed}},
  \bibinfo{journal}{International Journal for Numerical Methods in Fluids}
  \bibinfo{volume}{66}~(\bibinfo{number}{8}) (\bibinfo{year}{2011})
  \bibinfo{pages}{1019--1036}, ISSN \bibinfo{issn}{02712091},
  \doi{\bibinfo{doi}{10.1002/fld.2300}}.

\bibitem[{Meyer-Peter and M{\"u}ller(1948)}]{MPM1948}
\bibinfo{author}{E.~Meyer-Peter}, \bibinfo{author}{R.~M{\"u}ller},
  \bibinfo{title}{Formulas for bed-load transport}, in:
  \bibinfo{booktitle}{IAHSR 2nd meeting, Stockholm, appendix 2},
  \bibinfo{organization}{IAHR}, \bibinfo{pages}{--}, \bibinfo{year}{1948}.

\bibitem[{Grass(1981)}]{Grass1981}
\bibinfo{author}{A.~J. Grass}, \bibinfo{title}{{Sediment transport by waves and
  currents}}, \bibinfo{publisher}{University College, London, Dept. of Civil
  Engineering}, \bibinfo{year}{1981}.

\bibitem[{Engelund and Hansen(1972)}]{Engelund1972}
\bibinfo{author}{F.~Engelund}, \bibinfo{author}{E.~Hansen}, \bibinfo{title}{A
  monograph on sediment transport in alluvial streams},
  \bibinfo{publisher}{Teknisk Forlag, Copenhagen}, \bibinfo{year}{1972}.

\bibitem[{Smart and Jaeggi(1983)}]{Smart1983}
\bibinfo{author}{G.~M. Smart}, \bibinfo{author}{M.~N.~R. Jaeggi},
  \bibinfo{title}{{Sediment Transport on Steep Slopes}},
  \bibinfo{type}{VAW-Mitteilung} \bibinfo{number}{64},
  \bibinfo{institution}{Versuchsanstalt f{\"{u}}r Wasserbau,Hydrologie und
  Glaziologie (VAW). Z{\"{u}}rich, ETH Z{\"{u}}rich.}, \bibinfo{year}{1983}.

\bibitem[{van Rijn(1989)}]{VanRijn1989}
\bibinfo{author}{L.~C. van Rijn}, \bibinfo{title}{{Handbook Sediment Transport
  by Current and Waves}}, \bibinfo{publisher}{Delft Hydraulics Laboratory},
  \bibinfo{address}{Delft, The Netherlands}, \bibinfo{year}{1989}.

\bibitem[{Chen et~al.(2010)Chen, Ma, and Dey}]{Chen2010}
\bibinfo{author}{X.~Chen}, \bibinfo{author}{J.~Ma}, \bibinfo{author}{S.~Dey},
  \bibinfo{title}{{Sediment Transport on Arbitrary Slopes: Simplified Model}},
  \bibinfo{journal}{Journal of Hydraulic Engineering}
  \bibinfo{volume}{136}~(\bibinfo{number}{5}) (\bibinfo{year}{2010})
  \bibinfo{pages}{311--317}, ISSN \bibinfo{issn}{0733-9429},
  \doi{\bibinfo{doi}{10.1061/(ASCE)HY.1943-7900.0000175}}.

\bibitem[{Talmon et~al.(1995)Talmon, Struiksma, and {Van Mierlo}}]{Talmon1995}
\bibinfo{author}{A.~M. Talmon}, \bibinfo{author}{N.~Struiksma},
  \bibinfo{author}{M.~{Van Mierlo}}, \bibinfo{title}{{Laboratory measurements
  of the direction of sediment transport on transverse alluvial-bed slopes}},
  \bibinfo{journal}{Journal of Hydraulic Research}
  \bibinfo{volume}{33}~(\bibinfo{number}{4}) (\bibinfo{year}{1995})
  \bibinfo{pages}{495--517}, ISSN \bibinfo{issn}{0022-1686},
  \doi{\bibinfo{doi}{10.1080/00221689509498657}}.

\bibitem[{Ikeda(1982)}]{Ikeda82r}
\bibinfo{author}{S.~Ikeda}, \bibinfo{title}{{Lateral bed-load transport on side
  slopes}}, \bibinfo{journal}{Journal of the Hydraulics Division, ASCE}
  \bibinfo{volume}{108}~(\bibinfo{number}{11}) (\bibinfo{year}{1982})
  \bibinfo{pages}{1369--1373}.

\bibitem[{Engelund(1974)}]{Engelund1974}
\bibinfo{author}{F.~Engelund}, \bibinfo{title}{{Flow and bed topography in
  channel bends.}}, \bibinfo{journal}{Journal of the Hydraulics Division ASCE}
  \bibinfo{volume}{100}~(\bibinfo{number}{11}) (\bibinfo{year}{1974})
  \bibinfo{pages}{1631--1648}.

\bibitem[{Vanzo et~al.(2016)Vanzo, Siviglia, and Toro}]{Vanzo2016}
\bibinfo{author}{D.~Vanzo}, \bibinfo{author}{A.~Siviglia},
  \bibinfo{author}{E.~F. Toro}, \bibinfo{title}{{Pollutant transport by shallow
  water equations on unstructured meshes: Hyperbolization of the model and
  numerical solution via a novel flux splitting scheme}},
  \bibinfo{journal}{Journal of Computational Physics} \bibinfo{volume}{321}
  (\bibinfo{year}{2016}) \bibinfo{pages}{1--20}.

\bibitem[{Hirano(1971)}]{Hirano1971}
\bibinfo{author}{M.~Hirano}, \bibinfo{title}{River-bed degradation with
  armoring}, in: \bibinfo{booktitle}{Proceedings of the Japan society of civil
  engineers}, \bibinfo{number}{195}, \bibinfo{organization}{Japan Society of
  Civil Engineers}, \bibinfo{pages}{55--65}, \bibinfo{year}{1971}.

\bibitem[{Einstein(1950)}]{Einstein1950}
\bibinfo{author}{H.~A. Einstein}, \bibinfo{title}{The bedload function for
  sediment transport in open channel flows, Tech}, \bibinfo{journal}{Bull}
  \bibinfo{volume}{1026} (\bibinfo{year}{1950}) \bibinfo{pages}{1--71}.

\bibitem[{Bertoldi et~al.(2014)Bertoldi, Siviglia, Tettamanti, Toffolon,
  Vetsch, and Francalanci}]{bertoldi2014}
\bibinfo{author}{W.~Bertoldi}, \bibinfo{author}{A.~Siviglia},
  \bibinfo{author}{S.~Tettamanti}, \bibinfo{author}{M.~Toffolon},
  \bibinfo{author}{D.~Vetsch}, \bibinfo{author}{S.~Francalanci},
  \bibinfo{title}{{Modeling vegetation controls on fluvial morphological
  trajectories}}, \bibinfo{journal}{Geophysical Research Letters}
  \bibinfo{volume}{41}~(\bibinfo{number}{20}) (\bibinfo{year}{2014})
  \bibinfo{pages}{7167--7175}, ISSN \bibinfo{issn}{19448007},
  \doi{\bibinfo{doi}{10.1002/2014GL061666}}.

\bibitem[{Caponi et~al.(2020)Caponi, Vetsch, and Siviglia}]{Caponi2020}
\bibinfo{author}{F.~Caponi}, \bibinfo{author}{D.~F. Vetsch},
  \bibinfo{author}{A.~Siviglia}, \bibinfo{title}{A model study of the combined
  effect of above and below ground plant traits on the ecomorphodynamics of
  gravel bars}, \bibinfo{journal}{Scientific reports}
  \bibinfo{volume}{10}~(\bibinfo{number}{1}) (\bibinfo{year}{2020})
  \bibinfo{pages}{1--14}, \doi{\bibinfo{doi}{10.1038/s41598-020-74106-9}}.

\end{thebibliography}


\end{document}


\begin{frontmatter}

\title{\textit{Supplementary Material of:} \mytitle}


\author[eth,eawag]{Davide Vanzo\corref{cor1}}
\cortext[cor1]{corresponding author: vanzo@vaw.baug.ethz.ch}
\author[axpo]{Samuel Peter}
\author[tk]{Lukas Vonwiller}
\author[eth]{Matthias Buergler}
\author[sis]{Manuel Weberndorfer}
\author[unitn]{Annunziato Siviglia}
\author[eth]{Daniel Conde}
\author[eth]{David F. Vetsch}

\address[eth]{Laboratory of Hydraulics, Hydrology and Glaciology. ETH, Swiss Federal Institute of Technology, Z\"urich, Switzerland.}
\address[eawag]{Dept. of Surface Waters - Research and Management. Eawag, Swiss Federal Institute of Aquatic Science and Technology, Kastanienbaum, Switzerland.}
\address[unitn]{Dept. Civil, Environmental and Mechanical Engineering. University of Trento, Trento, Italy.}
\address[sis]{ID Scientific IT Services. ETH, Swiss Federal Institute of Technology, Z\"urich, Switzerland.}
\address[axpo]{Axpo Group, Baden, Switzerland.}
\address[tk]{TK CONSULT AG , Z\"urich, Switzerland.}

\end{frontmatter}

This document collects the supplementary material for the manuscript entitled  "\mytitle "

{\renewcommand\arraystretch{1.25}}
\begin{center}
\begin{longtable}{cll}
\caption{Description of available features in \BM\ v3.}\label{tab:features}\\ \hline
\multicolumn{1}{c}{\textbf{Module}} & \multicolumn{1}{l}{\textbf{Feature}} & \multicolumn{1}{l}{\textbf{Description} }\\ \hline 
\endfirsthead
\multicolumn{3}{c}{\tablename\ \thetable{} -- continued from previous page} \\\hline
\multicolumn{1}{c}{\textbf{Module}} & \multicolumn{1}{l}{\textbf{Feature}} & \multicolumn{1}{l}{\textbf{Description}}\\\hline 
\endhead
\hline\multicolumn{3}{c}{\textit{Continued on next page}}\\
\endfoot
\hline
\endlastfoot

\multirow{1}{*}{Hydrodynamics}& Numerical Solver & \multicolumn{1}{p{10cm}}{HLLC, with hydrostatic reconstruction based on modified states.}\\
\cline{2-3}
& Parameters & \multicolumn{1}{p{10cm}}{ \rom{1} CFL, \rom{2} Minimum water depth, \rom{3} Fluid density, \rom{4} Upper limit of the time step.}\\
\cline{2-3}
& Initial Conditions & \multicolumn{1}{p{10cm}}{ \rom{1} Dry, \rom{2} Continue; \rom{3} Region defined (regiondef) for water surface elevation or water depth, u and v.}\\
\cline{2-3}
& Boundary Conditions & \multicolumn{1}{p{10cm}}{\rom{1} WALL : inviscid (default). \rom{2} STANDARD: a) INFLOW: uniform (discharge\textsuperscript{*}, slope), froude (discharge, froude number), zhydrograph (water surface elevation); b) OUTFLOW: uniform (slope), zero\_gradient, weir (weir height, Poleni factor), hq\_relation (H-Q relation), dynamic wall (collapse time), zhydrograph (water surface elevation). \rom{3} INTERNAL: dynamic wall (collapse time), internal wall, hq\_relation (H-Q relation). \rom{4} LINKED: hq\_relation (H-Q relation), 2 way hq\_relation (2 H-Q relations, time lag, water surface elevation upstream and downstream), weir (weir height, Poleni factor), zhydrograph (water surface elevation), zhydrograph\_linked\_kinE (water surface elevation). *\textit{(in parenthesis required input data)}}\\
\cline{2-3}
& Friction & \multicolumn{1}{p{10cm}}{Semi-implicit Runge-Kutta 2\textsuperscript{nd} order integration. Closure types: Manning, Strickler, Chezy, Bezzola. All cells require a default (or index based) friction value.} \\
\cline{2-3}
& Flood & \multicolumn{1}{p{10cm}}{ Flood tracking of water front arrival time, maximum water depth, maximum flow velocity, maximum specific discharge, maximum bed shear stress (tracking time step).} \\
\cline{2-3} 
& External source & \multicolumn{1}{p{10cm}}{ \rom{1} Type: a) total (as discharge [\si{m^3/s}]), b) distributed (as rainfall [mm/h]); \rom{2} Sink behavior: a) Exact (as prescribed), b) Available (as prescribed or less), c) Infinity (as much as possible).} \\
\hline
\multirow{1}{*}{Morphodynamics}& Numerical Solver & \multicolumn{1}{p{10cm}}{HLL-type Approximate Riemann Solver (modified from \cite{Soares2011}).} \\
\cline{2-3}
& Parameters & \multicolumn{1}{p{10cm}}{\rom{1} Morphodynamic start time, \rom{2} Sediment porosity, \rom{3} Sediment density, \rom{4} Local bed gradient calculation method: a) Secondary mesh: plane defined by element centers of neighbouring elements (default), b) Area weighted: area weighted gradients to neighbouring elements.} \\
\cline{2-3}
& Initial Conditions & \multicolumn{1}{p{10cm}}{\rom{1} From mesh file, \rom{2} Continue (from previous simulation).} \\
\cline{2-3}
& Boundary Conditions & \multicolumn{1}{p{10cm}}{\rom{1} WALL: inviscid (default). \rom{2} STANDARD: a) INFLOW: equilibrium (reference bed elevation\textsuperscript{*}), sedimentograph (sediment discharge), transport capacity (boundary factor); b) OUTFLOW: equilibrium (reference bed elevation). \rom{3} LINKED: equilibrium\_linked. Different weighting scheme available for transport capacity and sedimentograph distribution: geometrical (default), wetted area, conveyance. *\textit{(in parenthesis required input data)}} \\
\cline{2-3}
& Bedload Closures & \multicolumn{1}{p{10cm}}{\rom{1} MPM-like \cite{MPM1948} (coefficient*, exponent, critical threshold), \rom{2} GRASS-like \cite{Grass1981} (coefficient, exponent, critical threshold), \rom{3} \citet{Engelund1972}, \rom{4} \citet{Smart1983} (critical threshold and characteristic sediment diameters $d_{30}$ and $d_{90}$). *\textit{(in parenthesis required input data)}} \\
\cline{2-3}
& Incipient Motion & \multicolumn{1}{p{10cm}}{\rom{1} Type: \citet{VanRijn1989} or \citet{Chen2010}, \rom{2} Angle of repose.} \\
\cline{2-3}
& Bedload direction & \multicolumn{1}{p{10cm}}{\rom{1} Lateral bed slope effect [e.g.  \cite{Talmon1995} or \cite{Ikeda82r}] (factor and exponent are adaptable), \rom{2} Curvature effect \cite{Engelund1974} (factor and update timestep are adaptable).}\\
\cline{2-3}
& Bed Material& \multicolumn{1}{p{10cm}}{\rom{1} Grain class diameter; \rom{2} Fixed bed elevation, e.g. due to the bed rock horizon: a) Region defined (fixed bed elevation relative to bed elevation), b) Mesh  (mesh file providing absolute fixed bed elevation).}\\
\cline{2-3}
& Gravitational Transport & \multicolumn{1}{p{10cm}}{Bedload transport due to slope failure: \rom{1} critical angle for dry material, \rom{2} critical angle for wet material, \rom{3} minimum bed elevation change, \rom{4} maximum rate of bed elevation change [m/s], \rom{5} update timestep.} \\
\cline{2-3}
& External Source & \multicolumn{1}{p{10cm}}{\rom{1} sediment input   over a region [m\textsuperscript{3}/s]; \rom{2} Sink behaviors: a) Exact (as prescribed),  b) Available (as prescribed or less), c) Infinity (as much as possible).} \\
\hline
\multirow{1}{*}{Scalar transport}
&  Numerical solver & \multicolumn{1}{p{10cm}}{HLLC Approximate Riemann Solver with SVT solver from \cite{Vanzo2016}.}\\
\cline{2-3}
& Parameters & \multicolumn{1}{p{10cm}}{\rom{1} Scalar transport start time, \rom{2} scalar diffusivity (each species) and \rom{3} relaxation time (global).} \\
\cline{2-3}
& Initial Conditions & \multicolumn{1}{p{10cm}}{ \rom{1} Zero, \rom{2} Constant, \rom{3} Continue and \rom{4} Region defined (regiondef) for each scalar concentration.} \\
\cline{2-3}
& Boundary Conditions & \multicolumn{1}{p{10cm}}{ \rom{1} WALL: inviscid (default), \rom{2} STANDARD (in parenthesis user-required data): a) INFLOW: discharge\_in (constant value or time series, evenly distributed or wet-area weighted), concentration\_in (constant value or time series), b) OUTFLOW: zero\_gradient\_out.} \\
\cline{2-3}
& External Source & \multicolumn{1}{p{10cm}}{\rom{1} scalar input over a region (constant value or time series, discharge or concentration), \rom{2} Sink behaviors: a) Exact (as prescribed), b) Available (as prescribed or less), c) Infinity (as much as possible).} \\
\end{longtable}
\end{center}

\clearpage

\begin{center}
\begin{longtable}{cll}
\caption{Overview of features under development.}\label{tab:features_dev}\\ \hline
\multicolumn{1}{c}{\textbf{Module}} & \multicolumn{1}{l}{\textbf{Feature}} & \multicolumn{1}{l}{\textbf{Description} }\\ \hline 
\endfirsthead
\multicolumn{3}{c}{\tablename\ \thetable{} -- continued from previous page} \\\hline
\multicolumn{1}{c}{\textbf{Module}} & \multicolumn{1}{l}{\textbf{Feature}} & \multicolumn{1}{l}{\textbf{Description}}\\\hline 
\endhead
\hline\multicolumn{3}{c}{\textit{Continued on next page}}\\
\endfoot
\hline
\endlastfoot

\multirow{1}{*}{Hydrodynamics} & Turbulence model & \multicolumn{1}{p{10cm}}{Simplified constitutive equations to predict the statistical evolution of depth-averaged turbulent flows by modelling the Reynolds tensor with the mixing-length, $\kappa$-$\epsilon$ and ASM models.}\\
\hline
\multirow{1}{*}{Morphodynamics} & Non-uniform transport & \multicolumn{1}{p{10cm}}{Implementation of a solver for the Hirano-Exner-model \cite{Hirano1971}, together with hiding and exposure effects \cite{Einstein1950}.}\\
\cline{2-3}
& Suspended sediment & \multicolumn{1}{p{10cm}}{Advection-diffusion equation solver based on a relaxation approach \citep{Vanzo2016} coupled with standard transport capacity relations.} \\
\hline
\multirow{1}{*}{Temperature}& Transport and mixing & \multicolumn{1}{p{10cm}}{ Advection-diffusion equation solver based on a relaxation approach \citep{Vanzo2016}. } \\
\cline{2-3}
& Thermal sources & \multicolumn{1}{p{10cm}}{Simulation of diffuse (e.g. atmospheric forcing terms) and local (e.g. point-release) external thermal sources.}\\
\hline
\multirow{1}{*}{Vegetation}& \multicolumn{1}{p{4.0cm}}{Hydro-morphodynamic effects} & \multicolumn{1}{p{10cm}}{Modification of friction term, bed shear stress, and threshold for bed load incipient motion, depending on vegetation characteristics \citep{bertoldi2014}.}\\
\cline{2-3}
& Vegetation dynamics & \multicolumn{1}{p{10cm}}{Implementation of a water table-dependent vegetation growth module and plant mortality mechanisms by uprooting and burial \citep{Caponi2020}.}\\
\end{longtable}
\end{center}


\bibliographystyle{elsarticle-num-names}
\bibliography{bib_BM3}